\documentclass[preprint,amsmath,amssymb,aps]{revtex4-1}
\usepackage{graphicx}
\usepackage{dcolumn}
\usepackage{bm}
\usepackage{subfigure}
\usepackage{multirow}
\usepackage{amsmath}
\usepackage{float}

\begin{document}

\preprint{APS/123-QED}

\title{A lattice Boltzmann method for binary fluids based on mass-conserved quasi-incompressible phase-field theory}

\author{Kang Yang}
\author{Zhaoli Guo}%
 \email{zlguo@hust.edu.cn}
\affiliation{%
 State Key Laboratory of Coal Combustion, Huazhong University of Science and Technology, Wuhan 430074, China}%
\date{\today}
\begin{abstract}
In this paper, a lattice Boltzmann equation (LBE) model is proposed for binary fluids based on a quasi-incompressible phase-field model [J. Shen et al, Comm. Comp. Phys. 13, 1045 (2013)]. Compared with the other incompressible LBE models based on the incompressible phase-field theory, the quasi-incompressible model conserves mass locally. A series of numerical simulations are performed to validate the proposed model, and comparisons with an incompressible LBE model [H. Liang et al, Phys. Rev. E 89, 053320 (2014)] are also carried out. It is shown that the proposed model can track the interface accurately, and the predictions by the quasi-incompressible and incompressible models agree qualitatively well as the distribution of chemical potential is uniform, otherwise differ significantly.
\begin{description}
\item[PACS numbers]
02.70.-c, 47.11.-j, 47.55.-t
\end{description}
\end{abstract}

\maketitle


\section{Introduction}
Multiphase fluid flows are ubiquitous in engineering problems and natural processes.
  Generally, a phase interface can be described by sharp interface approach \cite{Hirt,Tryggvason01,Sussman94} or diffuse interface approach \cite{Chen98,Jacqmin99,Ding,Shen,Lowengrub,He99,Zheng05,Lee05}. In the sharp interface approach, the fluid is separated into some sub-domains by sharp interfaces in which each sub-domain contains only one phase, and the fluid properties such as density and viscosity are discontinuous across the interfaces. On the contrary, in diffuse interface approach, the fluid is treated continuously in the whole domain and the fluid properties vary smoothly across interfaces. An attractive feature of the diffuse interface method is it can model the complex interfacial dynamics without explicitly tracking the interfaces, and this feature makes it an ideal basis for developing efficient numerical schemes.

   In the diffuse interface approach, usually a phase-field variable (or order parameter) is used to distinguish different phases. The variable takes two distinct constant values in the bulk regions of the two phases, respectively, and changes smoothly across the interface. Based on the phase-field variable and its gradient, the free-energy of the system can be modelled, from which one can obtain a transport equation for the order parameter. The dynamic change of the phase interface can then be described by this equation coupled with the governing equations of the flow.  In most of previous works on immiscible binary mixtures of incompressible fluids, the flow is usually assumed to be governed by the incompressible Navier-Stokes equations including the interfacial force.  However, as pointed out in \cite{Lowengrub}, the assumption that the mixture is incompressible in the whole region is inconsistent with the conservation of mass as the densities of the fluids are unequal. To remedy this physical problem, a quasi-incompressible phase-field model, which assumes the mixture is incompressible in bulk regions but compressible in the mixing layer, has been proposed \cite{Shen}.

A number of numerical schemes have been developed based on phase-field models including spectral methods \cite{Chen98,Chen13,Liu02}, finite element methods \cite{Chen00,Du08,Feng06} and LBE methods \cite{He99,Zheng05,Lee05}. Among these methods, the LBE method has received particular attentions due to some distinctive features \cite{SD}.
 The first phase-field LBE model was proposed by He \emph{et~al.}~\cite{He99} which adopts an order parameter to track the interface of two incompressible fluids. However, there exist some differences between the derived governing equations and the phase-field theory for incompressible two-phase flows \cite{Zu13,Li12}, and numerical instability can be produced for systems with a large density ratio. Later some improved LBE models based on phase-field theory have been developed from different viewpoints. For instance, in order to improve numerical stability, Lee and Lin \cite{Lee05} designed a three-stage discretization multiphase lattice Boltzmann (LB) scheme by discretizing the gradient terms in different manners before and after the steaming step. Later, Fakhari \emph{et al.} \cite{Fakhari} further generalized the model \cite{Lee05} by employing a multi-relaxation-time collision operator. Zheng \cite{Zheng05} and Zu \cite{Zu13} respectively proposed the modified LBE models in order to recover the correct Cahn-Hilliard (CH) equation.
  However, the extra terms in the both models will produce a large error in the interface capturing and the computation will become unstable as the dimensionless relaxation time equals to 1\cite{Zu13}. To overcome these problems, Liang \emph{et al.} \cite{Liang14} proposed a new LBE model by introducing a time-dependent source term in the evolution equation. Recently, Zheng \emph{et al.} \cite{Zheng15} presented an alternative model based on the kinetic theory to solve the problem. Although those LBE models \cite{Zheng05,Zu13,Zheng15} could recover the CH equation exactly, the recovered momentum equations are still inconsistent with the target momentum equations for the incompressible flows.  Li \emph{et al.} \cite{Li12}
noted this problem and proposed a correction method by introducing an artificial interfacial force.

All of the above LBE models were developed based on the incompressible phase-field models that do not conserve mass locally as the two fluids have different densities.
 In this work we aim to develop a LBE model based on the quasi-incompressible phase-field theory for two-phase flows, which can ensure the exact mass conservation.
 The rest of this paper is organized as follows. In Sec. 2, the quasi-incompressible phase field model is briefly reviewed, and a LBE model based on this theory is constructed in Sec. 3. In Sec. 4, some numerical simulations are carried out to validate the proposed model, and some comparisons with a recent incompressible LBE model are also made.
A brief summary is presented in section 5.

\section{Quasi-incompressible phase-field model}\label{phase}
In the phase field theory for a two-phase system, the thermodynamic behavior can be described by a free energy function with respect to an order parameter $\phi$
 \begin{equation}\label{FE}
 F(\phi) = \int_{\Omega} {[{\psi}(\phi) + \frac{\kappa }{2}|{\bm{\nabla}} \phi {|^2}]d{\Omega}},
 \end{equation}
 where $\phi$ is used to distinguish different phases, ${\psi}(\phi)$ is the bulk free-energy density, $\kappa$ is the coefficient of surface tension, and $\Omega$ is the control volume.

 For binary fluids, a double-well form of free-energy density \cite{Jacqmin99,Jacqmin2000} can be used
 \begin{equation}
 \psi (\phi ) = \beta {(\phi-\phi_A) ^2}{(\phi-\phi_B)^2},
 \end{equation}
 where $\phi_A$ and $\phi_B$ are the equilibrium values of the order parameters for fluids A and B, respectively, $\beta$ is a constant related to the interfacial thickness $W$ \cite{Jacqmin99,Jacqmin2000,YZu} and the surface tension $\sigma$  \cite{YZu,JSR},
 \begin{equation}
W = \frac{1}{{|{\phi _A} - {\phi _B}|}}\sqrt {\frac{{8\kappa }}{\beta }},
 \end{equation}
 and
 \begin{equation}\label{surface}
\sigma  = \frac{{|{\phi _A} - {\phi _B}{|^3}}}{6}\sqrt {2\kappa \beta }.
 \end{equation}
 With the bulk free energy, the chemical potential $\mu$ \cite{Jacqmin99,Jacqmin2000,YZu} can be obtained
 \begin{align}
{\mu} &= \frac{{\delta F}}{{\delta \phi }} = \frac{{\partial \psi }}{{\partial \phi }} - \kappa {\nabla ^2}\phi \notag\\
&= 4\beta (\phi-\phi_A) (\phi  - \phi_B)(\phi  - \frac{\phi_A+\phi_B}{2})-\kappa {\nabla ^2}\phi,
\end{align}
and the order-parameter profile across the equilibrium interface can be obtained by solving $\mu(\phi)=0$ \cite{YZu},
\begin{equation}
\phi (\zeta ) = \frac{{{\phi _A} + {\phi _B}}}{2} + \frac{{{\phi _A}
- {\phi _B}}}{2}\tanh \left( {\frac{{2\zeta }}{W}}\right),
\end{equation}
where $\zeta$ is the coordinate normal to the interface.
The evolution of the order parameter can be described by the Cahn-Hilliard (CH) equation \cite{Jacqmin99,Jacqmin2000,CH58,CH59}
\begin{equation}\label{CH}
\frac{{\partial \phi }}{{\partial t}}  + \bm{\nabla}  \cdot (\phi{\bf{u}}) = \bm{\nabla}  \cdot ( {\lambda\bm{\nabla} \mu } ),
\end{equation}
where $\lambda$ is the mobility coefficient and $\bf{u}$ is the fluid velocity.

In the incompressible phase-field model, the fluid is assumed to be incompressible everywhere, and the flow can be described by the incompressible Navier-Stokes equations with an interfacial force \cite{Ding,BHB},
\begin{equation}\label{orderin}
\frac{{\partial \phi }}{{\partial t}}  + \bm{\nabla}  \cdot (\phi{\bf{u}}) = \bm{\nabla}  \cdot ( {\lambda\bm{\nabla} \mu } ),
\end{equation}
\begin{equation}\label{M1NS}
 \rho \left( {\frac{{\partial {\bf{u}}}}{{\partial t}} + {\bf{u}} \cdot {\bm{\nabla}} {\bf{u}}} \right) =  - {\bm{\nabla}} p + {\bm{\nabla}} \cdot  \left[ {\rho \nu \left( {{\bm{\nabla}} {\bf{u}} + {\bm{\nabla}} {{\bf{u}}^T}} \right)} \right] + {\bf{F}},
\end{equation}
\begin{equation}\label{M1Mass}
{\bm{\nabla}} \cdot {\bf{u}} =0,
\end{equation}
with
\begin{equation}\label{density}
\rho  = \frac{{\phi  - {\phi _B}}}{{{\phi _A} - {\phi _B}}}({\rho _A} - {\rho _B}) + {\rho _B},
\end{equation}
where $\rho_A$ and $\rho_B$ are the densities of fluids A and B, respectively.
From Eqs.~\eqref{orderin}, \eqref{M1Mass} and \eqref{density}, we can obtain
\begin{equation}\label{massequation}
\frac{{\partial \rho }}{{\partial t}} + {\bm{\nabla}} \cdot \left( {\rho {\bf{u}}} \right) = \frac{{d\rho }}{{d\phi }}{\bm{\nabla}} \cdot \left[ {\lambda {\bm{\nabla}} \mu } \right],
\end{equation}
where
\begin{equation}
\frac{{d\rho }}{{d\phi }} = \frac{{{\rho _A} - {\rho _B}}}{{{\phi _A} - {\phi _B}}}.
\end{equation}
It is obvious that the mass conservation is constrained by the $d\rho/d\phi$ and ${\bm{\nabla}} \cdot \left[ {\lambda {\bm{\nabla}} \mu } \right]$.
 In generally, ${\bm{\nabla}} \cdot \left[ {\lambda {\bm{\nabla}} \mu } \right]$ is nonzero in the interfacial region. Hence, as long as $\rho_A \neq \rho_B$, the mass is not locally conserved in the incompressible phase-field model.

In the quasi-incompressible phase field model \cite{Shen13}, the governing equations are expressed as
\begin{equation}\label{order}
\frac{{\partial \phi }}{{\partial t}}  + \bm{\nabla}  \cdot (\phi{\bf{u}}) = \bm{\nabla}  \cdot ( {\lambda\bm{\nabla} \mu } ),
\end{equation}
\begin{equation}\label{quasi-momentum}
 \rho \left( {\frac{{\partial {\bf{u}}}}{{\partial t}} + {\bf{u}} \cdot {\bm{\nabla}} {\bf{u}}} \right) =  - {\bm{\nabla}} p + {\bm{\nabla}} \cdot  \left[ {\rho \nu \left( {{\bm{\nabla}} {\bf{u}} + {\bm{\nabla}} {{\bf{u}}^T}} \right)} \right] + {\bf{F}},
\end{equation}
\begin{equation}\label{diverge}
{\bm{\nabla}} \cdot {\bf{u}} =  - \gamma {\bm{\nabla}} \cdot \left[ {\lambda {\bm{\nabla}} \mu } \right],
\end{equation}
with
\begin{equation}\label{gamma}
\gamma  = \frac{{d\rho /d\phi }}{{\rho  - \phi d\rho /d\phi }}=\frac{{\rho _r - 1}}{{{\phi _A} - \phi_B\rho _r}},
\end{equation}
where $p$ is the hydrodynamic pressure, $\nu$ is the kinematic viscosity, ${{\bf{F}}}$ is the total force including  the surface tension force ${{\bf{F}}_s}(=-\phi{\bm{\nabla}}\mu)$ and other body forces ${{\bf{F}}_b}$, $\rho_r$ is the density ratio $\rho_A/\rho_B$.
From Eqs. \eqref{massequation}, \eqref{quasi-momentum} and \eqref{diverge}, one can obtain that
\begin{equation}
{\partial _t}\rho  + {\bm{\nabla}} \cdot (\rho {\bf{u}}) = 0,
\end{equation}
which means that the mass is conserved locally in the qusi-incompressible model. Furthermore, equation \eqref{diverge} suggests that the fluid is compressible in the mixing zone.
To investigate the effect of compressibility, we substitute Eq.~\eqref{diverge} into Eq.~\eqref{order} to get,
\begin{equation}\label{order-new}
\frac{{\partial \phi }}{{\partial t}} + {\bm{ u}}\cdot {\bm{\nabla}} \phi  = (1 + \gamma \phi ){\bm{\nabla}}\cdot(\lambda {\bm{\nabla}} \mu ).
\end{equation}
On the other hand, from Eqs.~\eqref{orderin} and \eqref{M1Mass} we can obtain
\begin{equation}\label{order-old}
\frac{{\partial \phi }}{{\partial t}} + {\bm{ u}}\cdot {\bm{\nabla}} \phi  = {\bm{\nabla}}\cdot(\lambda {\bm{\nabla}} \mu ).
\end{equation}
From Eqs.~\eqref{order-new} and \eqref{order-old}, we can see that the discrepancy between the two models is related to the term $\gamma \phi {\bm{\nabla}}\cdot(\lambda {\bm{\nabla}} \mu)$, which depends on the density ratio and
 the spatial distribution of the chemical potential. If the chemical potential is uniformly distributed, the additional term plays weak role in the results, otherwise the discrepancy between the two models is tremendous.
Eq.~\eqref{order-new} can be also expressed in dimensionless formulation as
\begin{equation}\label{nondimensional}
\begin{array}{l}
{\partial _t}\phi   + {\bf{u}}\cdot {\bm{\nabla}} \phi  = (1+\gamma \phi)P{e^{ - 1}}{\nabla ^2}\{ 4(\phi  - {\phi _A})(\phi  - {\phi _B})[\phi  - ({\phi _A} + {\phi _B})/2]\\\\
 - C{n^2}{({\phi _A} - {\phi _B})^2}{\nabla ^2}\phi /8\},
\end{array}
\end{equation}
where $Pe=U_cL_c/\lambda\beta$ is the Peclet number and $Cn=W/L_c$ is the Cahn number with the characteristic length $L_c$ and velocity $U_c$. This suggests that the distinctions between the two models are also related to the magnitudes of $Pe$ and $Cn$.
In the numerical simulations, we will investigate the difference between the two models by changing the dimensionless  parameters $Pe$, $Cn$ and $\gamma$.

\section{The quasi-incompressible LBE model}\label{LBM}
In this section, we will propose the LBE model based on the quasi-incompressible phase-field equations \cite{Shen13}.
The model consists of two LBEs, one for the CH equation, and one for the Navier-Stokes equations,
\begin{equation}\label{LB2}
{f _i}({\bf{x}} + {{\bf{c}}_i}{\delta t},t + {\delta t}) - {f _i}({\bf{x}},t) =  - \frac{1}{\tau_{f} }\left[ {{f _i}({\bf{x}},t) - f _i^{eq}({\bf{x}},t)} \right] + {\delta t}\left[ {1 - \frac{1}{{2{\tau _{f}}}}} \right]{F_i},
\end{equation}
\begin{equation}\label{LB1}
{\textrm g _i}({\bf{x}} + {{\bf{c}}_i}{\delta t},t + {\delta t}) - {\textrm g _i}({\bf{x}},t) =  - \frac{1}{\tau_{\textrm g} }\left[ {{\textrm g _i}({\bf{x}},t) - \textrm g _i^{eq}({\bf{x}},t)} \right] + {\delta t}\left[ {1 - \frac{1}{{2{\tau _{\textrm g}}}}} \right]{G_i},
\end{equation}
where ${f _i}({\bf{x}},t)$ and ${\textrm g _i}({\bf{x}},t)$ are the distribution functions for the hydrodynamics and order parameter fields, respectively, ${{\bf{c}}_i}$ is the discrete velocity in the $i$th direction, $\delta t$ is the time step, $\tau_{f}$ and $\tau_{\textrm g}$ are dimensionless  relaxation times related to the shear viscosity and mobility, respectively, $F_i$ and $G_i$ are the distribution functions for the force term, and $G_i$ is used for eliminating the extra term in the CH equation \cite{Chai13}.
The local equilibrium distribution functions ${f _i^{eq}}({\bf{x}},t)$ and ${\textrm g _i^{eq}}({\bf{x}},t)$
are respectively defined as
\begin{equation}\label{feq}
f_i^{eq} = {\omega _i}[p + c_s^2\rho {s_i}({\bf{u}})],
\end{equation}
\begin{equation}\label{geq}
\textrm g_i^{eq} = {H_i} + {\omega _i}\phi {s_i}({\bf{u}}),
\end{equation}
with
\begin{equation}
{s_i}({\bf{u}}) = \frac{{{{\bf{c}}_i} \cdot {\bf{u}}}}{{c_s^2}} + \frac{{{\bf{uu}}:({{\bf{c}}_i}{{\bf{c}}_i} - c_s^2{\bf{I}})}}{{2c_s^4}},
\end{equation}
\begin{equation}
{H_i} = \left\{ {\begin{array}{*{20}{c}}
{\phi  - (1 - {\omega _0})\alpha \mu, \;\;\;\;i = 0}\\
{{\omega _i}\alpha \mu, \;\;\;\;\;\;\;\;\;\;\;\;\;\;\;\;\;\;\;\;i \ne 0}
\end{array}} \right.
\end{equation}
where $\omega _i$ is the weighting coefficient, $D$ is the spatial dimension and ${c_s}$ is the sound speed for an ideal fluid, and $\alpha$ is an adjustable parameter for a given mobility.

In Eqs.~\eqref{LB2} and \eqref{LB1}, the source terms $F_i$ and $G_i$ are respectively given by
\begin{equation}
{F_i} = ({{\bf{c}}_i} - {\bf{u}}) \cdot \left[ {{\omega _i}{\bf{F}}{\Gamma _i}({\bf{u}}) + {\omega _i}{s_i}({\bf{u}})c_s^2{\bm{\nabla}} \rho } \right] - {\omega _i}c_s^2\rho \gamma {\bm{\nabla}}  \cdot (\lambda {\bm{\nabla}} \mu ),
\end{equation}
\begin{equation}\label{force}
{G_i} =  - \frac{\phi }{{c_s^2\rho }}\left( {{{\bf{c}}_i} - {\bf{u}}} \right)\cdot \left( {{\bm{\nabla}} p -{\bf{F}} } \right)\omega_i{\Gamma _i}({\bf{u}}).
\end{equation}
The macroscopic quantities, $\phi$, $\bf{u}$ and $p$, are computed evaluated as
\begin{equation}
\phi= \sum\limits_i {{\textrm g_i}},
\end{equation}
\begin{equation}\label{velocity-q}
{\bf{u}} = \frac{1}{{c_s^2\rho }}\left[ {\sum\limits_i {{{\bf{c}}_i}} {f_i} + \frac{{\delta t}}{2}c_s^2{\bf{F}}} \right],
\end{equation}
\begin{equation}\label{pressure-q}
p = \sum\limits_i {{f_i}}  + \frac{{\delta t}}{2}c_s^2\left( {{\bf{u}} \cdot {\bm{\nabla}} \rho  - \gamma \rho {\bm{\nabla}
} \cdot \left( {\lambda {\bm{\nabla}} \mu } \right)} \right).
\end{equation}
The kinetic viscosity $\nu$ and the mobility $\lambda$ are respectively given by
\begin{equation}
\nu  = c_s^2({\tau _f} - 0.5)\delta t,\;\;\;\;\;\;\;\;\;\lambda  = c_s^2({\tau _g} - 0.5)\alpha \delta t.
\end{equation}
Through the Chapman-Enskog analysis (see the Appendix for details), we can obtain the following macroscopic hydrodynamics equations
\begin{equation}
 \rho \left( {\frac{{\partial {\bf{u}}}}{{\partial t}} + {\bf{u}} \cdot {\bm{\nabla}} {\bf{u}}} \right) =  - {\bm{\nabla}} p + {\bm{\nabla}} \cdot  \left[ {\rho \nu \left( {{\bm{\nabla}} {\bf{u}} + {\bm{\nabla}} {{\bf{u}}^T}} \right)} \right] + {\bf{F}},
\end{equation}
\begin{equation}\label{divergence}
\frac{1}{{c_s^2\rho }}\frac{{\partial p}}{{\partial t}}+{\bm{\nabla}} \cdot {\bf{u}} =  - \gamma {\bm{\nabla}} \cdot \left[ {\lambda {\bm{\nabla}} \mu } \right],
\end{equation}
\begin{equation}
\frac{{\partial \phi }}{{\partial t}}  + \bm{\nabla}  \cdot (\phi{\bf{u}}) = \bm{\nabla}  \cdot ( {\lambda\bm{\nabla} \mu } ).
\end{equation}
In the limit of low Mach number ($Ma=|{\bf{u}}|/c_s$), the dynamic pressure is assumed to be $p\sim O(Ma^2)$, and the above set of equations reduce to quasi-incompressible model given by Eqs.~\eqref{order} to \eqref{diverge}.

In the present work, we consider two-dimensional cases, and the two-dimensional nine-velocity (D2Q9) LBE model is used without loss of generality, in which ${{\bf{c}}_0} = (0,0)$,
${{\bf{c}}_{i = 1 - 4}} = c\{ {\rm cos}[(i - 1)\pi /2],{\rm sin}[(i - 1)\pi /2]\}$, ${{\bf{c}}_{i = 5 - 8}} = \sqrt 2 c\{ {\rm cos}[(2i - 1)\pi /4],{\rm sin}[(2i - 1)\pi /4]\}$, and the corresponding weight coefficients are  $\omega_0=4/9$, $\omega_{1-4}=1/9$ and $\omega_{5-8}=1/36$.
The sound speed $c_s$ is given by $c_s=c/\sqrt{3}$, where $c = \delta x/\delta t$, with $\delta x$ representing the lattice space. For simplicity, we set the lattice space and time increment as the length and time units, i.e., $\delta x=\delta t=1$.
In the computations, the gradient operators are discretized with the isotropic central scheme \cite{Guo11}.
\section{Numerical simulations}
In this section, we will validate the accuracy of the proposed quasi-incompressible LBM, and compare it with a recent LBE model based on the incompressible phase-field theory given in \cite{Liang14} by a series of numerical simulations.
\subsection{One-dimensional flat interface}
\begin{figure}[H] \centering
\includegraphics[width=0.35\textwidth]{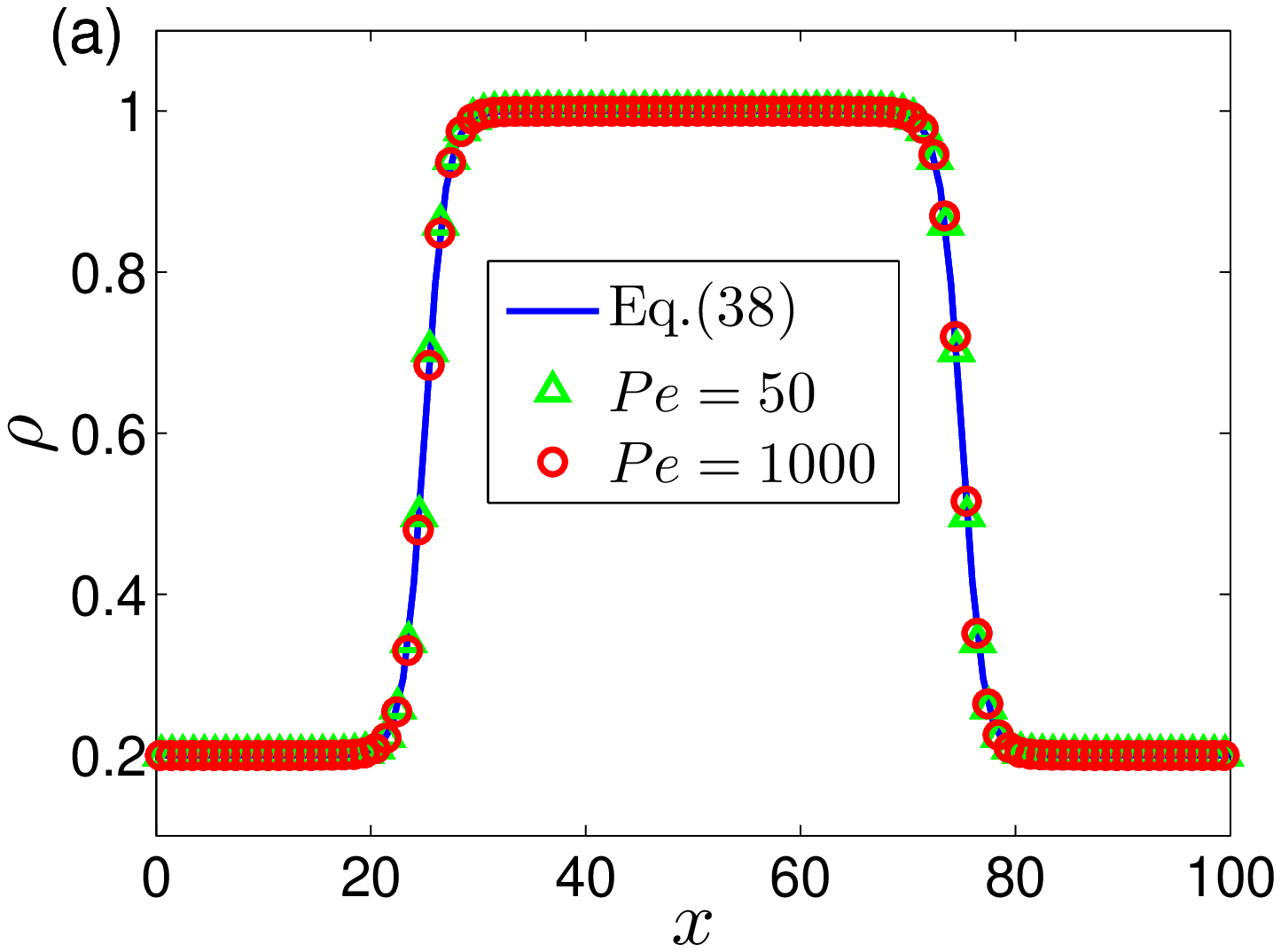}
\includegraphics[width=0.35\textwidth]{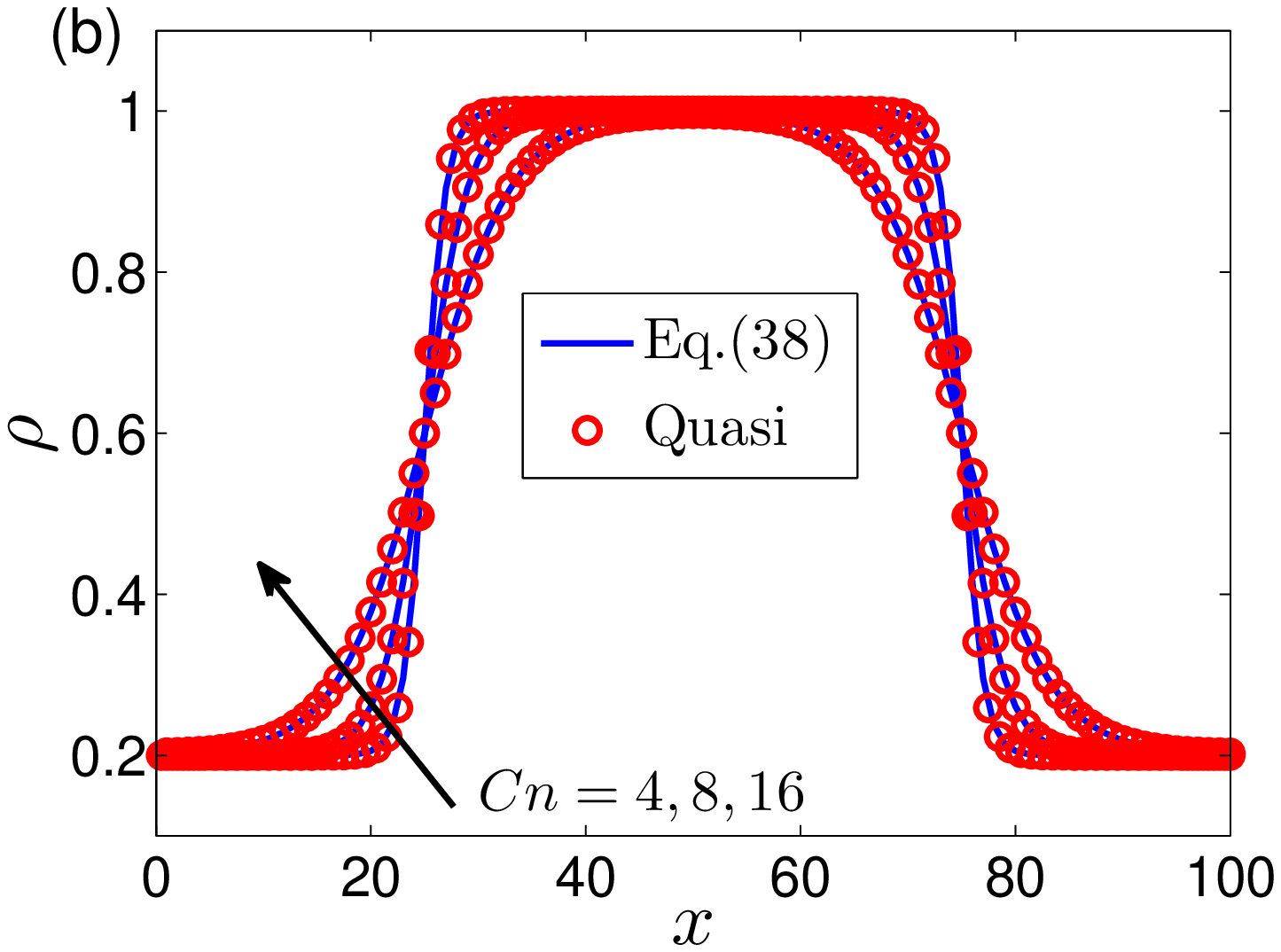}
\caption{(Color online) Density profiles across the interface with (a) different values of $Pe$ with $Cn=4$, and (b) different values of $Cn$ with $Pe=1000$.}\label{fig:figure1}
\end{figure}

 We firstly validate the proposed LBE model by a flat interface test.
Initially, the central region ($25\leq y \leq 75$) is filled with fluid A and the rest is occupied by fluid B. The order parameter and density profiles are respectively set to be at equilibrium, i.e.,
\begin{equation}
{\phi _0}(y) = \left\{ {\begin{array}{*{20}{c}}
{\frac{{{\phi _A} + {\phi _B}}}{2} + \frac{{{\phi _A} - {\phi _B}}}{2}\tanh y_1},\;\;\;y\leq50\\\\
{\frac{{{\phi _A} + {\phi _B}}}{2} - \frac{{{\phi _A} - {\phi _B}}}{2}\tanh y_2},\;\;\;y>50
\end{array}} \right.
\end{equation}
\begin{equation}
{\rho _0}(y) = \left\{ {\begin{array}{*{20}{c}}
{\frac{{{\rho _A} + {\rho _B}}}{2} + \frac{{{\rho _A} - {\rho _B}}}{2}\tanh y_1},\;\;\;y\leq50\\\\
{\frac{{{\rho _A} + {\rho _B}}}{2} - \frac{{{\rho _A} - {\rho _B}}}{2}\tanh y_2},\;\;\;y>50
\end{array}} \right.
\end{equation}
where $y_1=2(y-25)/W$ and $y_2=2(y-75)/W$.
The lattice used is $N_x \times N_y=10 \times 100$ and periodic boundary conditions are employed in both $x$ and $y$ directions.
The other parameters are fixed as $\rho_A=1$, $\rho_B=0.2$, $\tau_f=1$, $\tau_g=1$, $\phi_A=1$, $\phi_B=0$ and $\sigma=0.001$.
The effects of $Pe$ and $Cn$ numbers on the density distributions are investigated.
Note that the characteristic length and velocity in this paper take the values of lattice space $\delta x$ and velocity $c$, respectively.
Figure \ref{fig:figure1} shows the density distribution across the interface for the
quasi-incompressible (Quasi) model.
From the figure, it can be seen that the density profiles match the analytical profile which indicates the accuracy of the proposed model in the interface tracking.
\subsection{Stationary droplet}
\begin{figure}[H]\centering
\includegraphics[width=0.25\textwidth]{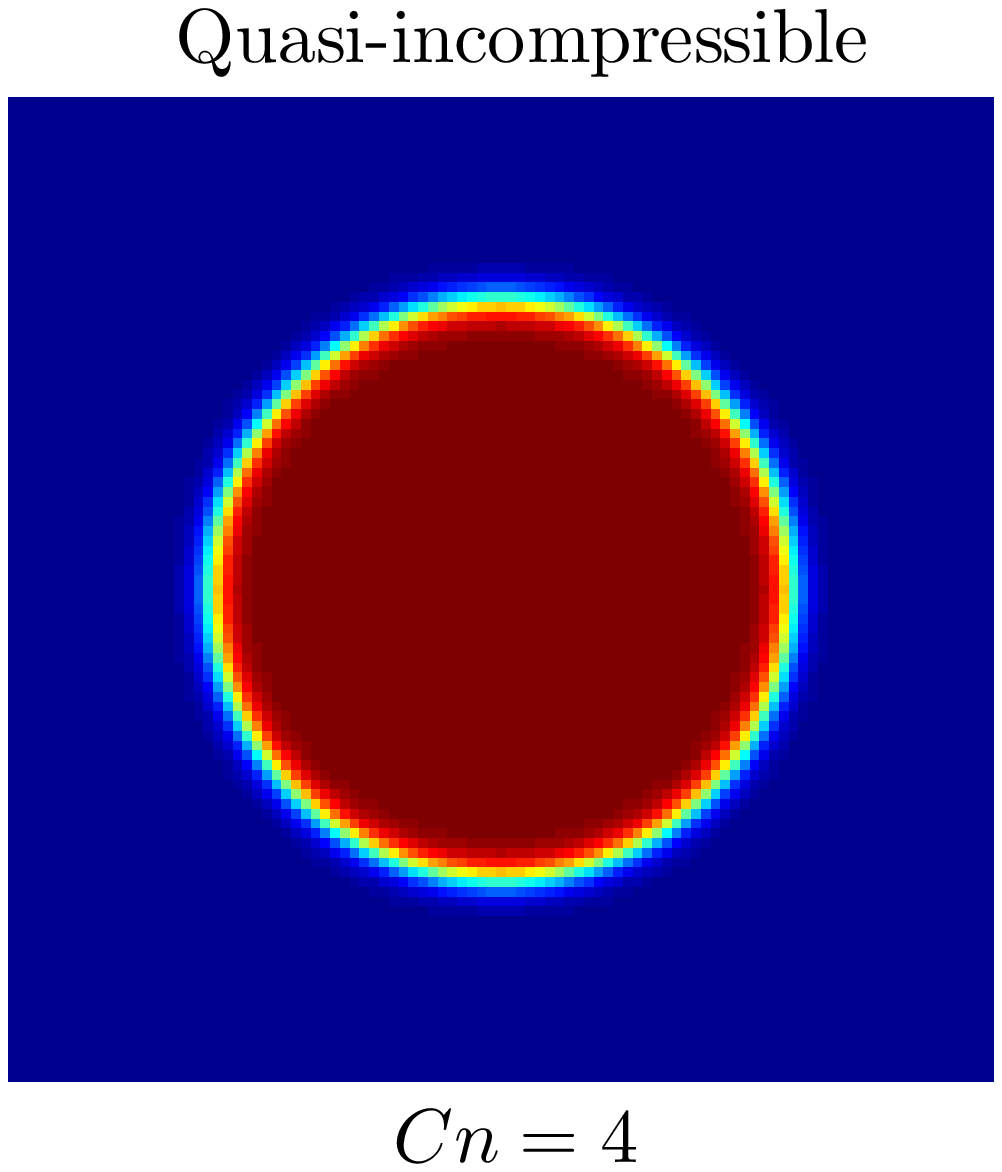}
\includegraphics[width=0.25\textwidth]{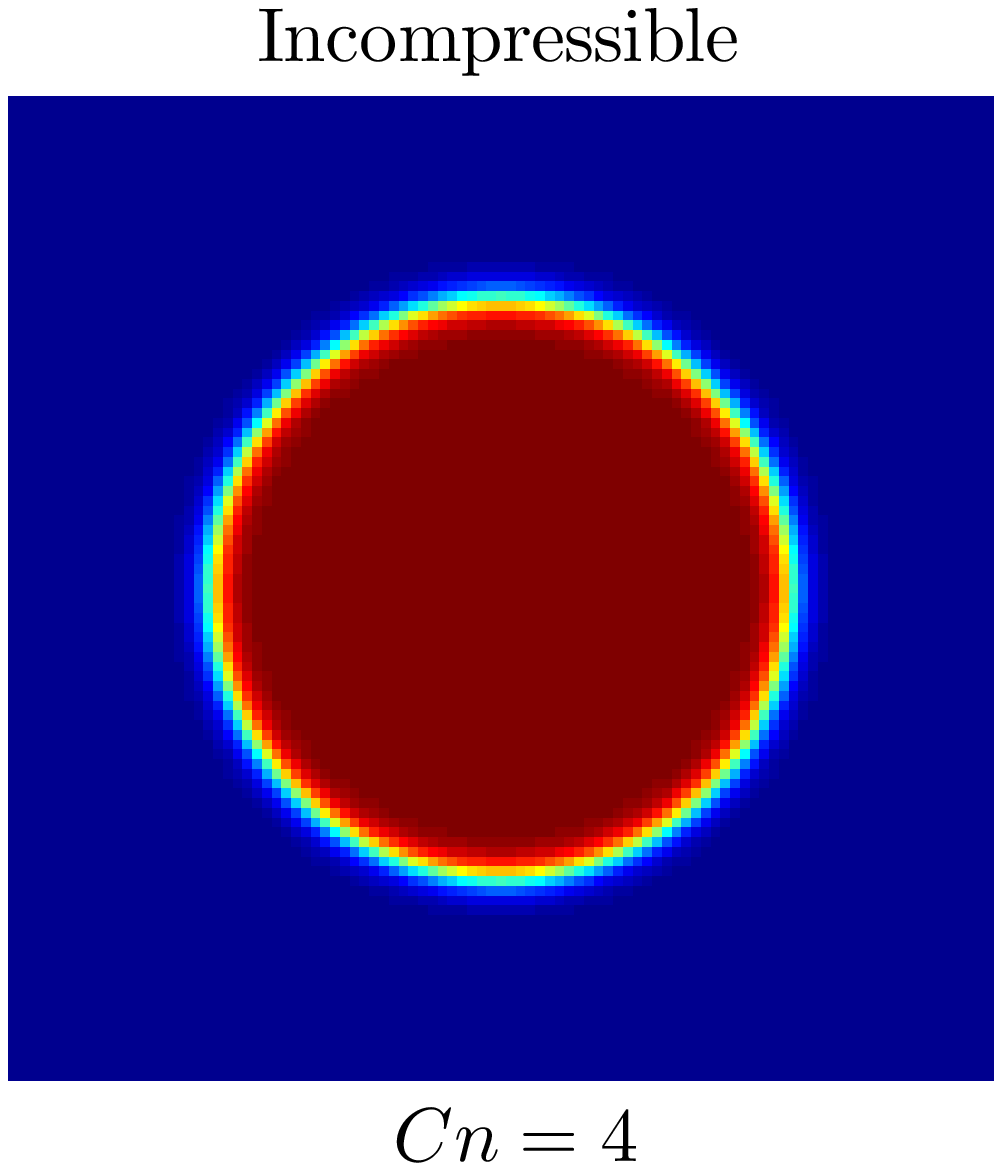}
\includegraphics[width=0.3\textwidth]{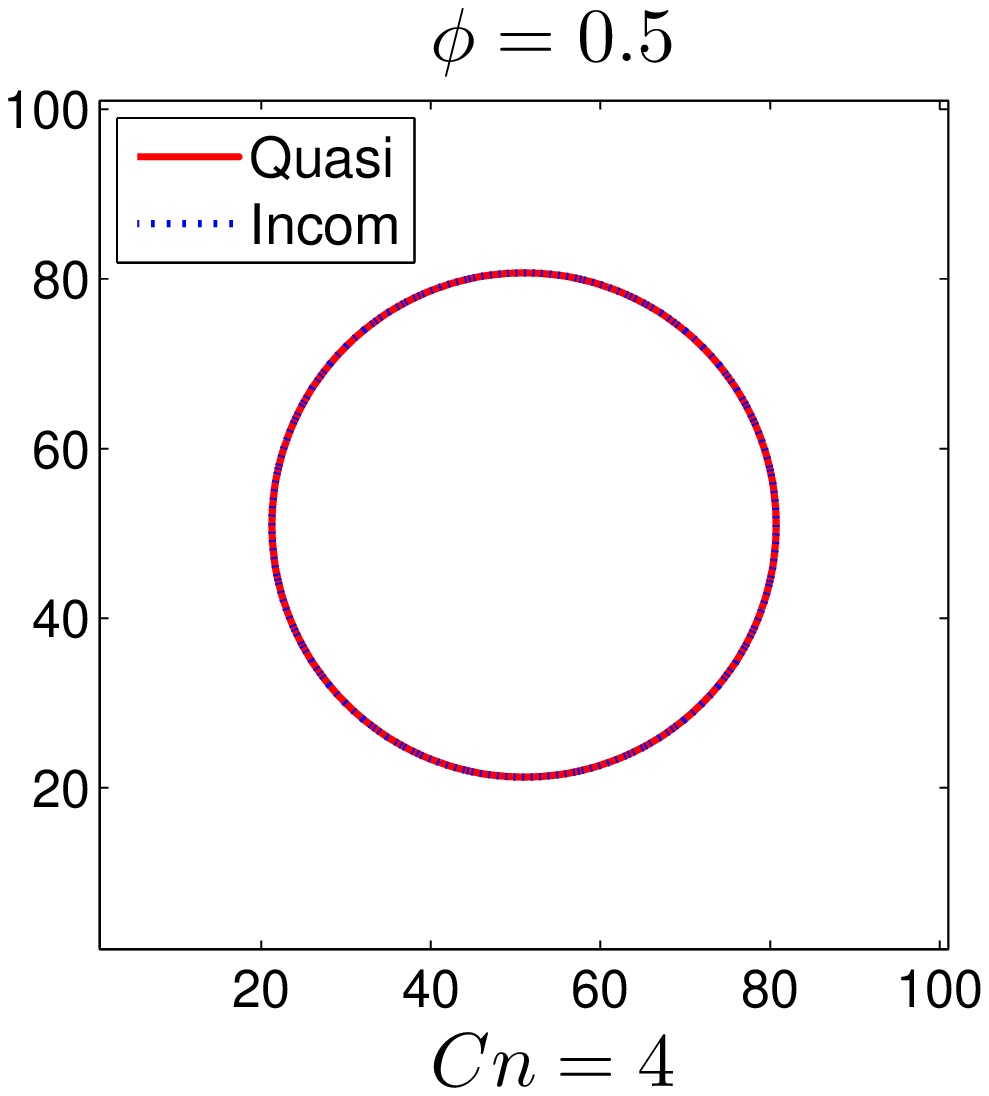}
\includegraphics[width=0.25\textwidth]{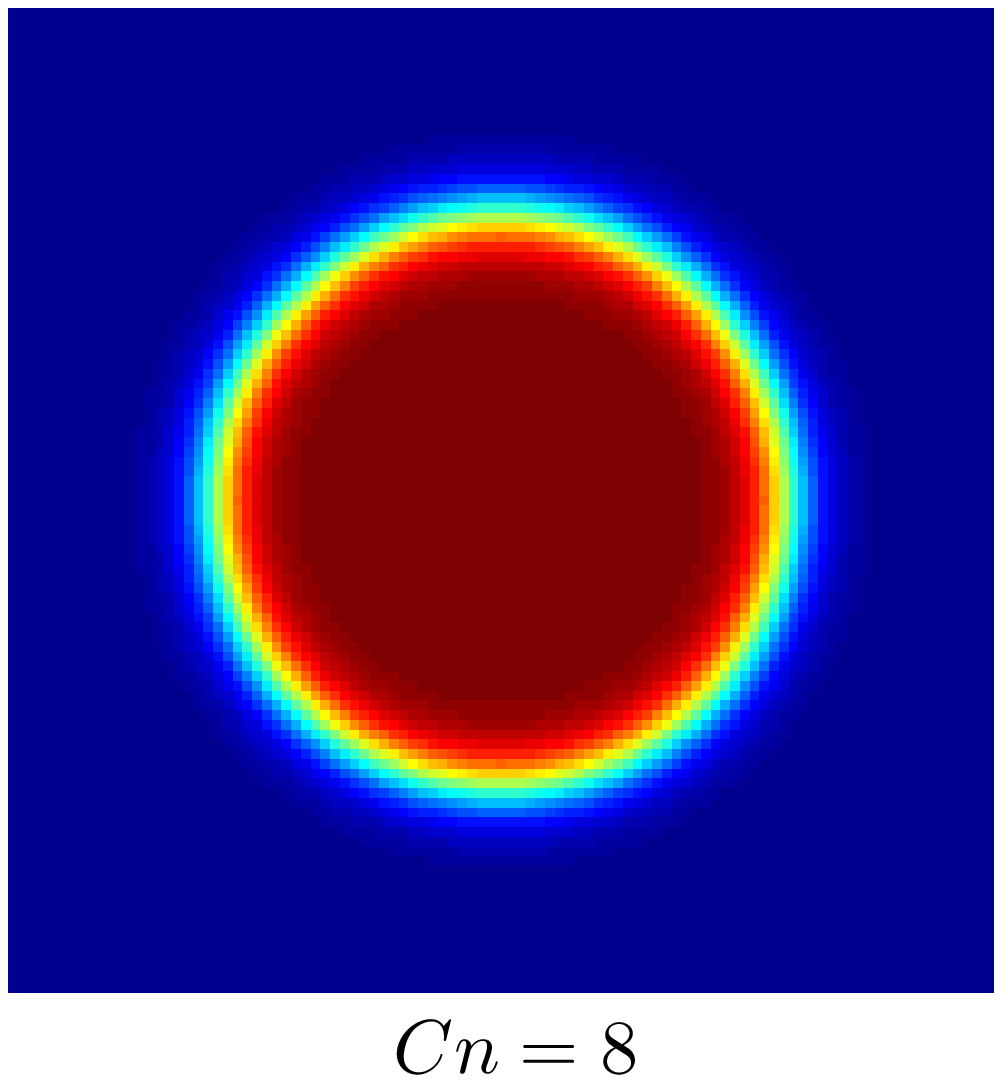}
\includegraphics[width=0.25\textwidth]{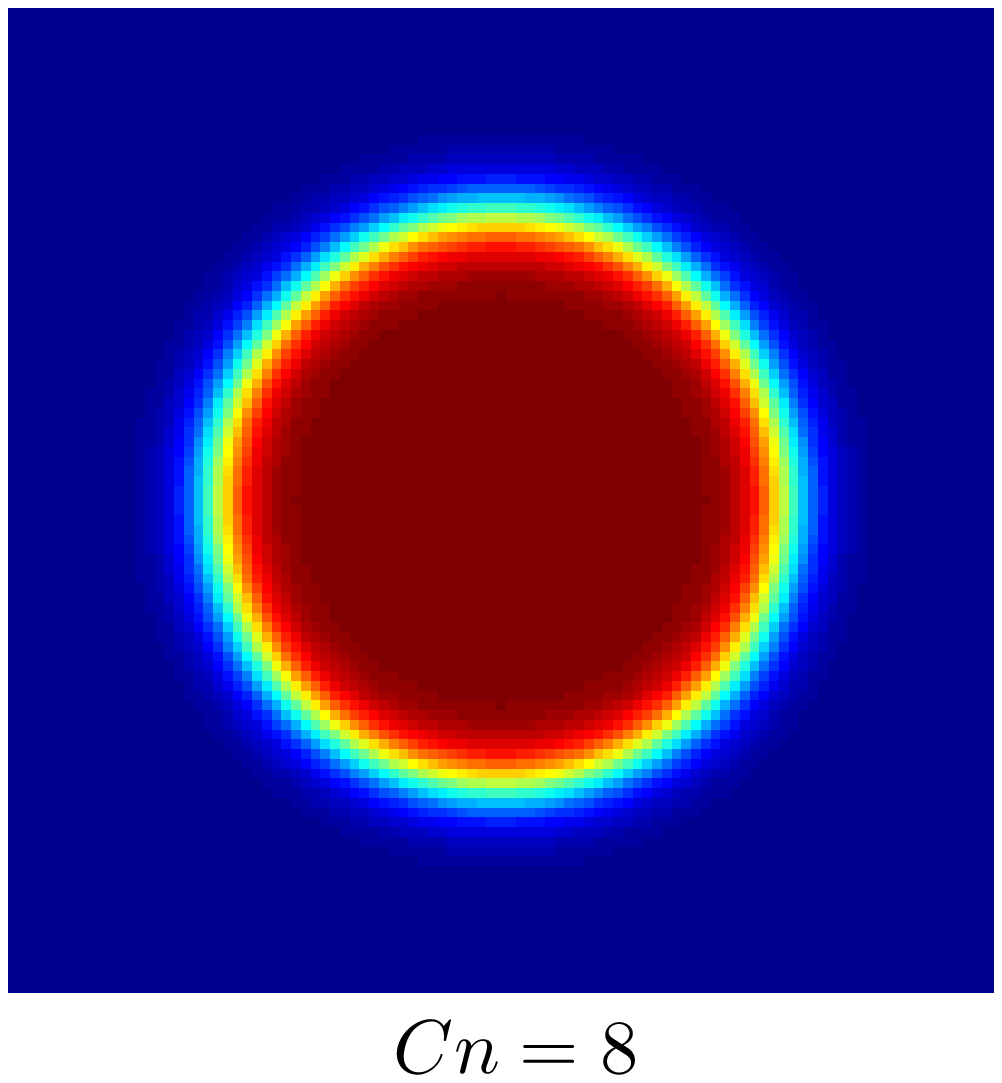}
\includegraphics[width=0.3\textwidth]{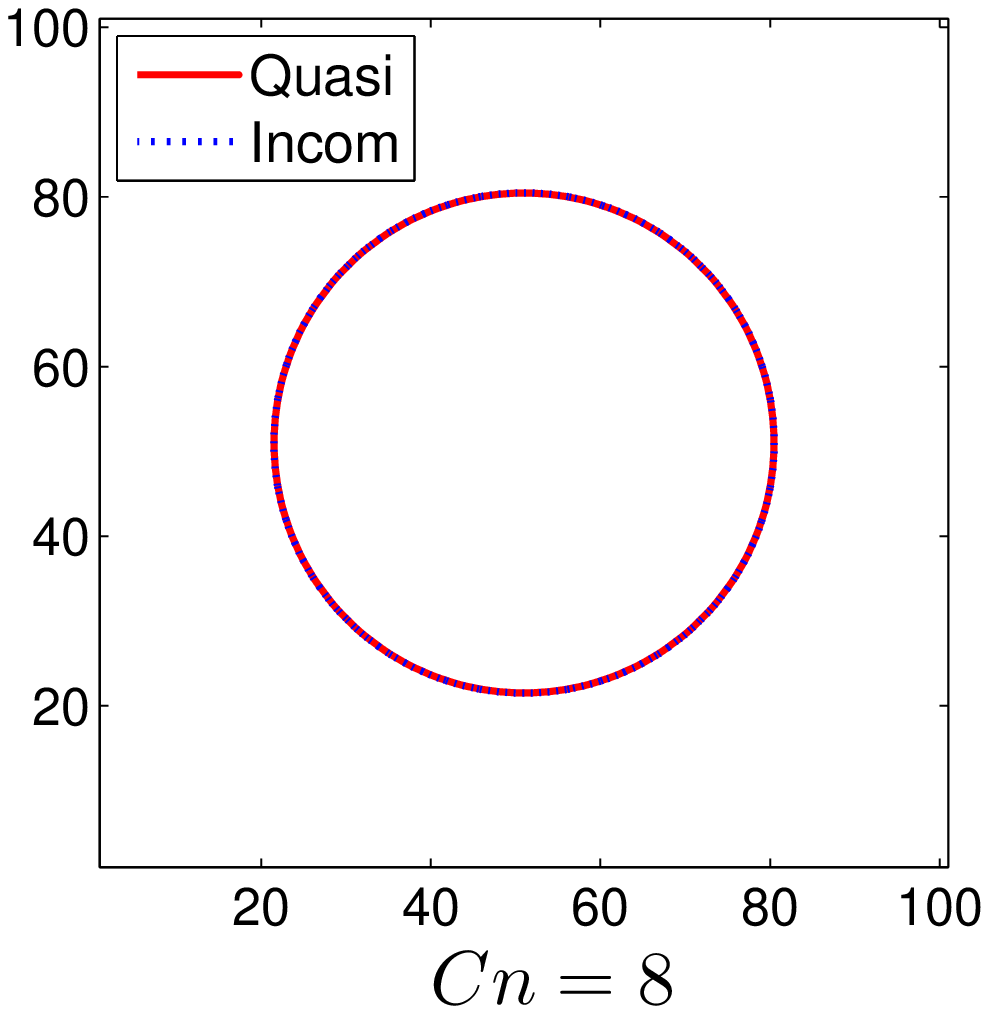}
\caption{(Color online) Order parameter configurations under different values of $Cn$ with $Pe=200$ fixed for the quasi-incompressible (Quasi) and incompressible (Incom) LBE models.}\label{fig:figure2}
\end{figure}

A 2D stationary droplet problem is further tested to verify the present model. Initially, a circular droplet with radius ranging from 20 to 40 is placed in the middle of the computational domain with $N_x \times N_y=100 \times 100$. The initial order parameter and density fields profile are given by
\begin{equation}
\phi_0 (x,y) = \frac{{{\phi _A} + {\phi _B}}}{2} + \frac{{{\phi _A} - {\phi _B}}}{2}\tanh \left( {2\frac{{R - \sqrt {{{(x - {x_c})}^2} + {{(y - {y_c})}^2}} }}{W}} \right),
\end{equation}
\begin{equation}
\rho_0 (x,y) = \frac{{{\rho _A} + {\rho _B}}}{2} + \frac{{{\rho _A} - {\rho _B}}}{2}\tanh \left( {2\frac{{R - \sqrt {{{(x - {x_c})}^2} + {{(y - {y_c})}^2}} }}{W}} \right),
\end{equation}
where $(x_c,y_c)$ is the center of the droplet. Different values of $Pe$ and $Cn$ are respectively investigated. The other parameters are fixed as $\rho_A=1$, $\rho_B=0.2$, $\tau_f=1$, $\tau_g=1$, $\phi_A=1$, $\phi_B=0$ and $\sigma=0.001$. When the droplet reaches the equilibrium state, the pressure difference $\Delta P$ between the inside and outside droplet should satisfy the Laplace law, i.e., $\Delta P=\sigma/R$, where $P$ is calculated by $P = {p_0} - \kappa \phi {\nabla ^2}\phi  + \kappa |\nabla \phi {|^2}/2 + p$ with the equation of state ${p_0} = \phi {\partial _\phi }\psi  - \psi$ \cite{Zu13,Zheng15}. Therefore, the surface tension can be calculated by $\sigma_{LBM}=R\Delta P$, and the numerical predictions and  theoretical values of the surface tension are shown in Table \ref{surfacetension}.
 It can be seen that the present quasi-incompressible LBE model satisfies the Laplace law. The distributions of the order parameter predicted by the both LBE models are shown in Fig.~\ref{fig:figure2}, and no obvious difference can be observed.
 In order to observe the distinctions between them, a moving interface problem is attempted to be simulated in the next section.
\begin{table}[htbp]
\caption{\label{surfacetension} Numerical and theoretical values of surface tension with different $Pe$ and $Cn$.}
\centering
\begin{tabular}{ccccccccccccc}
\\
\hline
\hline
\multirow{2}{*}{Surface tension}    &\multicolumn{3}{c}{$Pe{ \;(Cn=4)}$}&&\multicolumn{3}{c}{$ Cn{\;(Pe=200)}$} \\
\cline{2-4}\cline{6-8}
&$50$ & $125$ &200&&$4$ & $6$&8& \\
\hline
Numerical\;\;($\times 10^{-3}$)   &$0.992$ &$0.992 $&$0.992$&&$0.992$&$0.993$&$0.989$\\
Theoretical ($\times 10^{-3}$)    &$1$ &$1 $&$1 $&&$1 $&$1 $&$1 $\\
\hline
\hline
\end{tabular}
\end{table}

\subsection{Bubble rising under buoyancy}
\begin{figure}[H] \centering
\subfigure[Quasi-incompressible] {
\includegraphics[width=0.12\columnwidth]{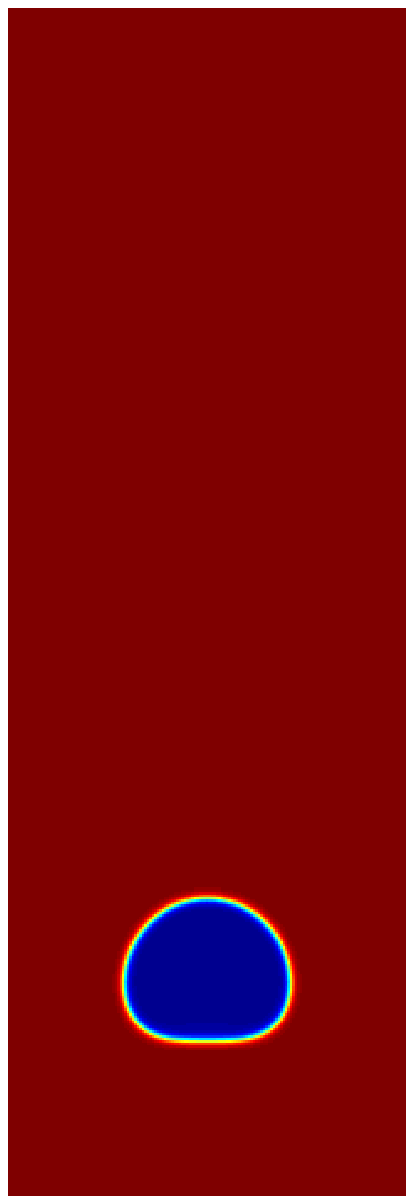}
\includegraphics[width=0.12\columnwidth]{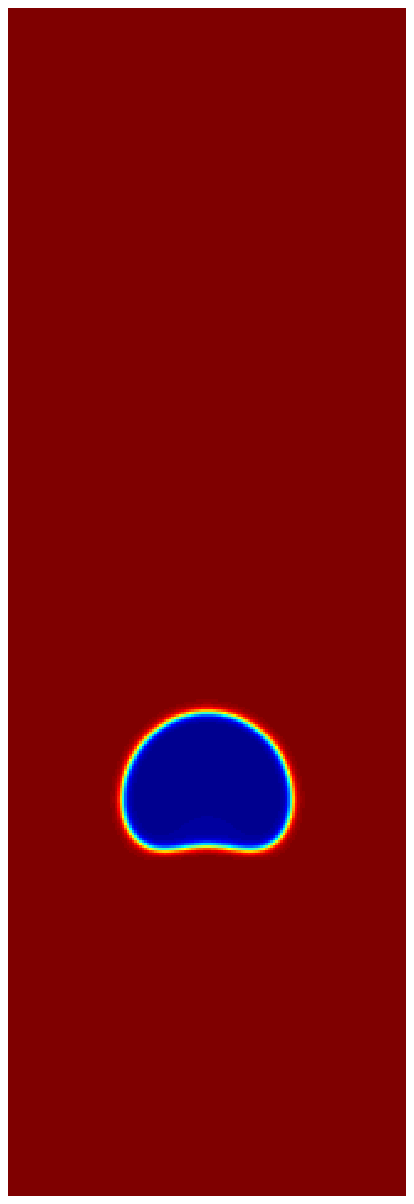}
\includegraphics[width=0.12\columnwidth]{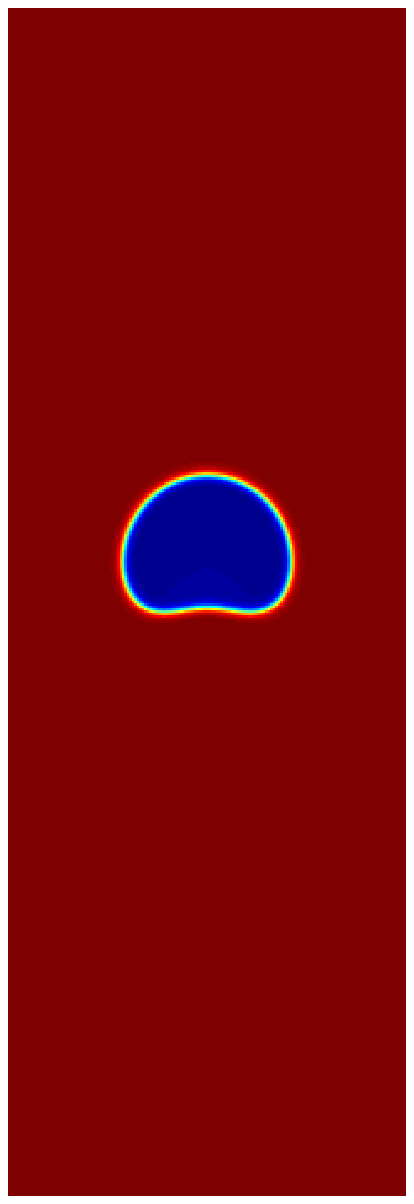}
\includegraphics[width=0.12\columnwidth]{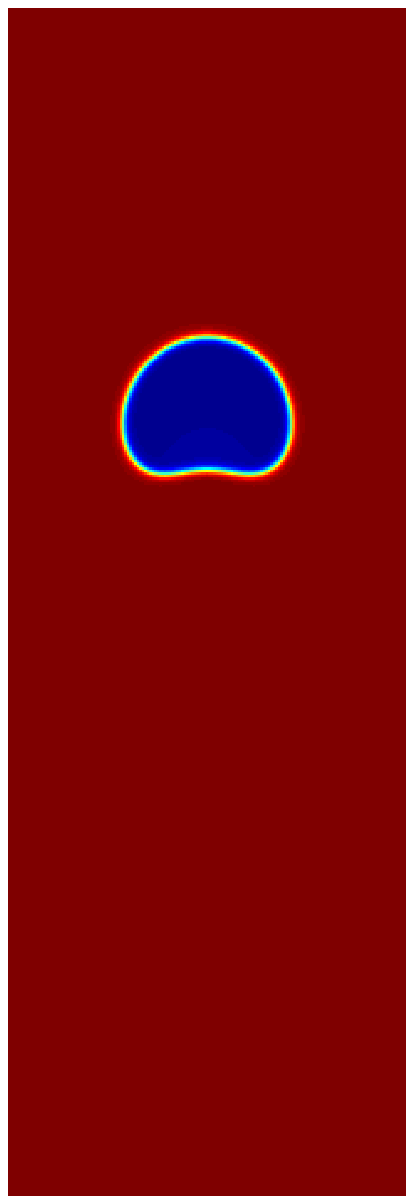}
\includegraphics[width=0.12\columnwidth]{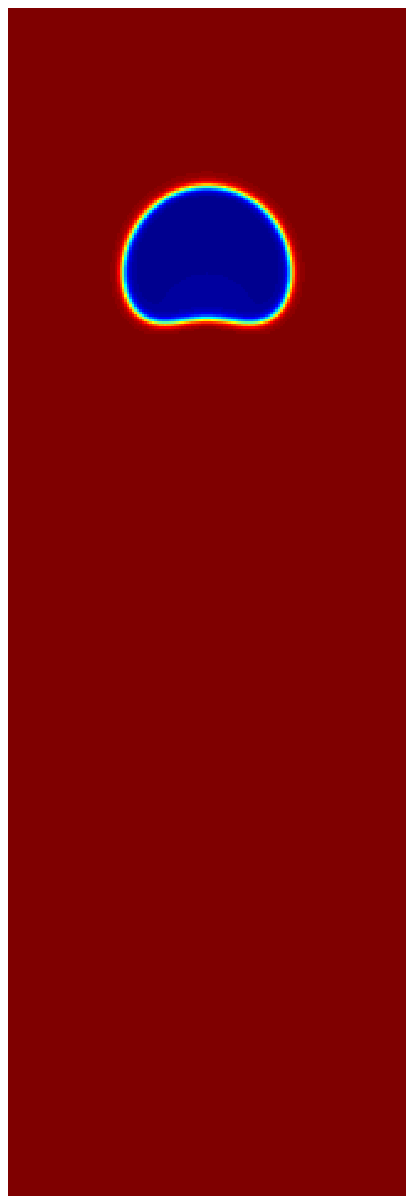}
}
\subfigure[Incompressible] {
\includegraphics[width=0.12\columnwidth]{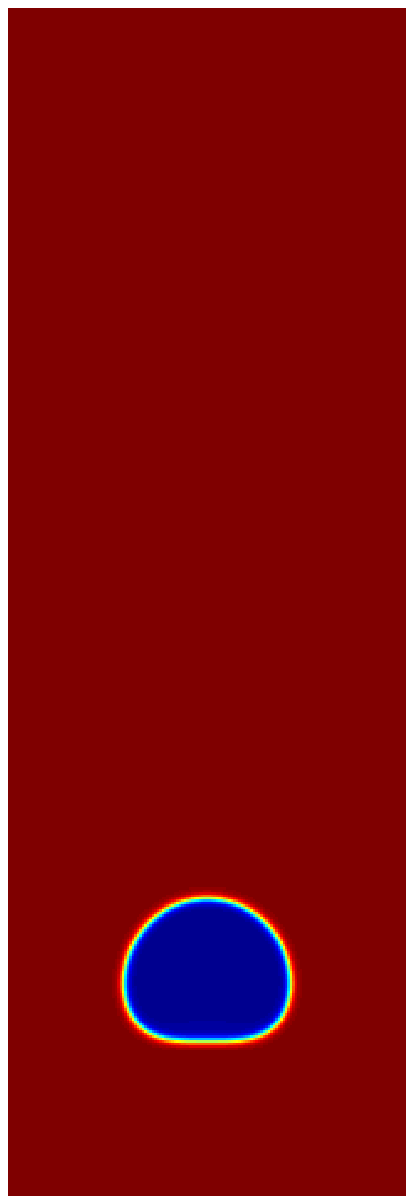}
\includegraphics[width=0.12\columnwidth]{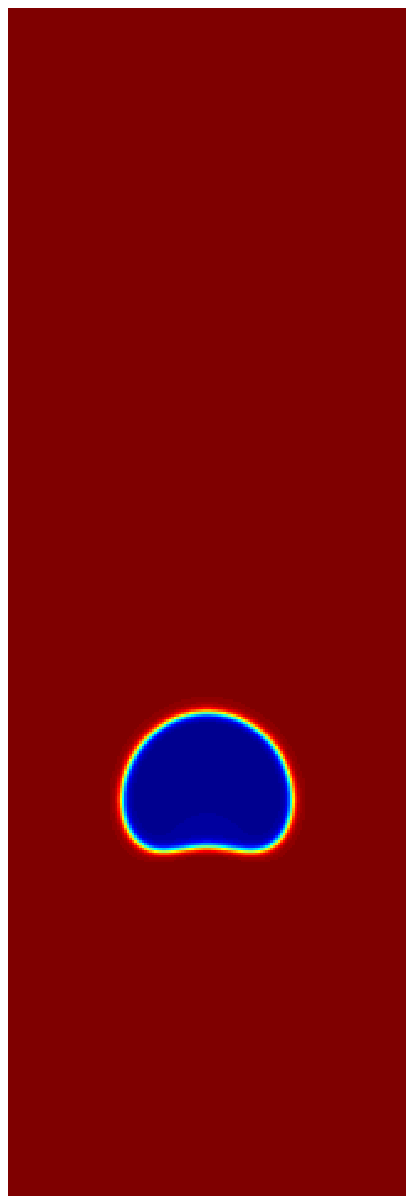}
\includegraphics[width=0.12\columnwidth]{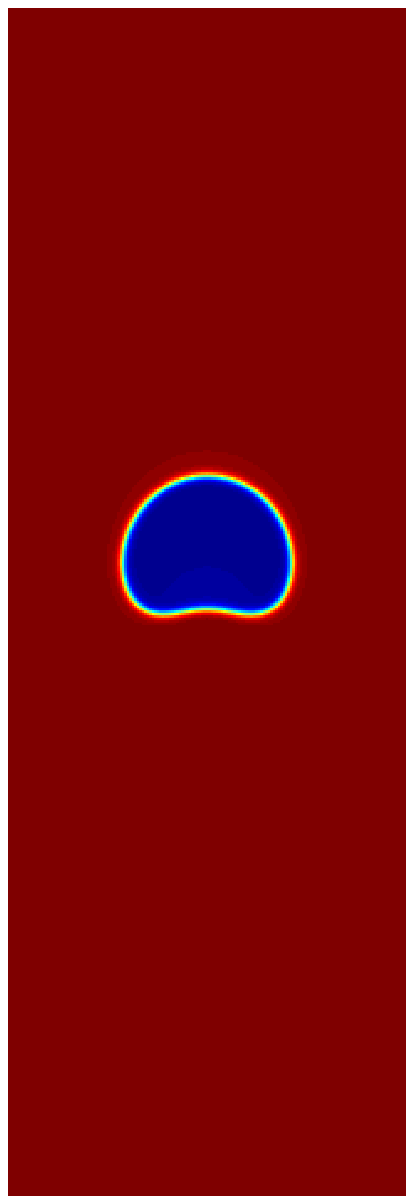}
\includegraphics[width=0.12\columnwidth]{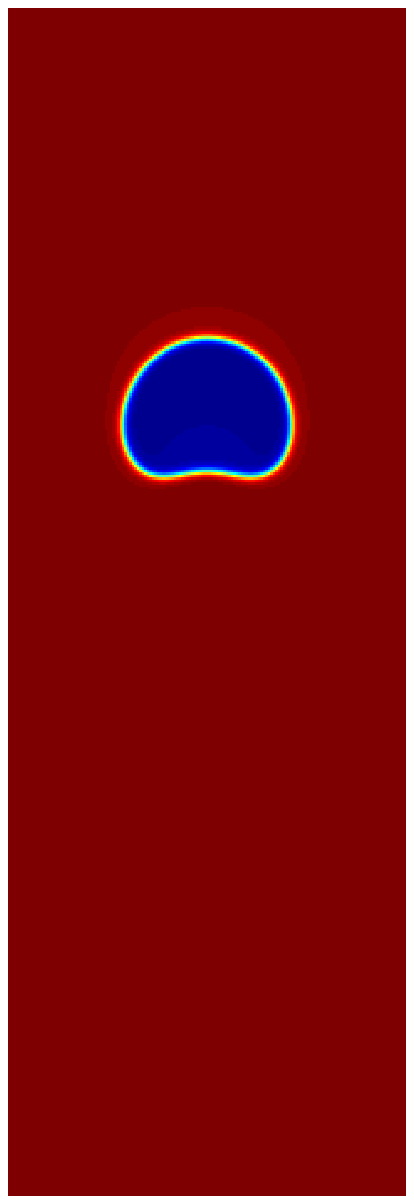}
\includegraphics[width=0.12\columnwidth]{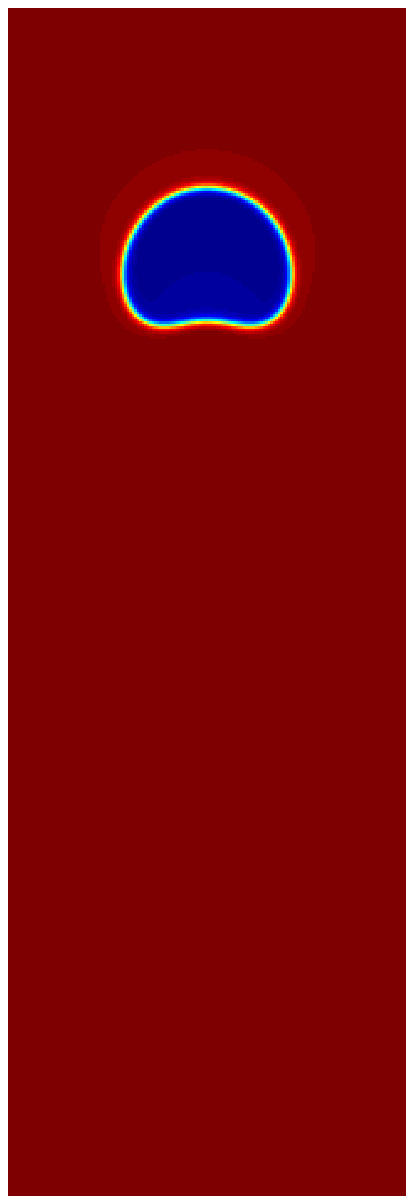}
}
\caption{(Color online) Density configuration of the rising bubble at $t/1000=10,20,30,35,40$ for the quasi-incompressible model (a) and incompressible model (b).}
\label{fig:figure3}
\end{figure}

\begin{figure}[H]\centering
\includegraphics[width=0.3\textwidth]{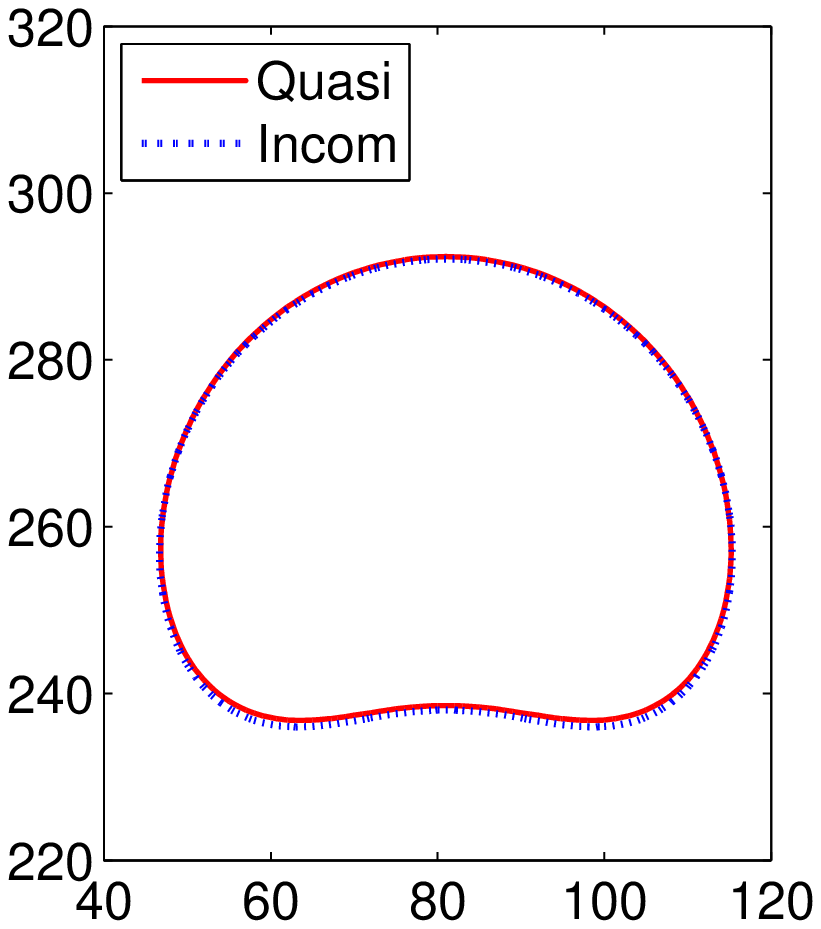}
\includegraphics[width=0.3\textwidth]{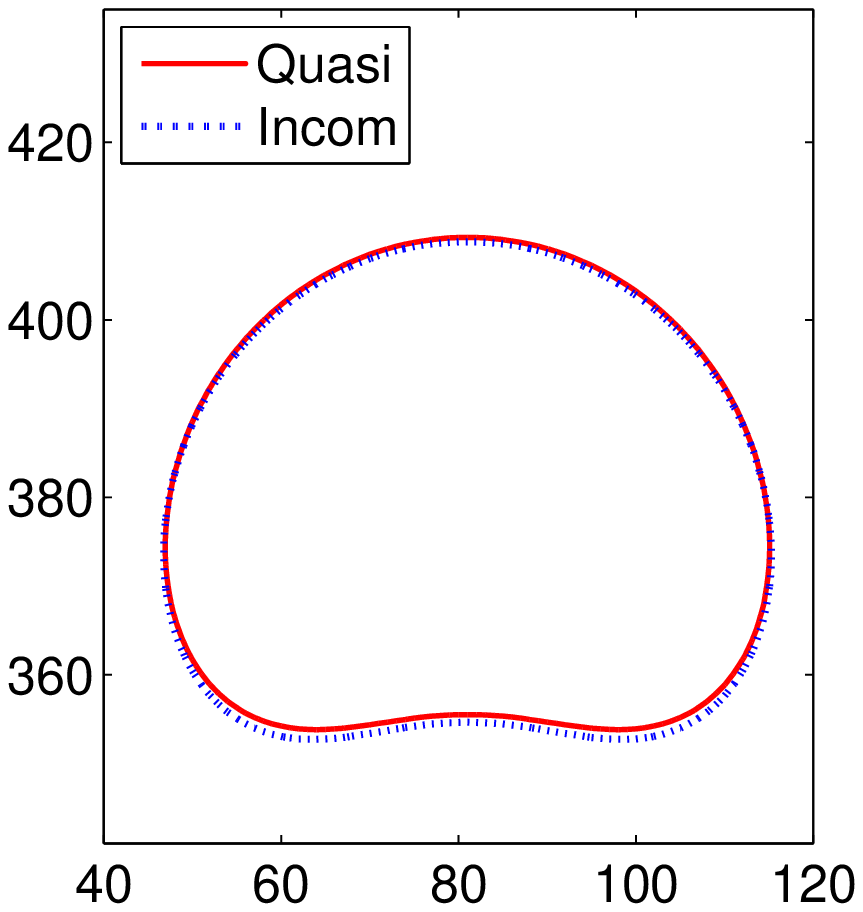}
\caption{(Color online) Density profiles predicted by the quasi-incompressible (Quais) and incompressible (Incom) LBE models at $t/1000=30,40$.}\label{fig:figure4}
\end{figure}

\begin{figure}[H] \centering
\subfigure[Quasi-incompressible] {
\includegraphics[width=0.12\columnwidth]{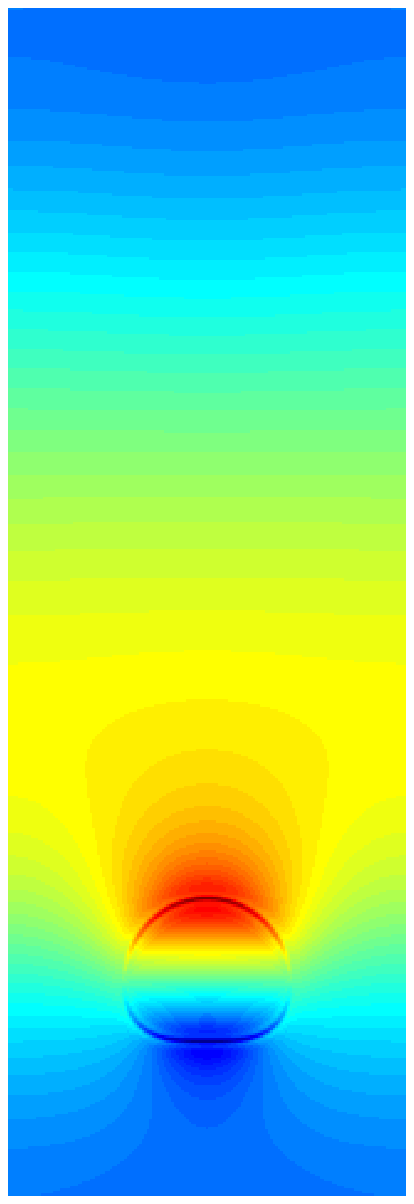}
\includegraphics[width=0.12\columnwidth]{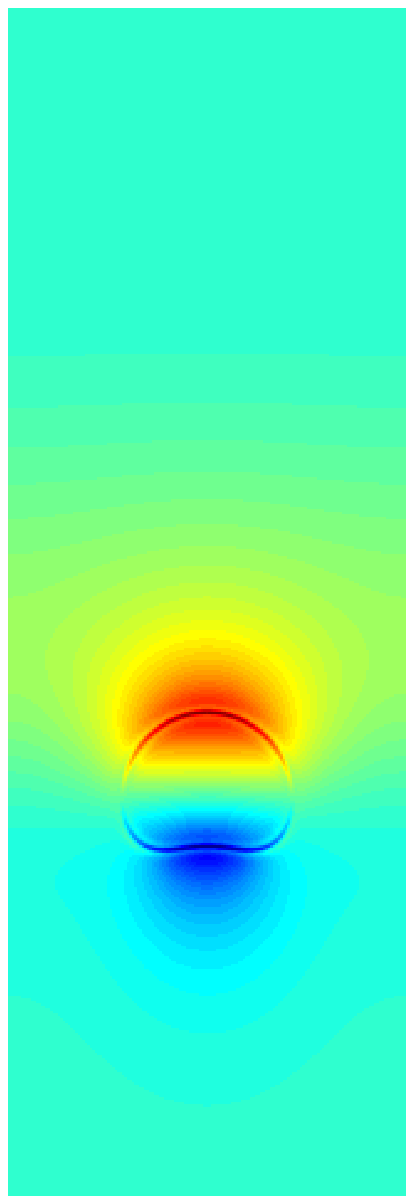}
\includegraphics[width=0.12\columnwidth]{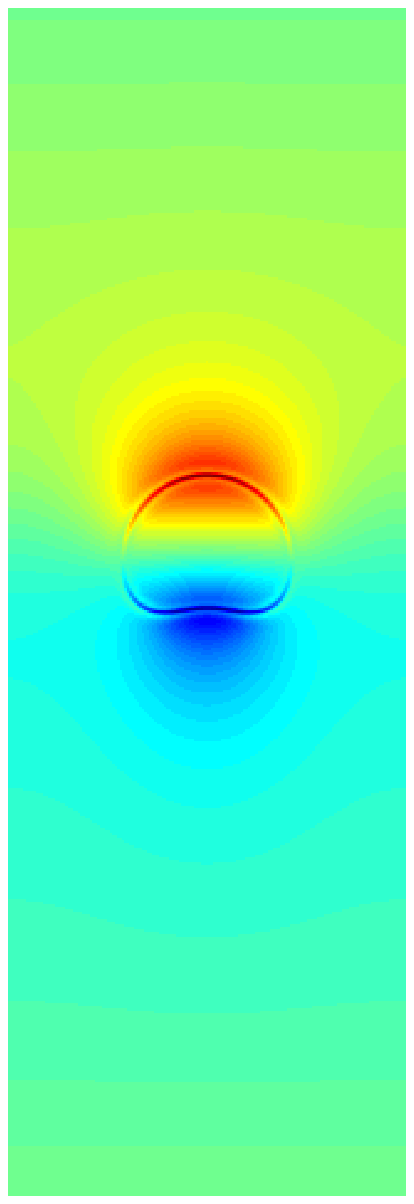}
\includegraphics[width=0.12\columnwidth]{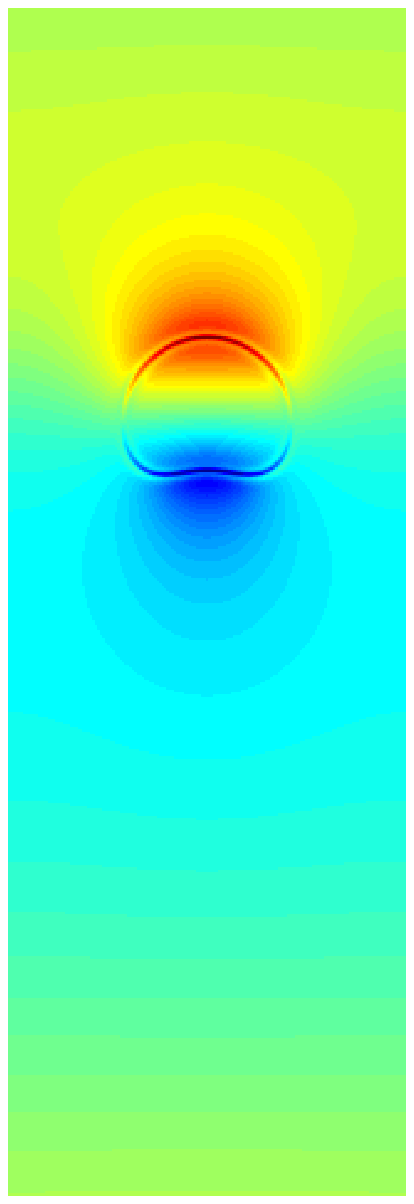}
\includegraphics[width=0.12\columnwidth]{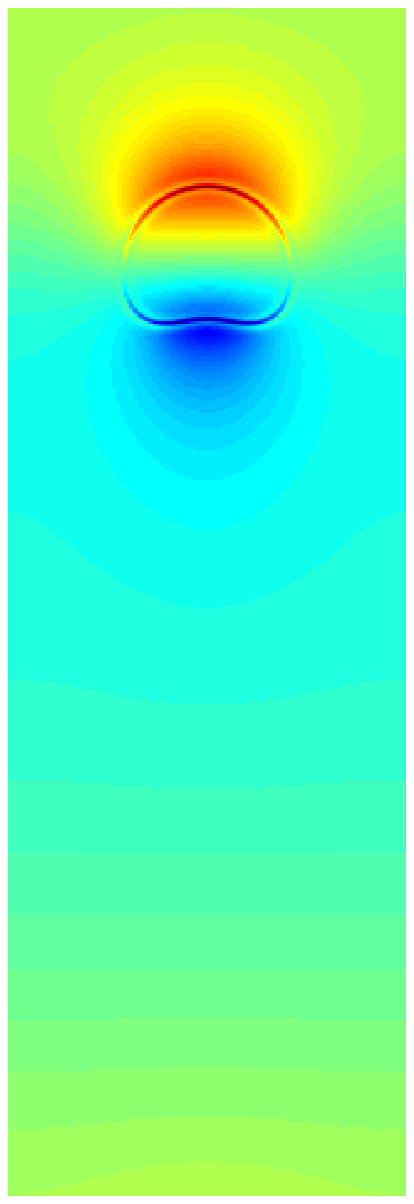}
}
\subfigure[Incompressible] {
\includegraphics[width=0.12\columnwidth]{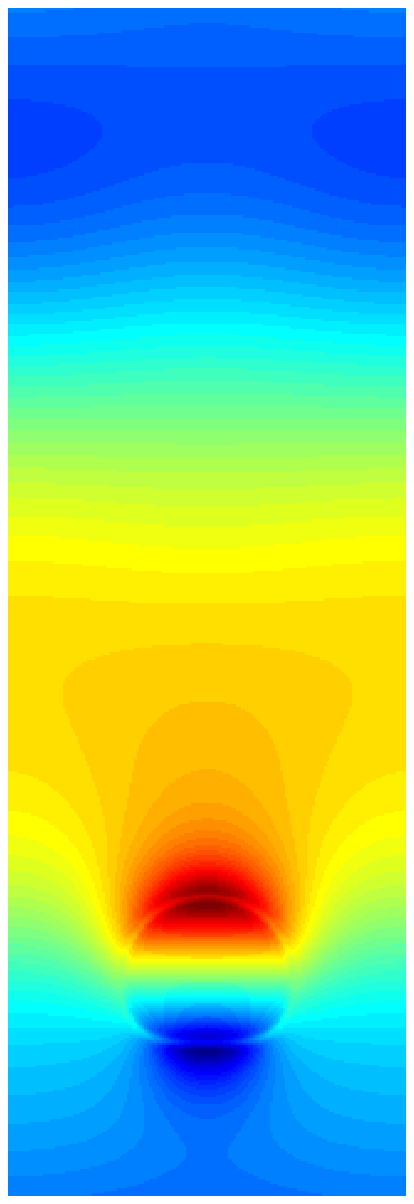}
\includegraphics[width=0.12\columnwidth]{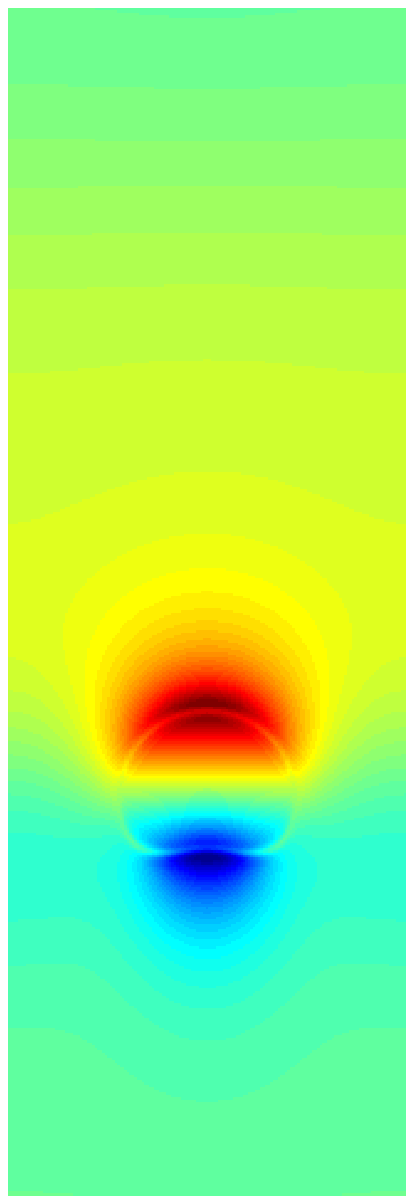}
\includegraphics[width=0.12\columnwidth]{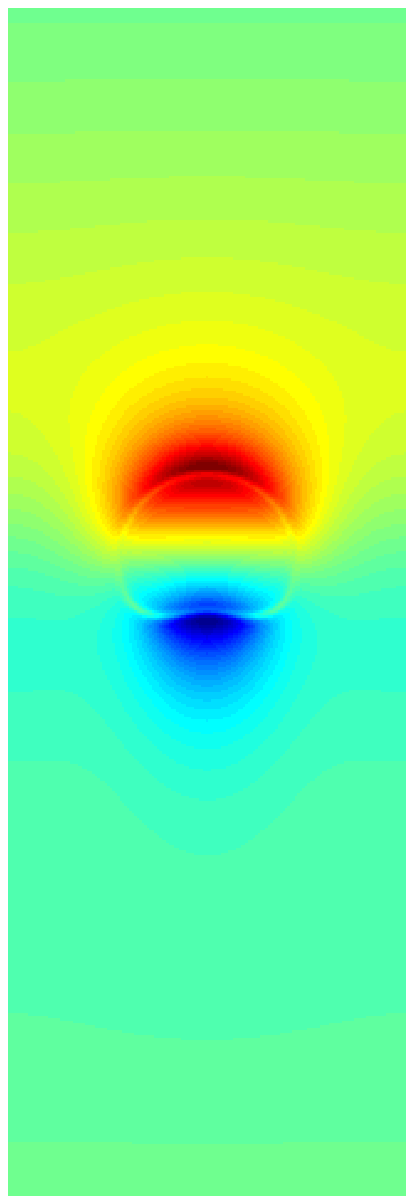}
\includegraphics[width=0.12\columnwidth]{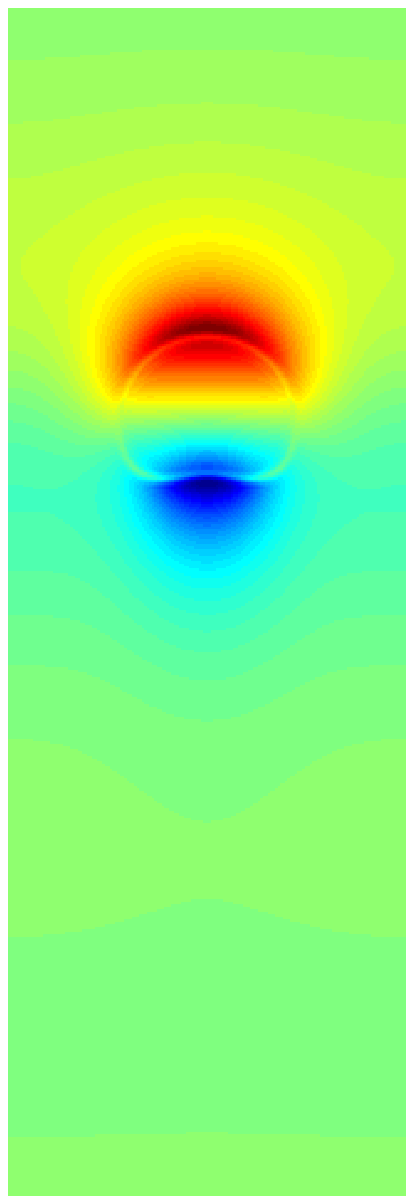}
\includegraphics[width=0.12\columnwidth]{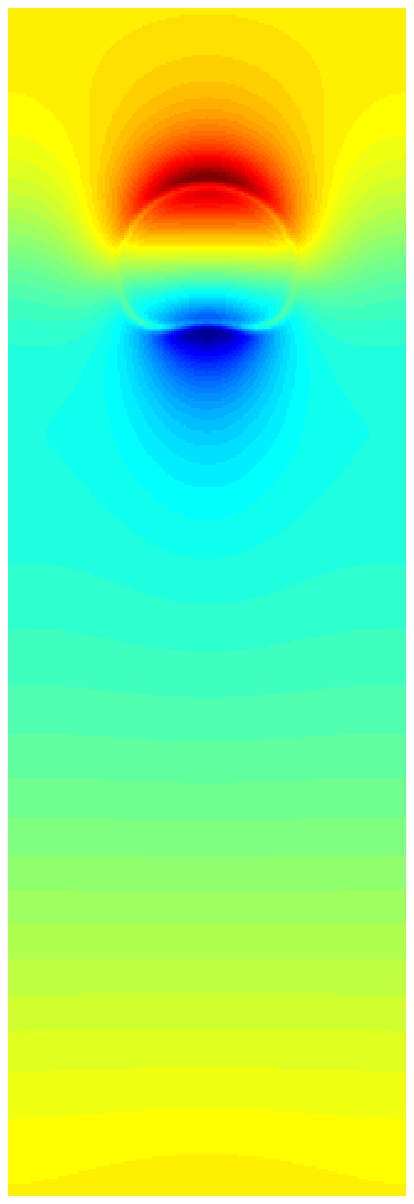}
}
\caption{(Color online) Dynamic pressure field of the rising bubble at $t/1000=10,20,30,35,40$ for the quasi-incompressible model (a) and incompressible model (b).}
\label{fig:figure5}
\end{figure}

\begin{figure}[H] \centering
\subfigure[Quasi-incompressible] {
\includegraphics[width=0.12\columnwidth]{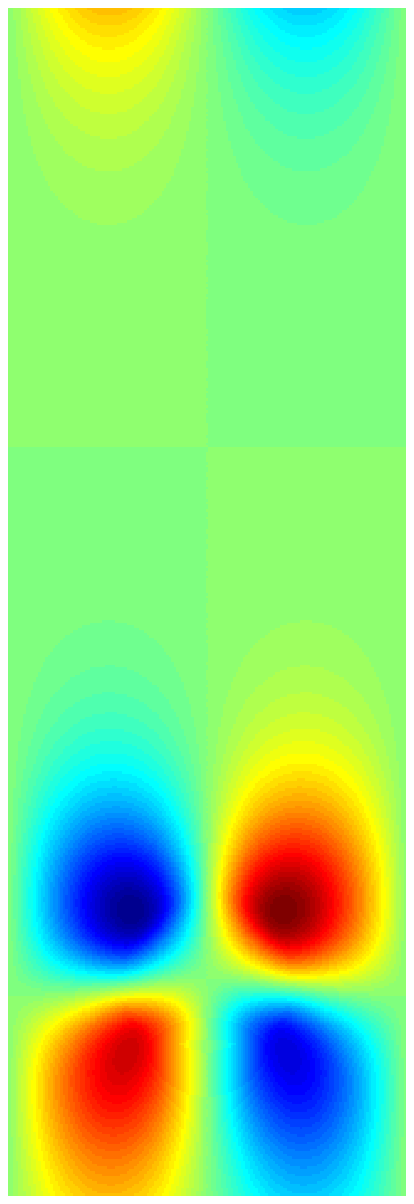}
\includegraphics[width=0.12\columnwidth]{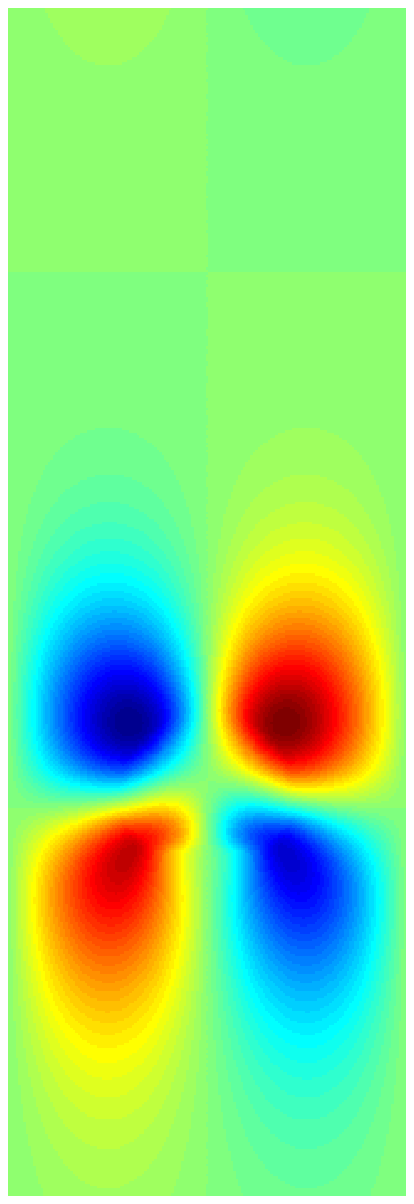}
\includegraphics[width=0.12\columnwidth]{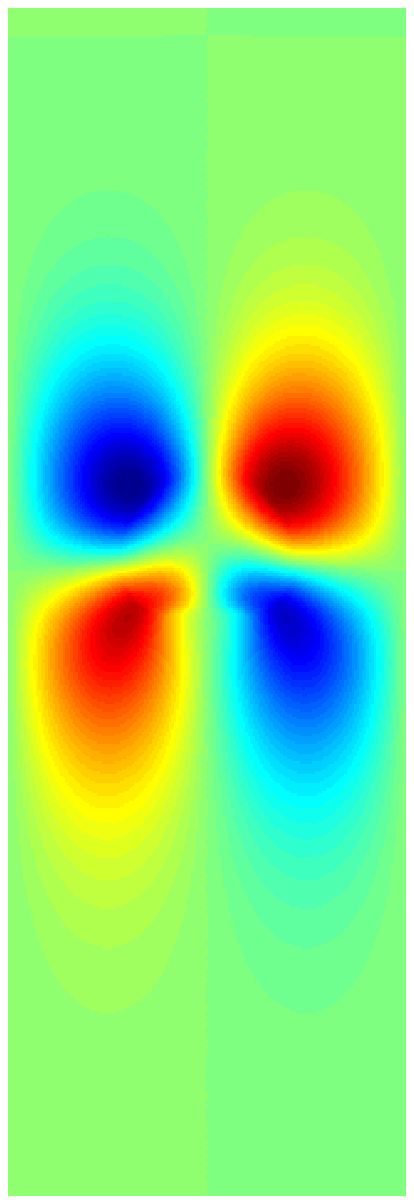}
\includegraphics[width=0.12\columnwidth]{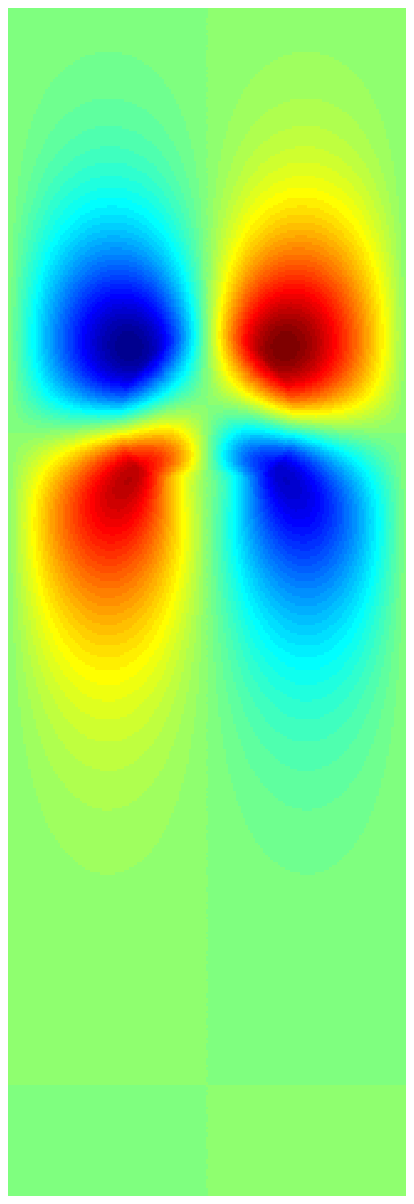}
\includegraphics[width=0.12\columnwidth]{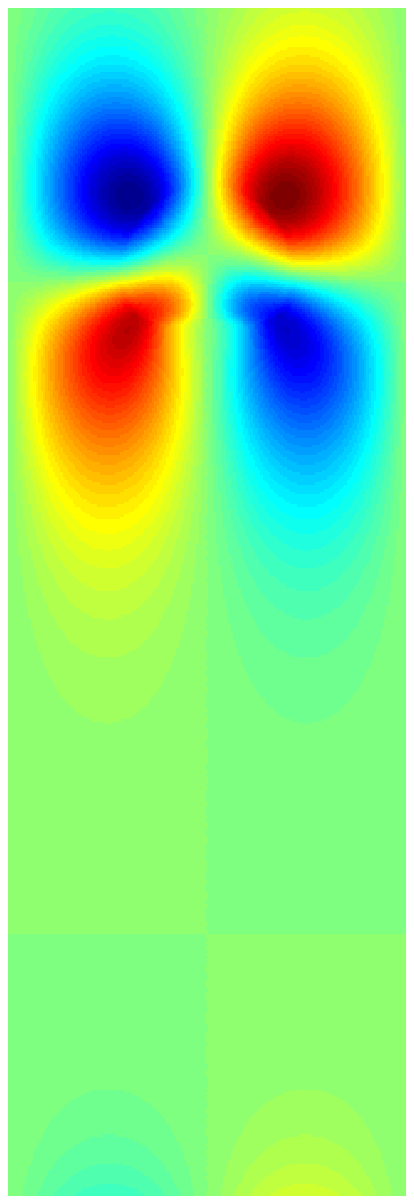}
}
\subfigure[Incompressible] {
\includegraphics[width=0.12\columnwidth]{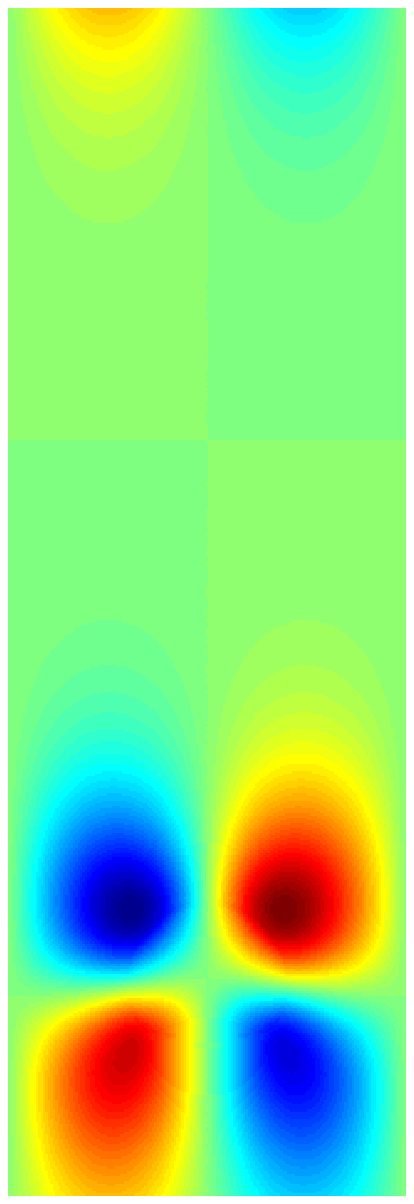}
\includegraphics[width=0.12\columnwidth]{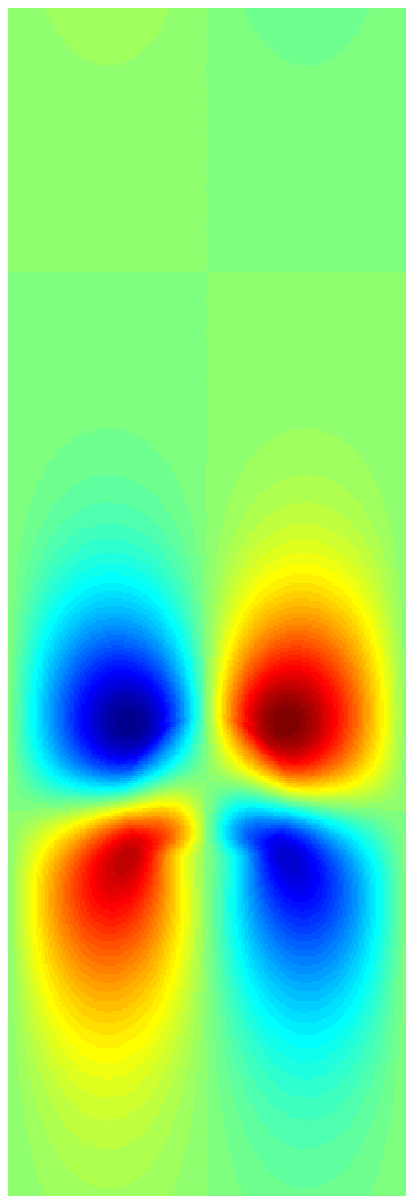}
\includegraphics[width=0.12\columnwidth]{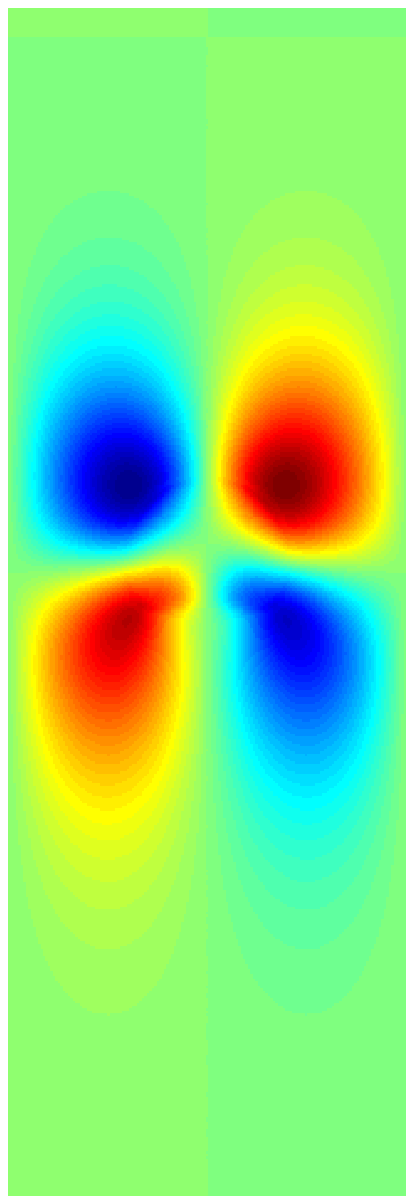}
\includegraphics[width=0.12\columnwidth]{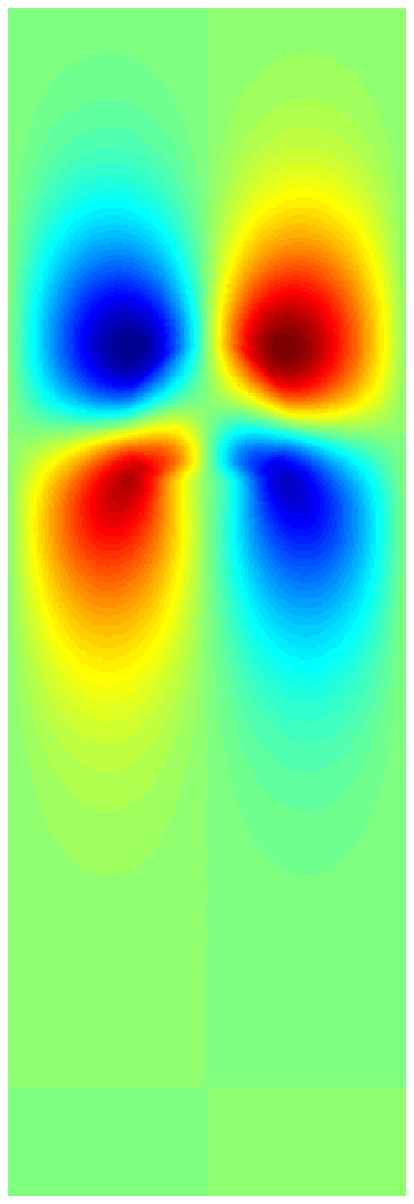}
\includegraphics[width=0.12\columnwidth]{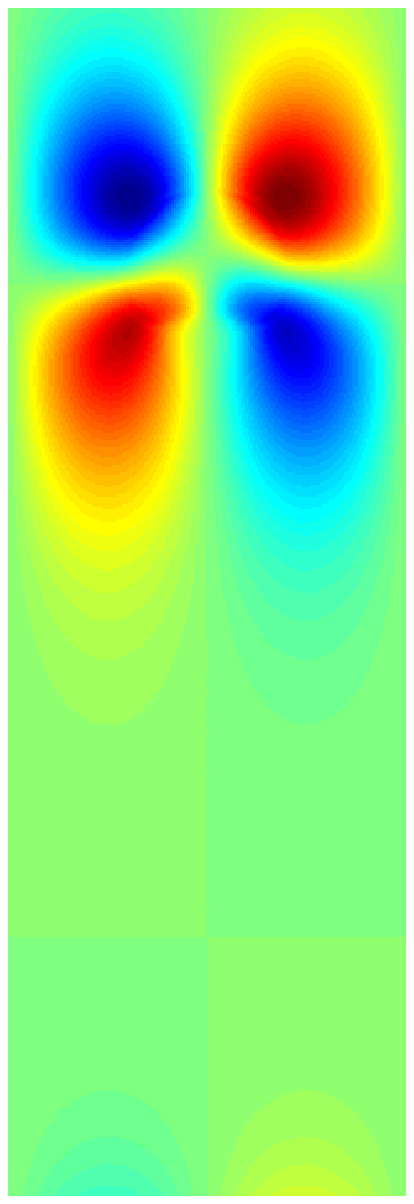}
}
\caption{(Color online) Horizontal velocity of the rising bubble at $t/1000=10,20,30,35,40$ for the quasi-incompressible model (a) and incompressible model (b).}
\label{fig:figure6}
\end{figure}

\begin{figure}[H] \centering
\subfigure[Quasi-incompressible] {
\includegraphics[width=0.12\columnwidth]{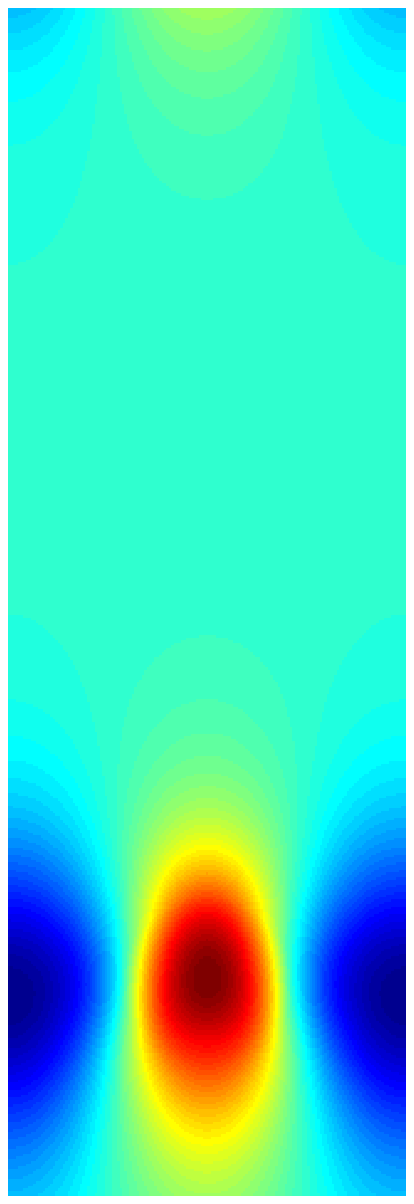}
\includegraphics[width=0.12\columnwidth]{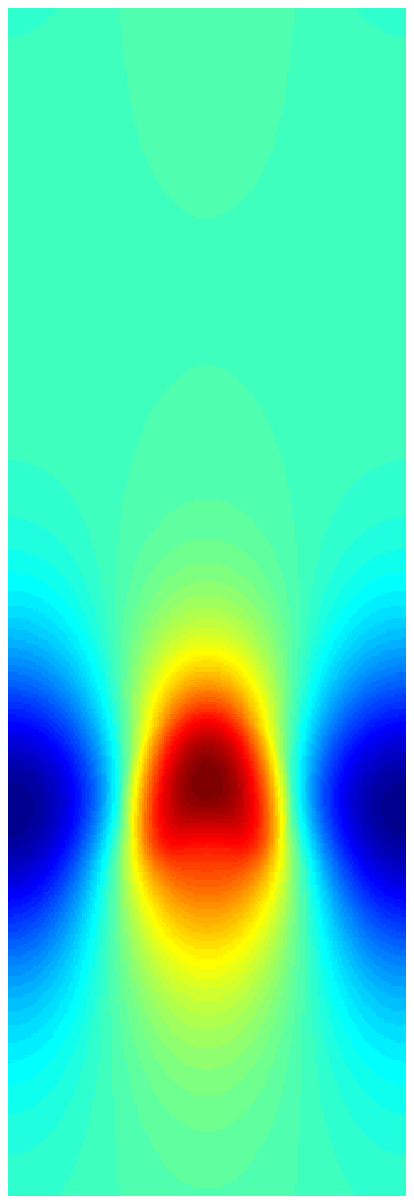}
\includegraphics[width=0.12\columnwidth]{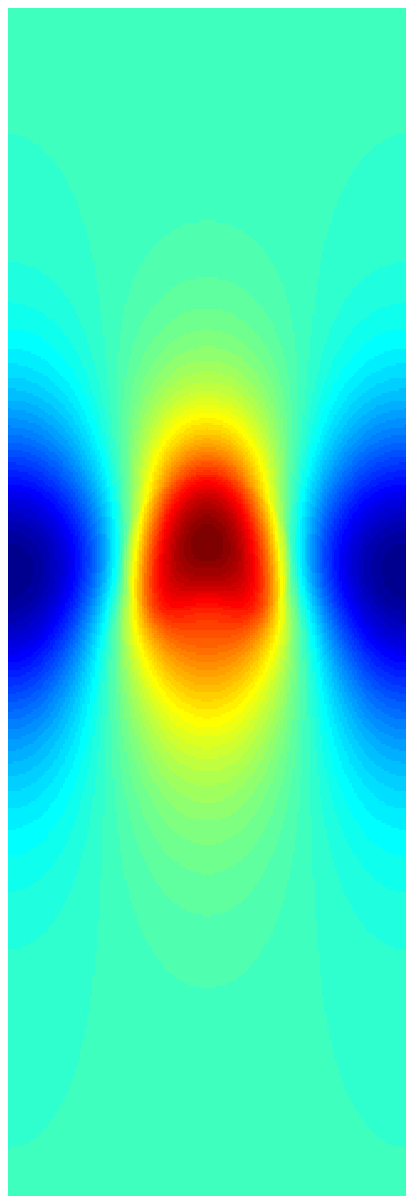}
\includegraphics[width=0.12\columnwidth]{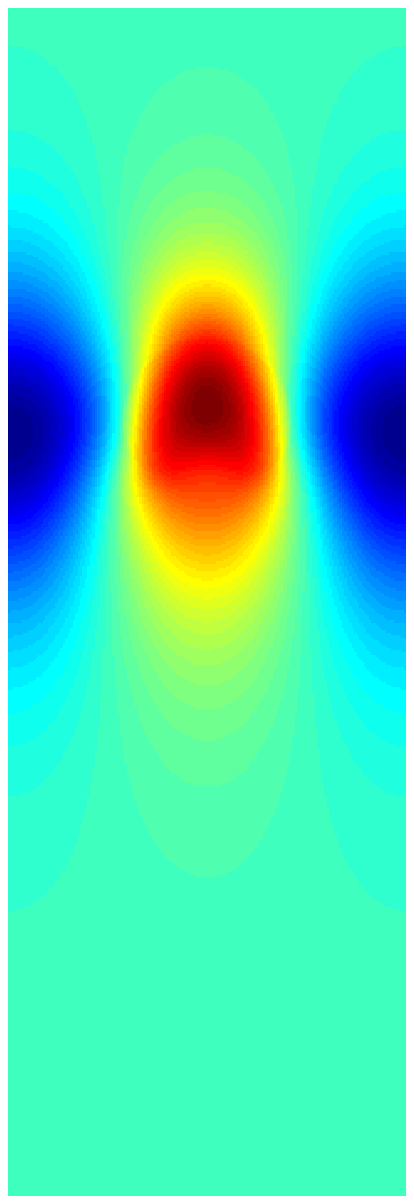}
\includegraphics[width=0.12\columnwidth]{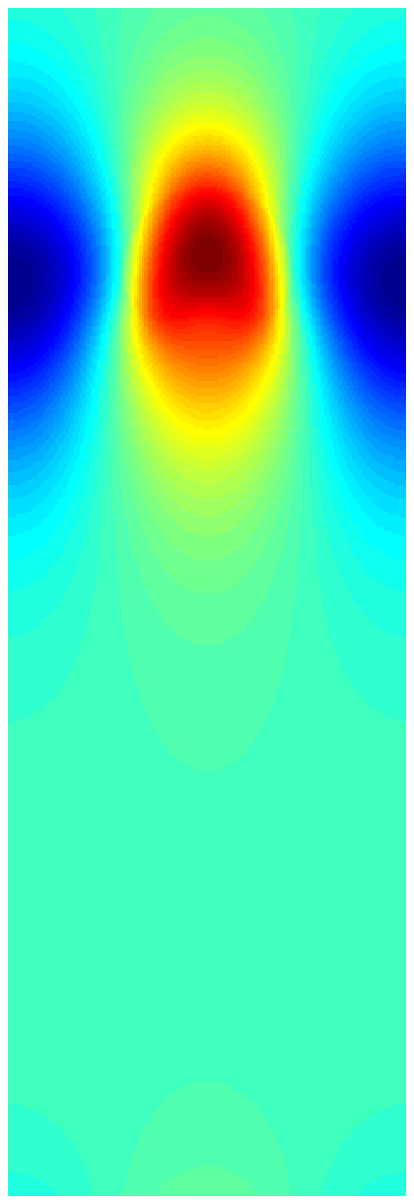}
}
\subfigure[Incompressible] {
\includegraphics[width=0.12\columnwidth]{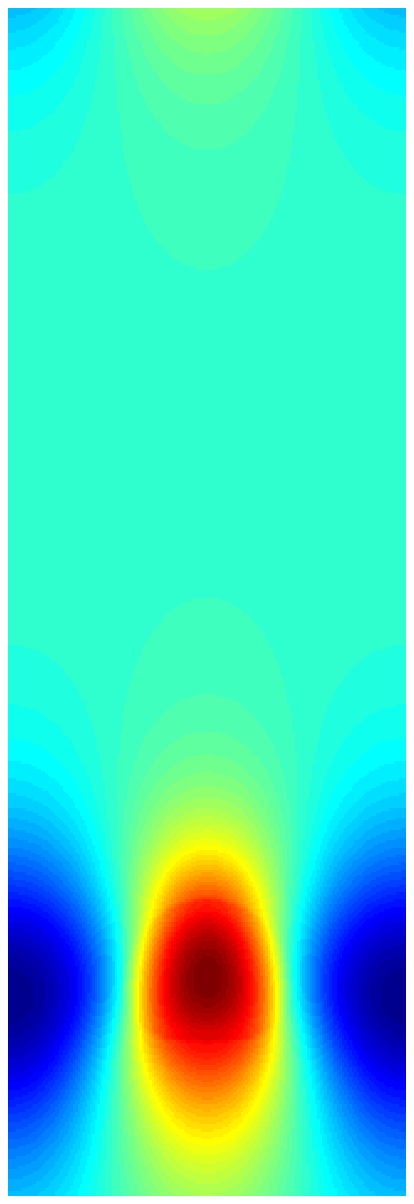}
\includegraphics[width=0.12\columnwidth]{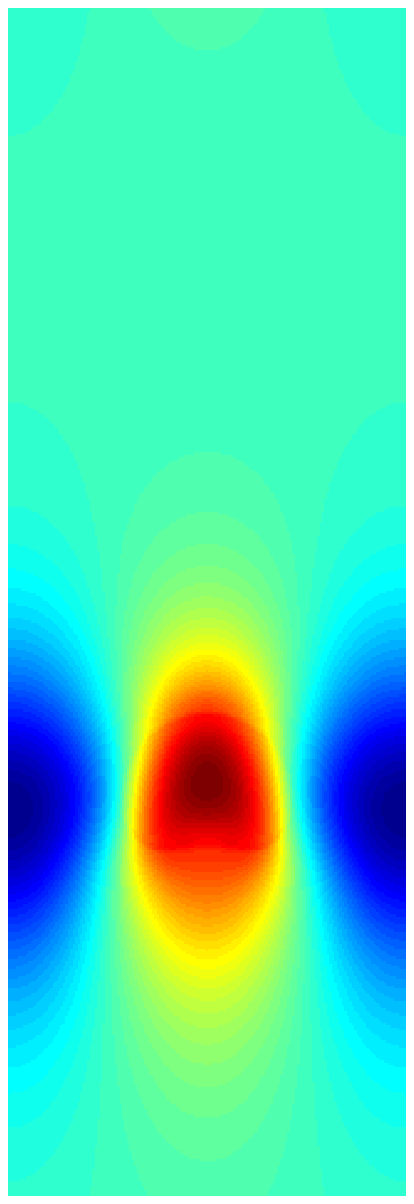}
\includegraphics[width=0.12\columnwidth]{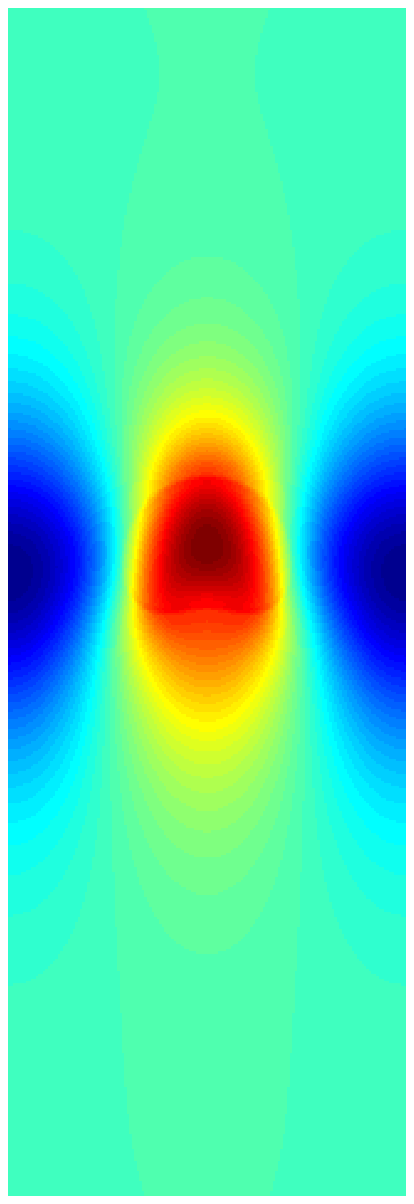}
\includegraphics[width=0.12\columnwidth]{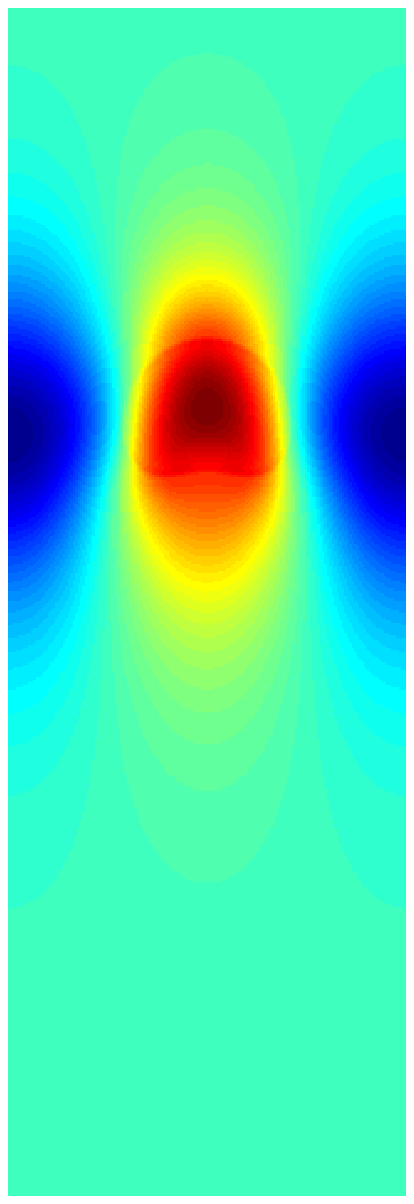}
\includegraphics[width=0.12\columnwidth]{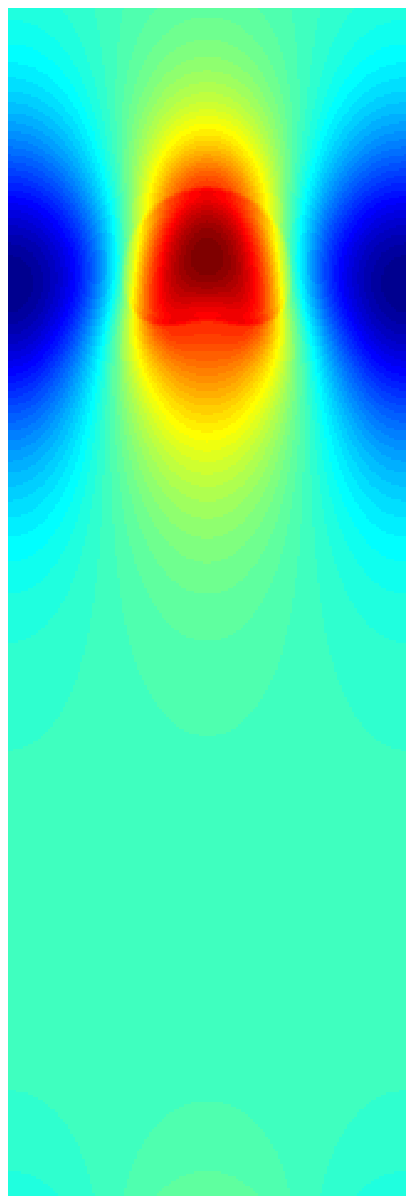}
}
\caption{(Color online) Vertical velocity of the rising bubble at $t/1000=10,20,30,35,40$ for the quasi-incompressible model (a) and incompressible model (b).}
\label{fig:figure7}
\end{figure}

\begin{figure}[H]\centering
\includegraphics[width=0.18\textwidth]{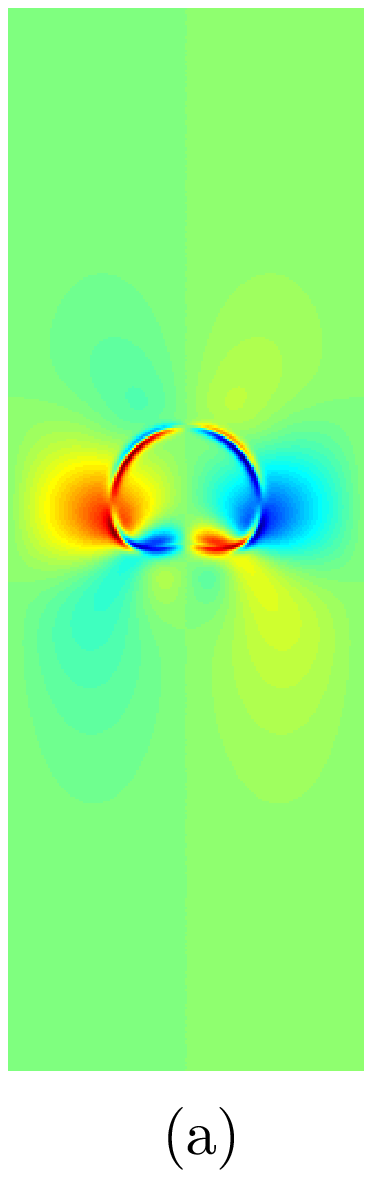}
\includegraphics[width=0.18\textwidth]{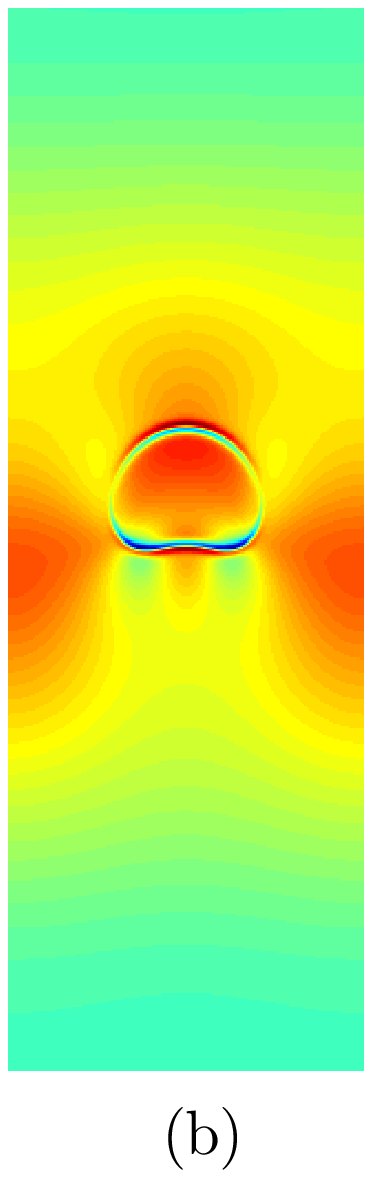}
\caption{Normalized velocity difference between the quasi-incompressible (Quais) model and incompressible (Incom) model at $t=3\times 10^4$, i.e,  (a)  $(u_{{\rm Quasi}}-u_{{\rm Incom}})/u_{{\rm max,  Quasi}}$ and (b) $(v_{{\rm Quasi}}-v_{{\rm Incom}})/v_{{\rm max, Quasi}}$, where $u_i$ and $v_i$ are the horizontal and vertical velocities of model $i$, respectively, and $u_{{\rm max, Quasi}}$ and $v_{\rm max, Quasi}$ are the maximum horizontal and vertical velocities of quasi-incompressible model, respectively.}\label{fig:figure8}
\end{figure}

 In this section, a bubble rising under buoyancy is simulated to compare the two LBE models. Initially, a light circular bubble (fluid B) with radius $R$ is immersed in another fluid (A) with higher density. To generate the buoyancy effect, a body force, $F_{b,y}=-(\rho-\rho_A)\textrm g$, is added to the fluid flow, where $\textrm g$ is the gravitational acceleration. In the simulations, the computational domain is set to be $160 \times 480$, and periodic boundary conditions are applied to all boundaries. The other parameters are set as follows: $\rho_A=1$, $\rho_B=0.5$, $\textrm g= 10^{-5}$, $\phi_A=1$, $\phi_B=0$, $\tau_f=1$, $\tau_g=1$, $\sigma=0.001$, $R=32$, $Pe=50$ and $Cn=4$.
Figure \ref{fig:figure3} shows the shape of the rising bubble at different times predicted by the two LBE models. It can be seen that the results are quite similar.  A comparison of the interface shapes at $t=10^4$ and $4\times 10^4$ confirms the similarity in Fig. \ref{fig:figure4}. Figure \ref{fig:figure5} shows the distributions of the dynamic pressure at different times, and the difference in the vicinity of the interface is more obvious. Figures \ref{fig:figure6} to \ref{fig:figure8} show the bubble velocity and the normalized velocity differences between the two LBE models.
It can be seen from Figs. \ref{fig:figure6} and \ref{fig:figure7}, the horizontal and vertical velocity components are nearly identical, but from Fig. \ref{fig:figure8}, we can observe some the normalized velocity differences with maximum magnitude of order $10^{-2}$. Based on the above observations, we can conclude that the predictions by the two LBE models yield almost the same results for this test case. According to the previous theoretical analysis, these phenomena are reasonable since the initial equilibrium order parameter yields the approximate uniform chemical potential
so that the term $\gamma \phi {\bm{\nabla}}\cdot(\lambda {\bm{\nabla}} \mu)$ exerts a weak influence on
the results.


\subsection{Phase separation}
In this subsection, the simulation of phase separation will be carried out to further compare the two LBE  models.
Initially, the order parameter with a small perturbation is set as $\phi = 0.5[1 + 0.1 {\rm {sin}}(4\pi x/Lx){\rm{cos}}(4\pi y/Ly)]$, where $x$, $y$ represent the Cartesian coordinates, and $Lx$ and $Ly$ are the length and width of the computational domain, respectively.
To consider the effects of $Pe$, $Cn$ and $\gamma$ on the phase separation, different mobilities ($\lambda \beta=0.02$ and 0.005), interfacial thicknesses ($W=4$ and 8) and density ratios ($\rho_r=2$ and $5$)
are taken into account in the simulations. The computational domain is set to be $100 \times 100$.
The other parameters are set as $\rho_A=1$, $\phi_A=1$, $\phi_B=0$, $\tau_f=1$, $\tau_g=1$ and $\sigma=0.001$. Periodic boundary conditions are applied to the all boundaries.

\begin{figure}[H] \centering
\subfigure[Quasi-incompressible] {
\includegraphics[width=0.16\columnwidth]{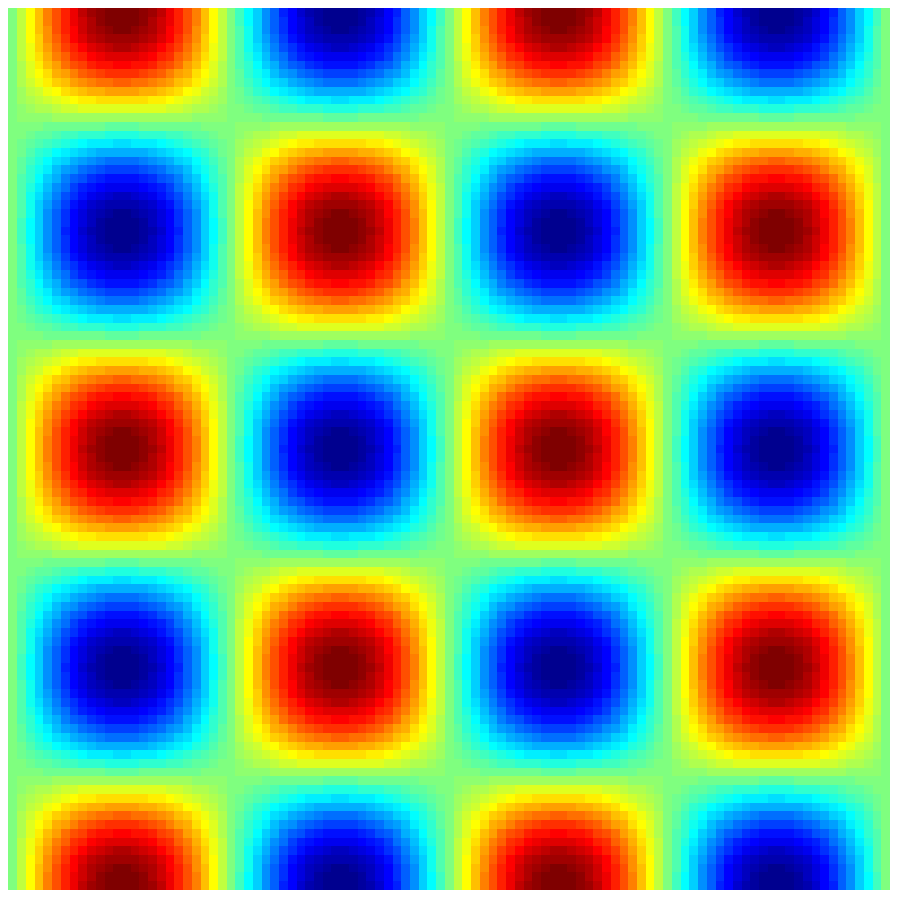}
\includegraphics[width=0.16\columnwidth]{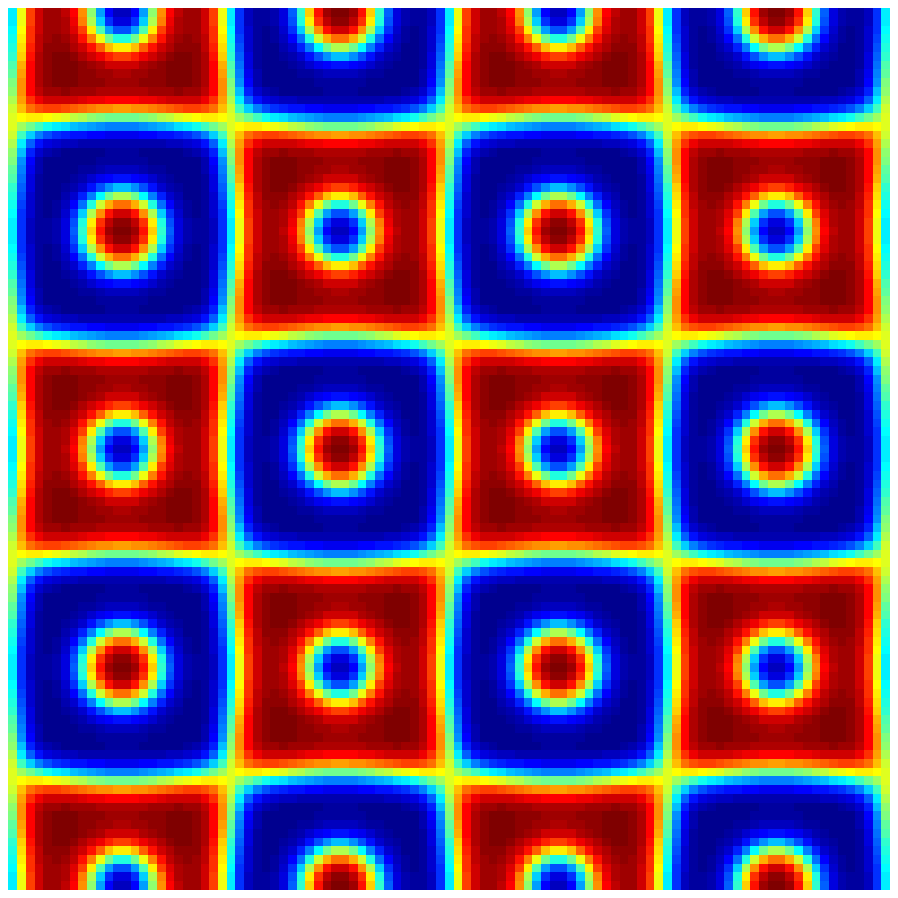}
\includegraphics[width=0.16\columnwidth]{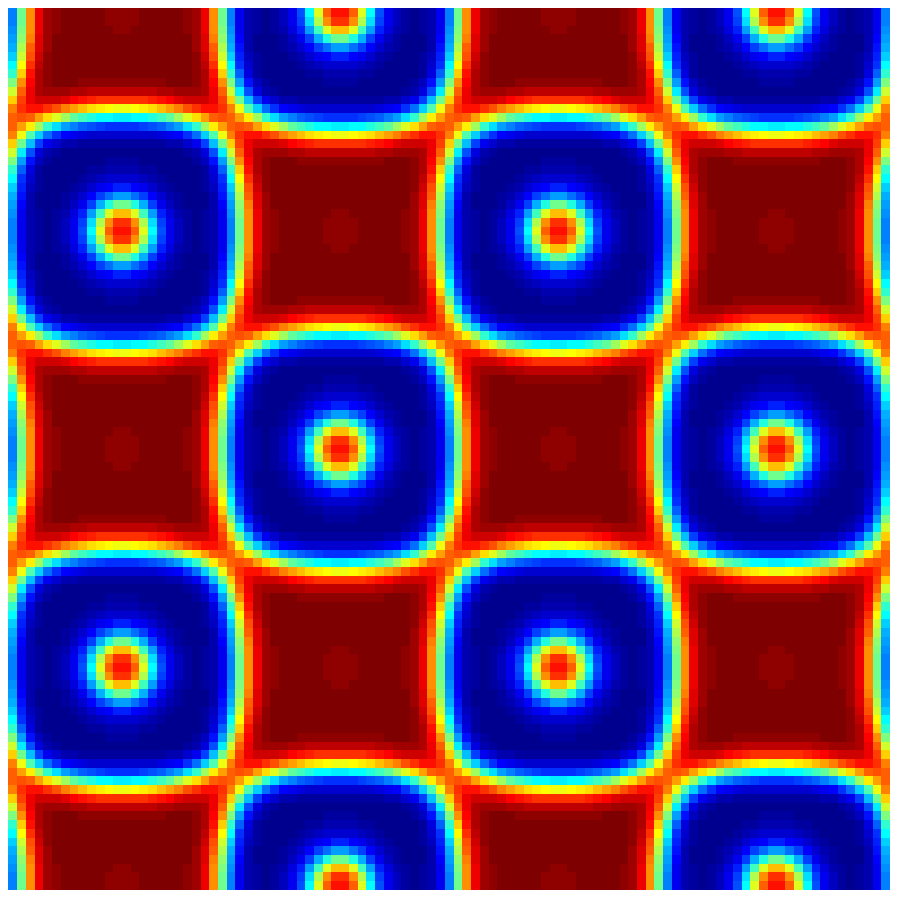}
\includegraphics[width=0.16\columnwidth]{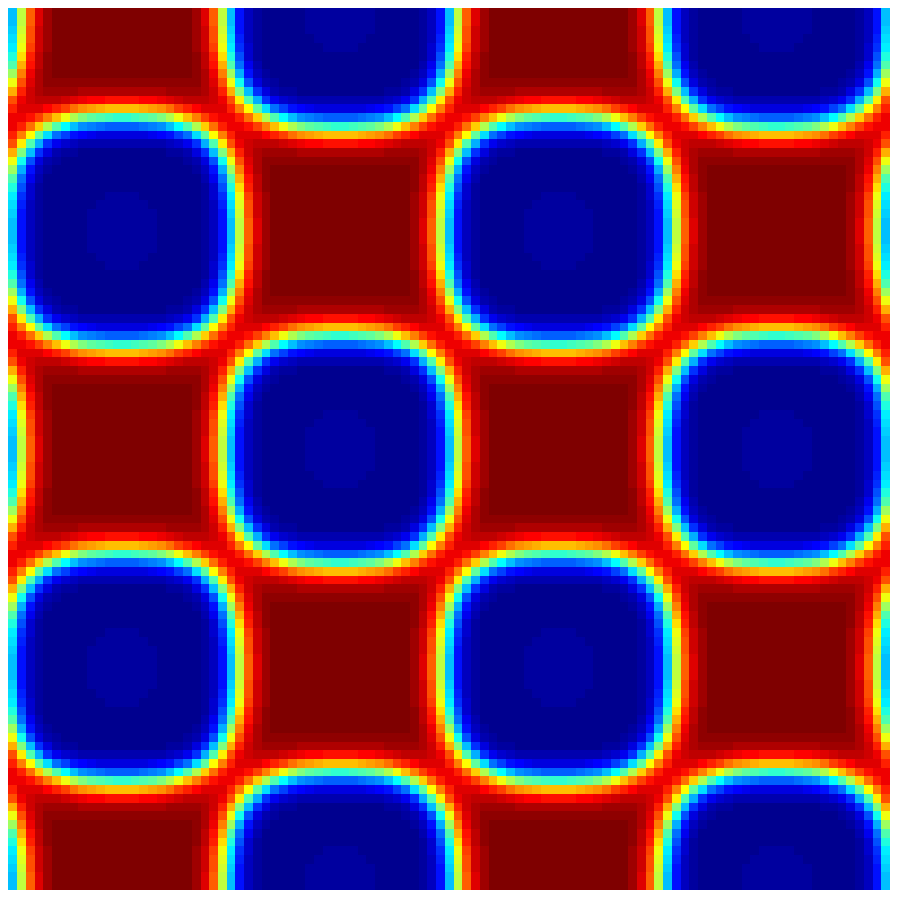}
\includegraphics[width=0.16\columnwidth]{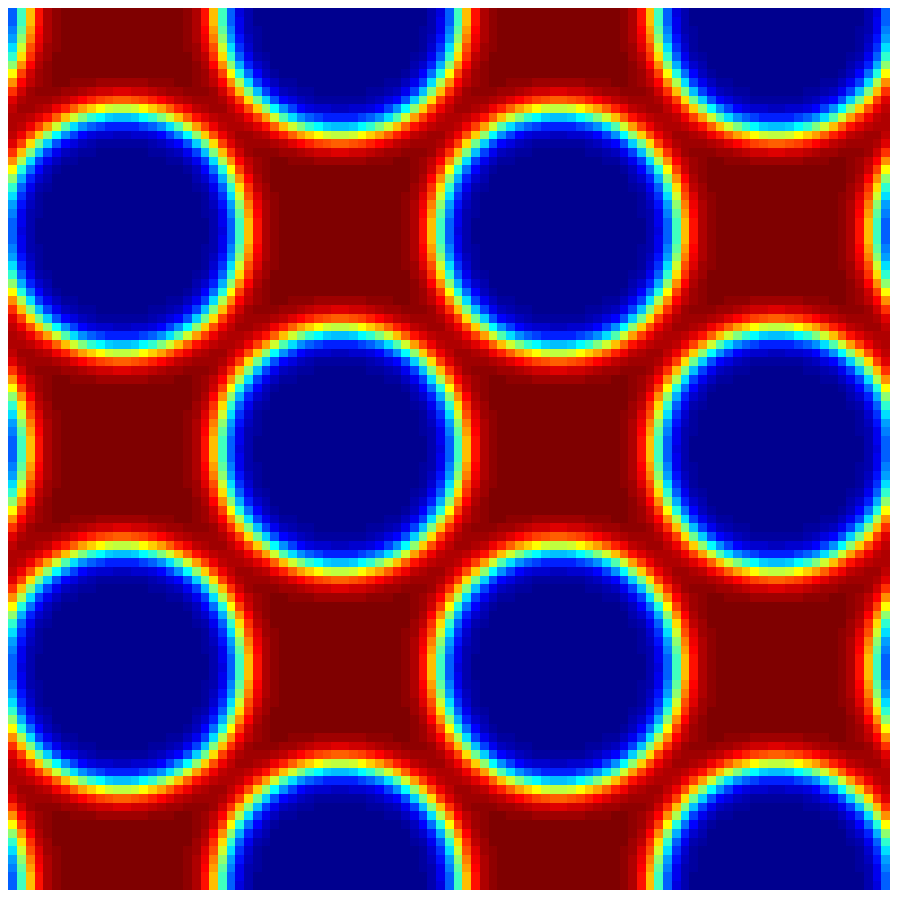}
}
\subfigure[Incompressible] {
\includegraphics[width=0.16\columnwidth]{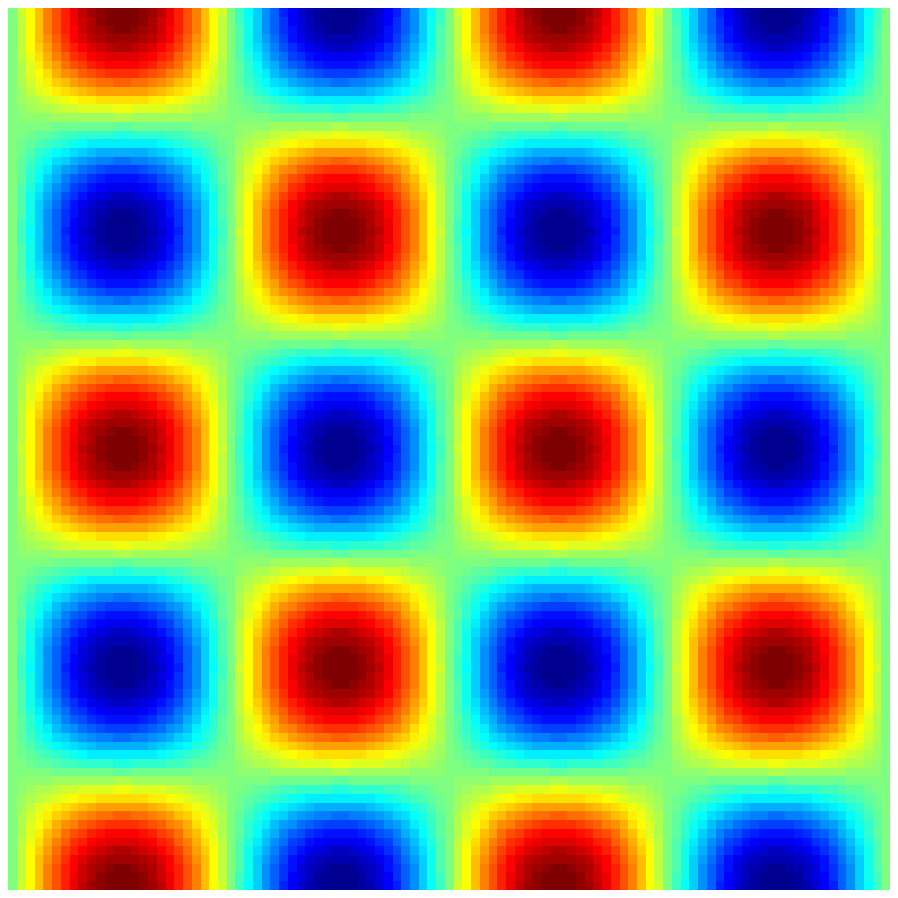}
\includegraphics[width=0.16\columnwidth]{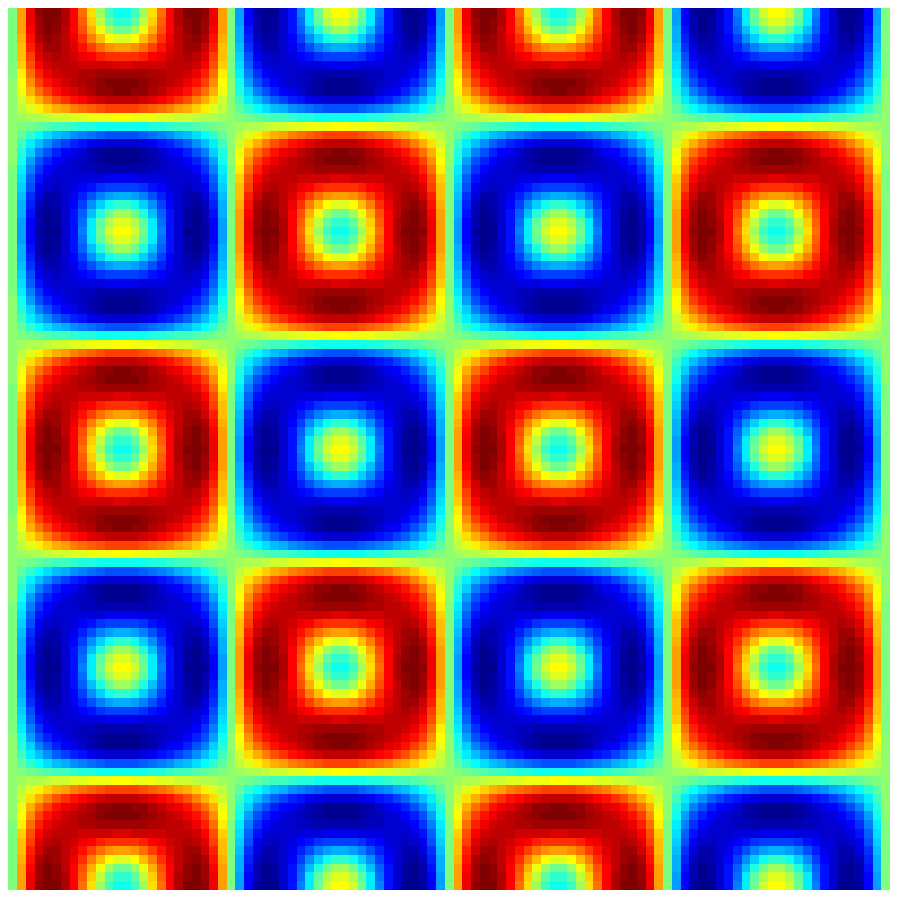}
\includegraphics[width=0.16\columnwidth]{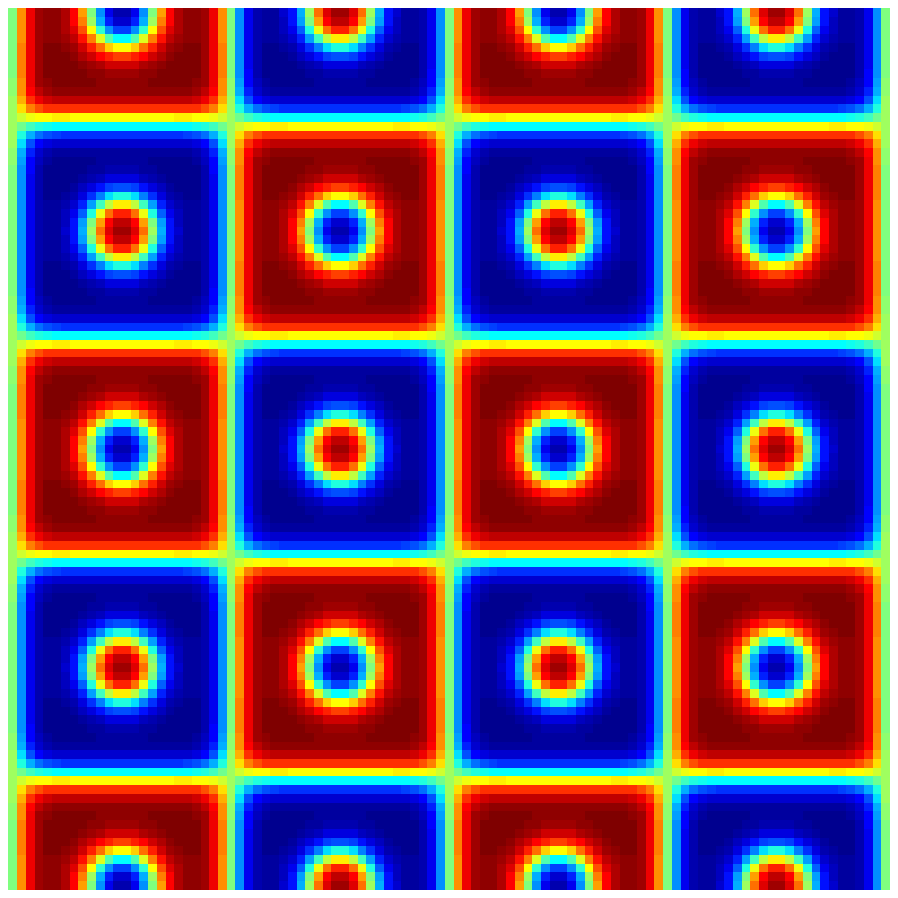}
\includegraphics[width=0.16\columnwidth]{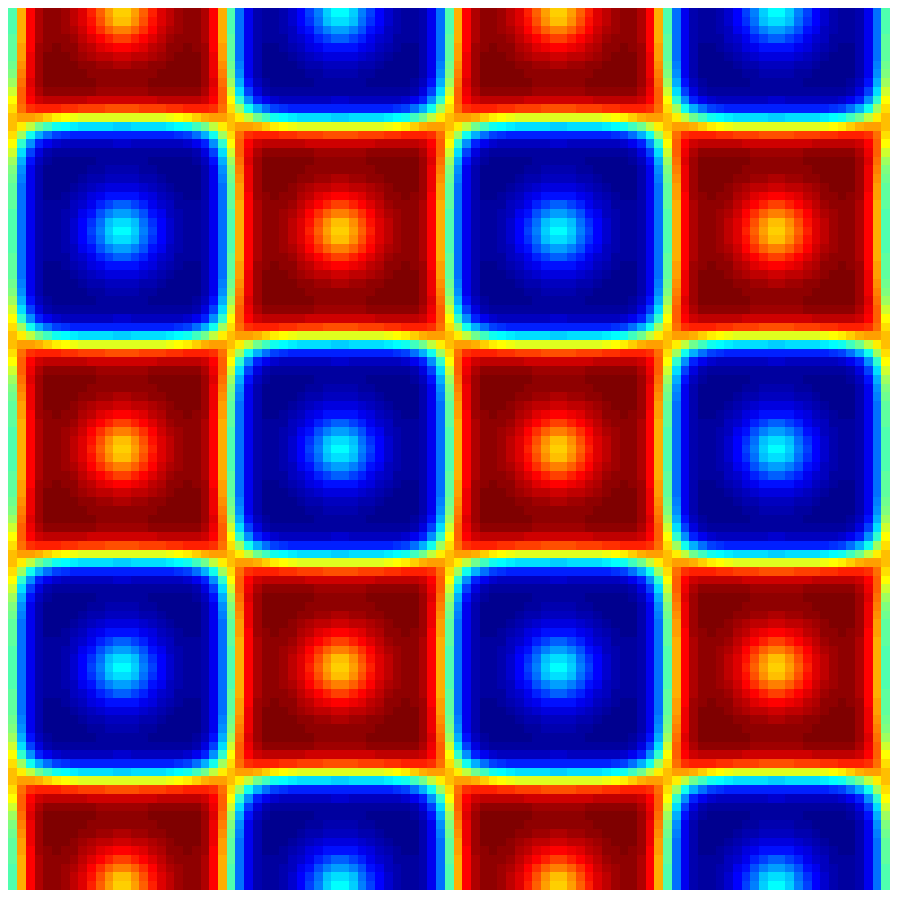}
\includegraphics[width=0.16\columnwidth]{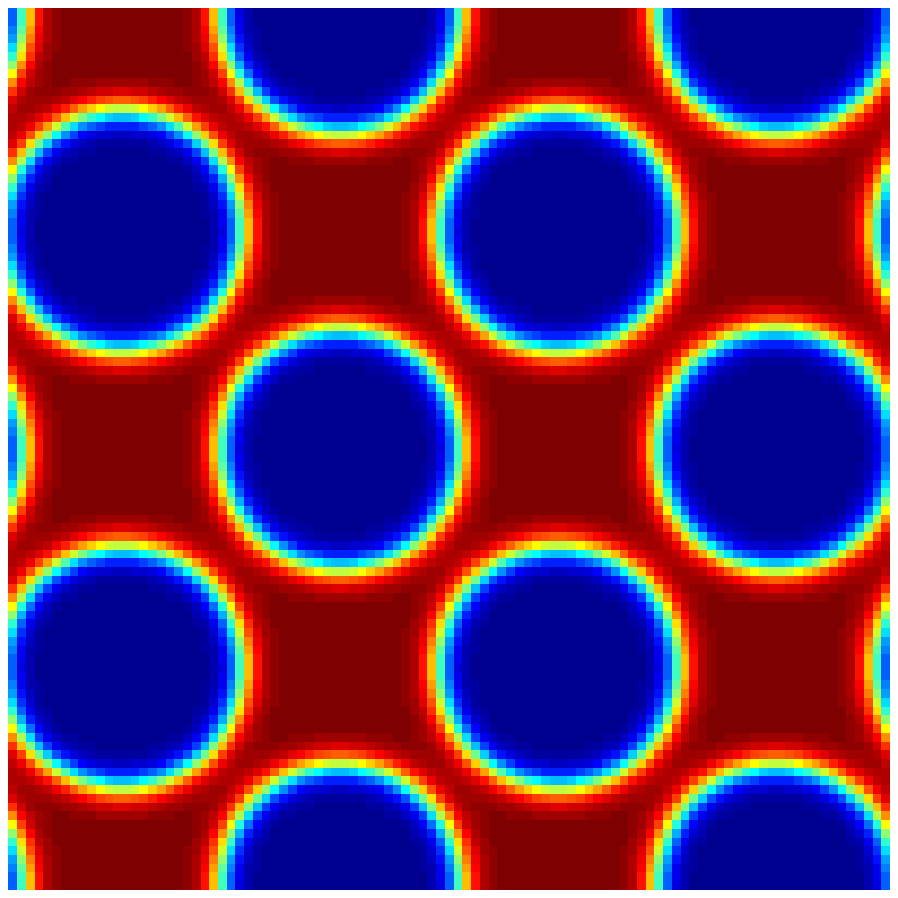}
}
\caption{(Color online) Density configuration with the minimum in blue and the maximum in red at $t/1000=0.1,3,5,7,20$ for the quasi-incompressible model (a) and incompressible model (b) as $\gamma=1$, $Pe=50$, and $Cn=4$.}
\label{fig:figure9}
\end{figure}

\begin{figure}[H] \centering
\subfigure[Quasi-incompressible] {
\includegraphics[width=0.16\columnwidth]{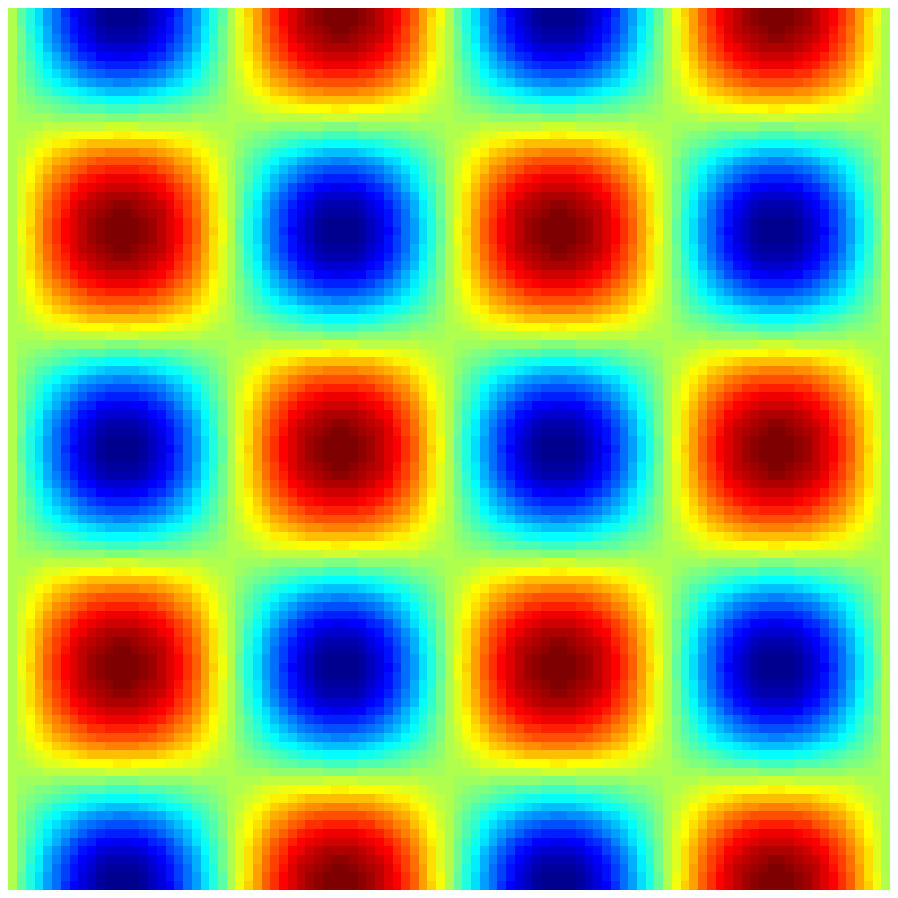}
\includegraphics[width=0.16\columnwidth]{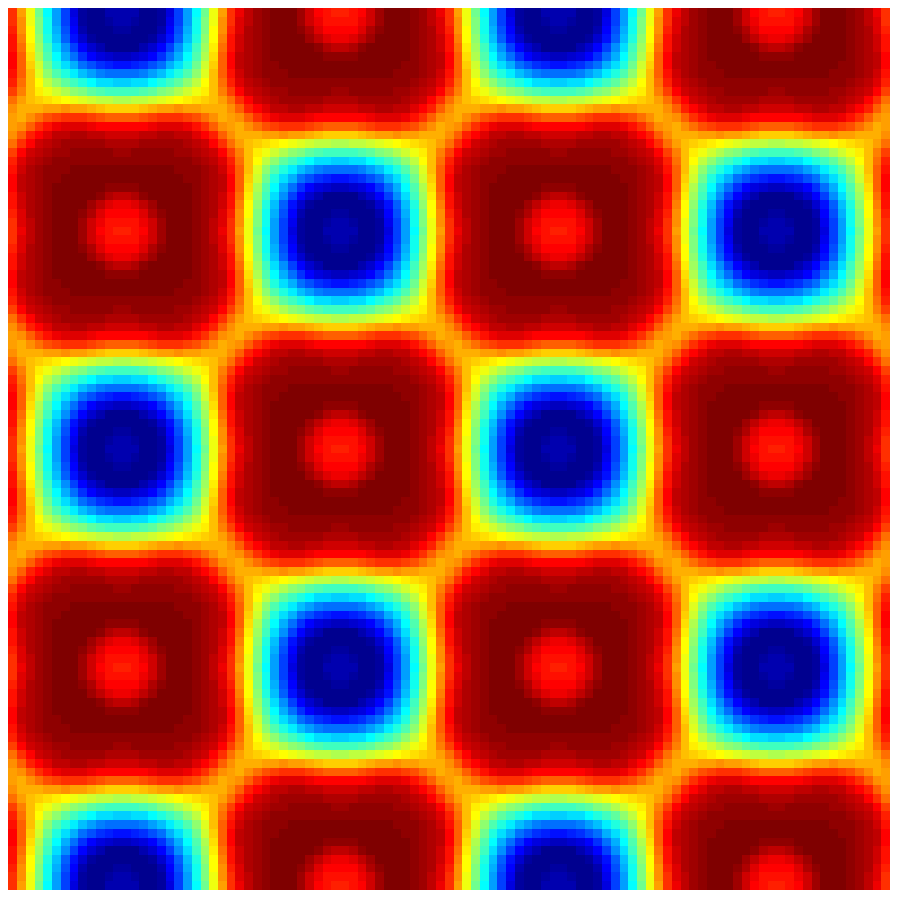}
\includegraphics[width=0.16\columnwidth]{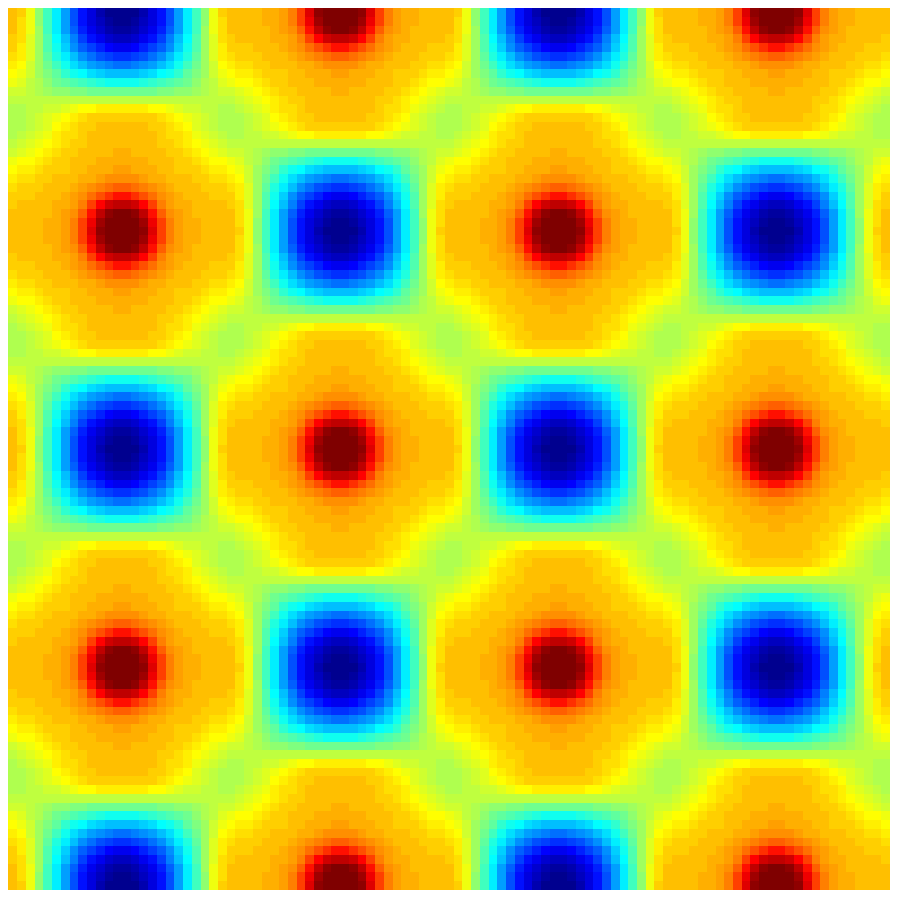}
\includegraphics[width=0.16\columnwidth]{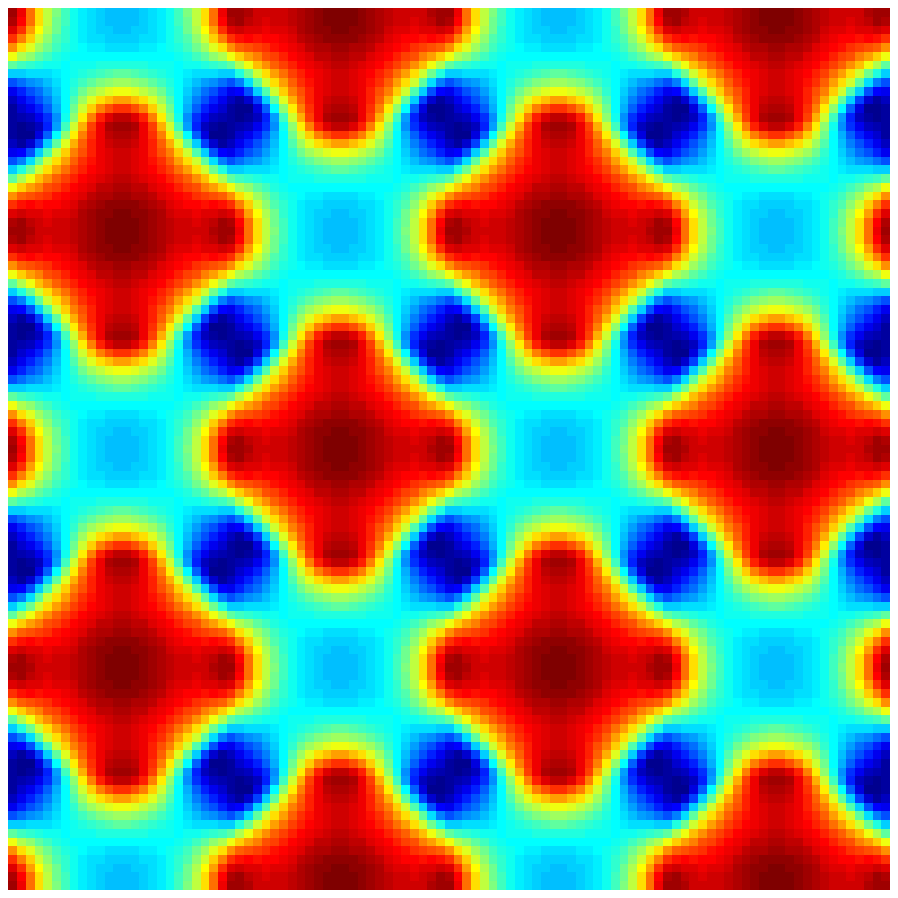}
\includegraphics[width=0.16\columnwidth]{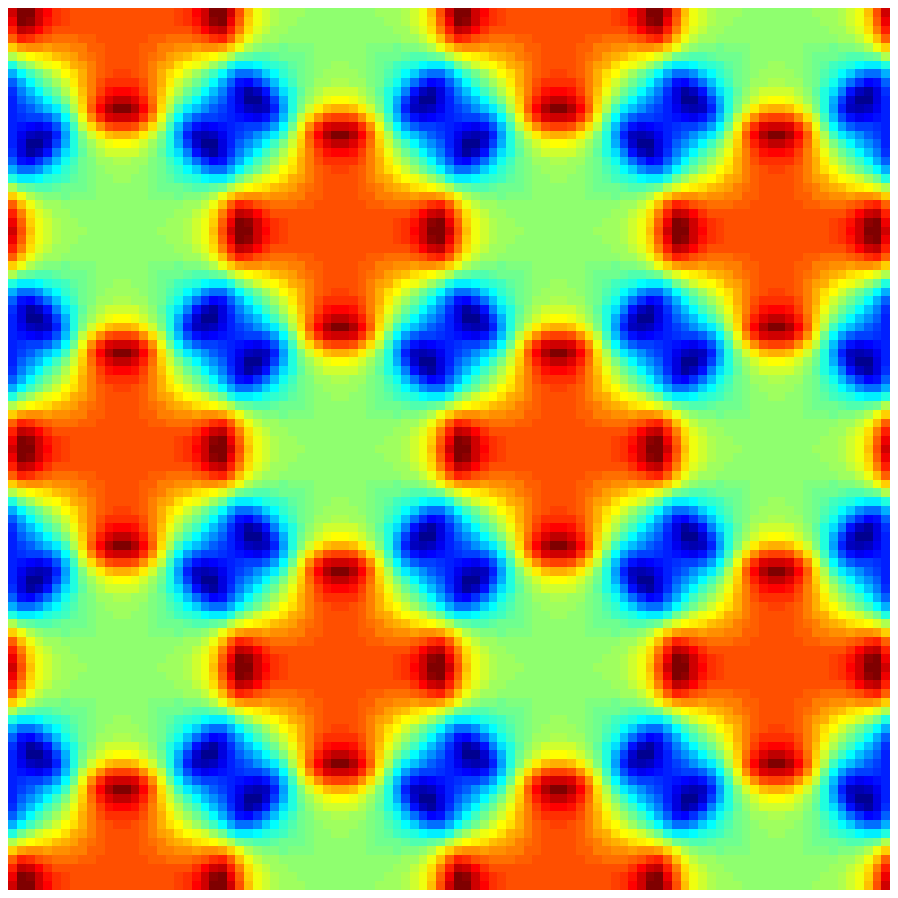}
}
\subfigure[Incompressible] {
\includegraphics[width=0.16\columnwidth]{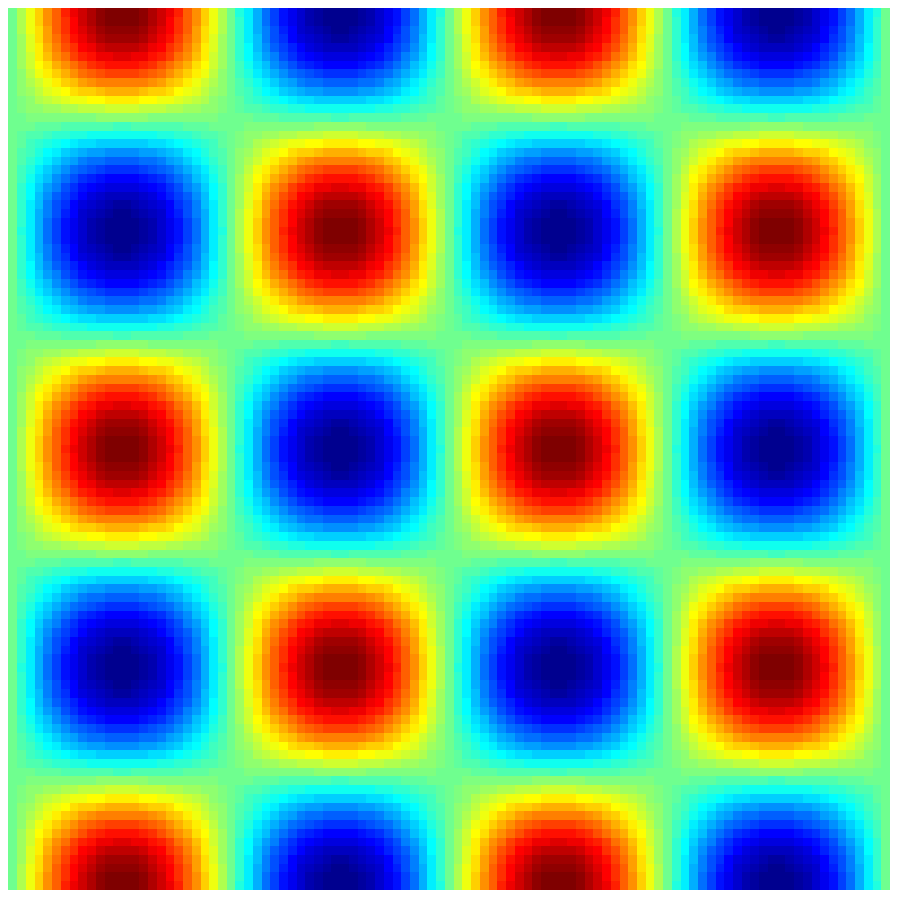}
\includegraphics[width=0.16\columnwidth]{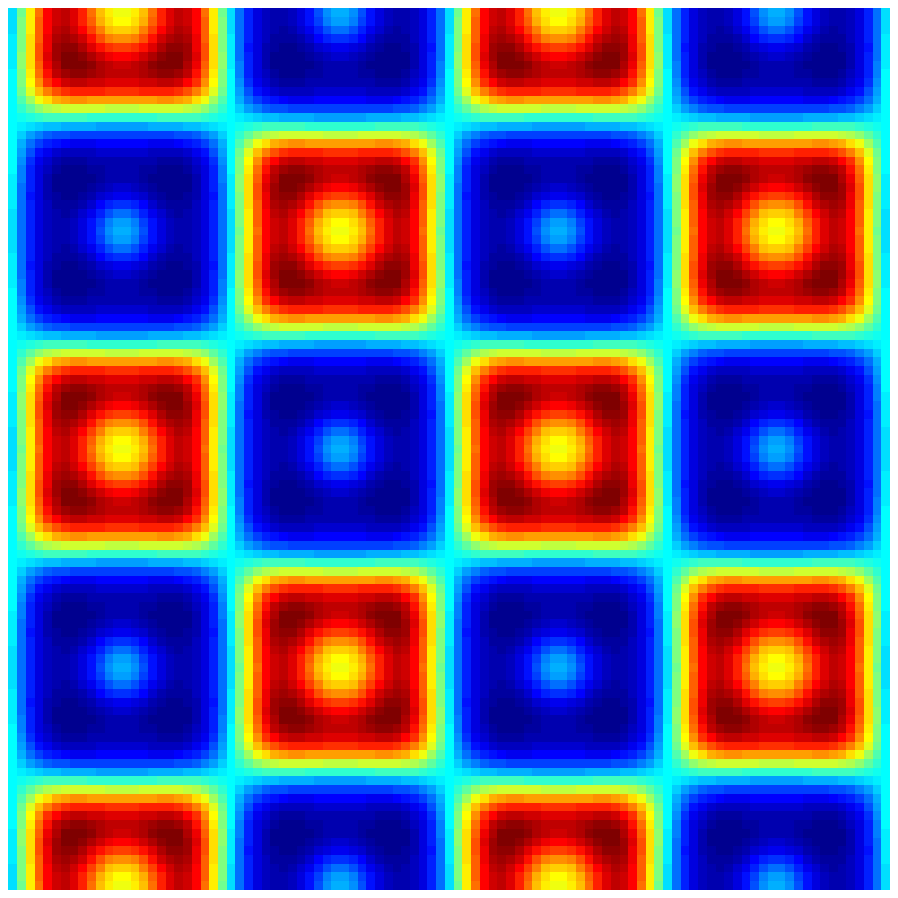}
\includegraphics[width=0.16\columnwidth]{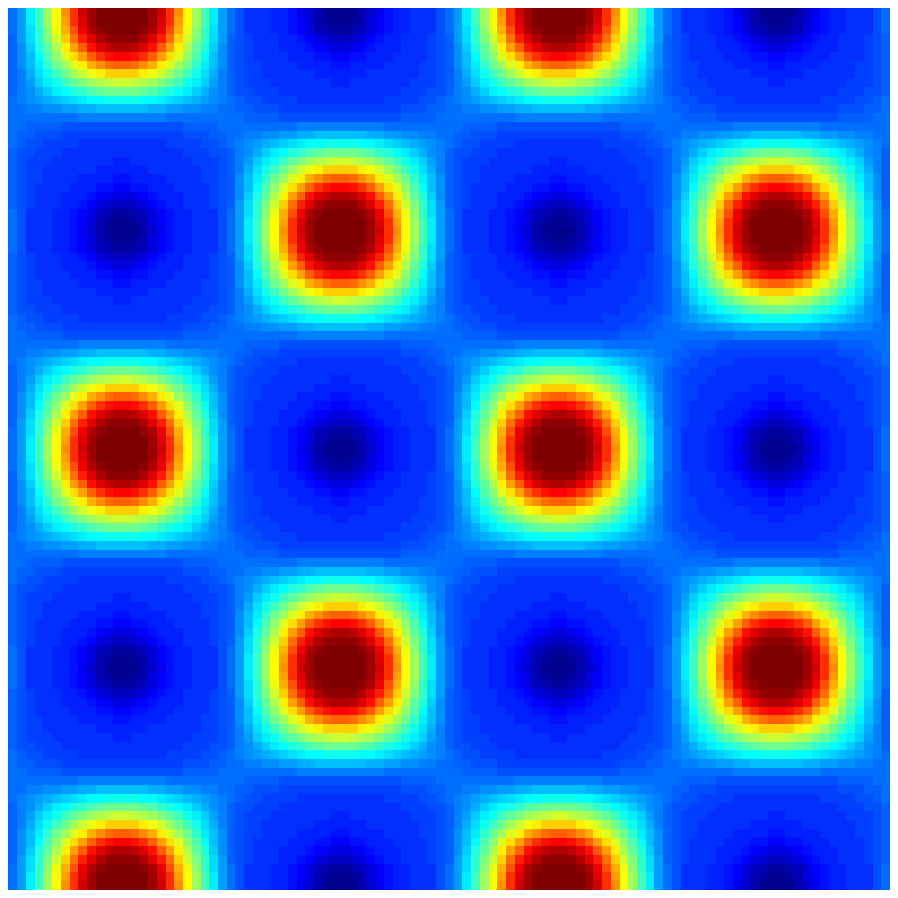}
\includegraphics[width=0.16\columnwidth]{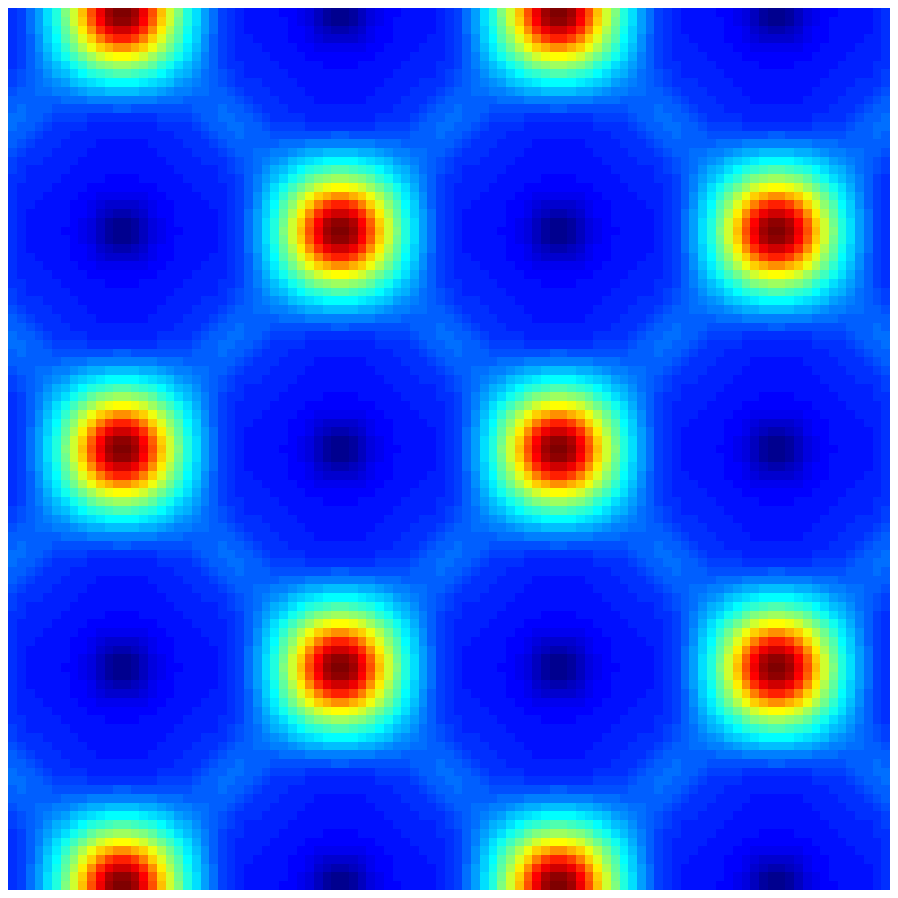}
\includegraphics[width=0.16\columnwidth]{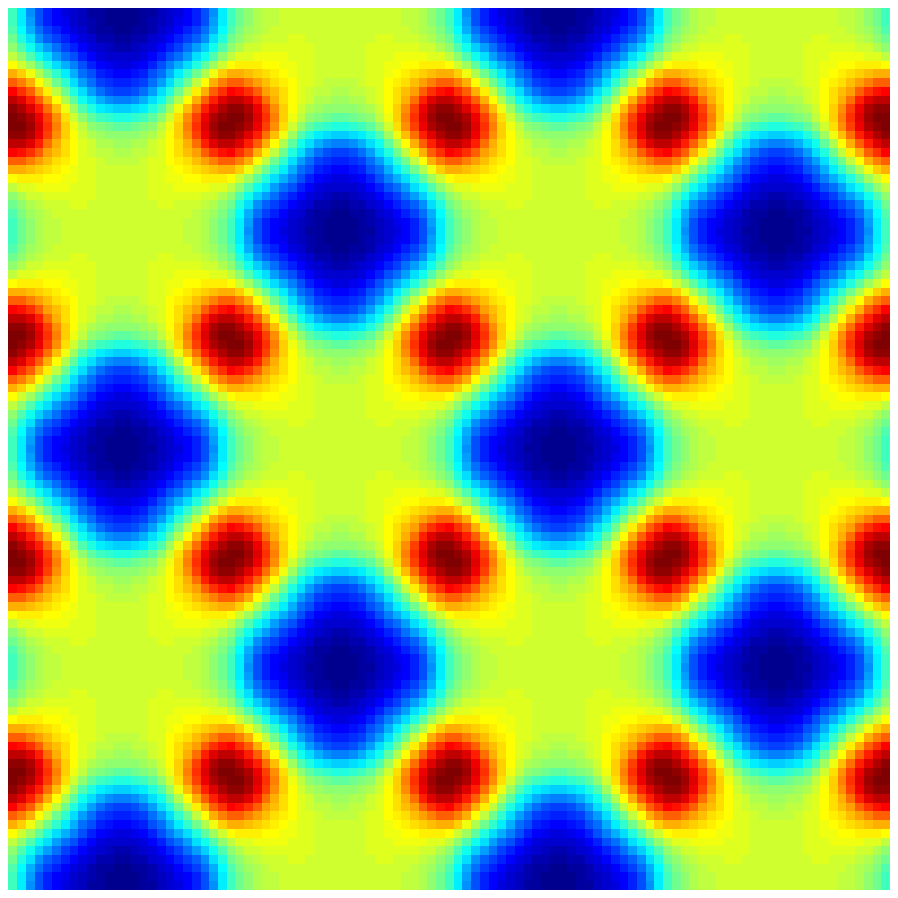}
}
\caption{(Color online) Dynamic pressure field with the minimum in blue and the maximum in red at $t/1000=0.1,3,5,7,20$ for the quasi-incompressible model (a) and incompressible model (b) as $\gamma=1$, $Pe=50$, and $Cn=4$.}
\label{fig:figure10}
\end{figure}

\begin{figure}[H] \centering
\subfigure[Quasi-incompressible] {
\includegraphics[width=0.16\columnwidth]{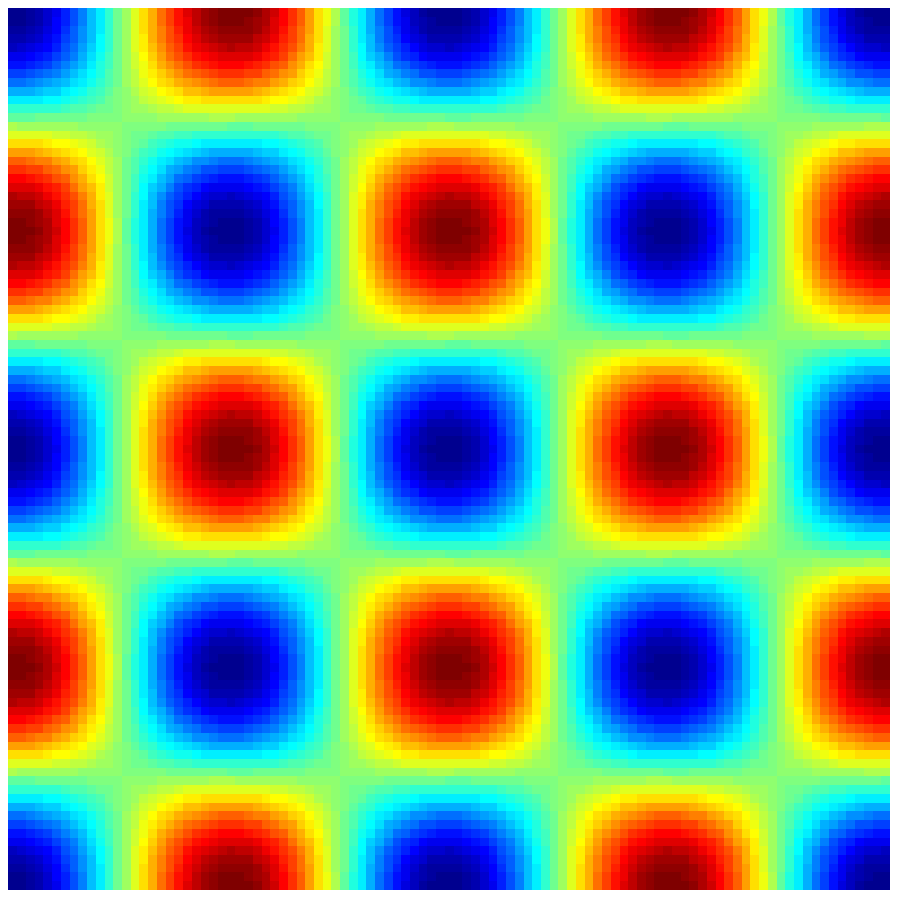}
\includegraphics[width=0.16\columnwidth]{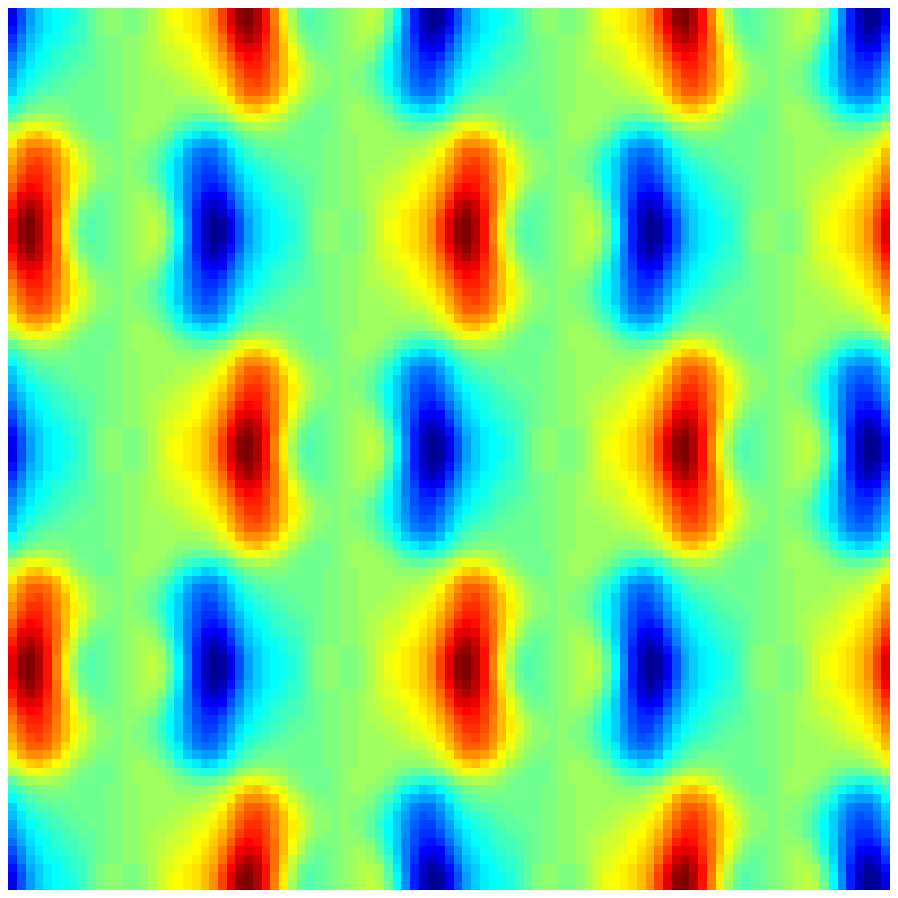}
\includegraphics[width=0.16\columnwidth]{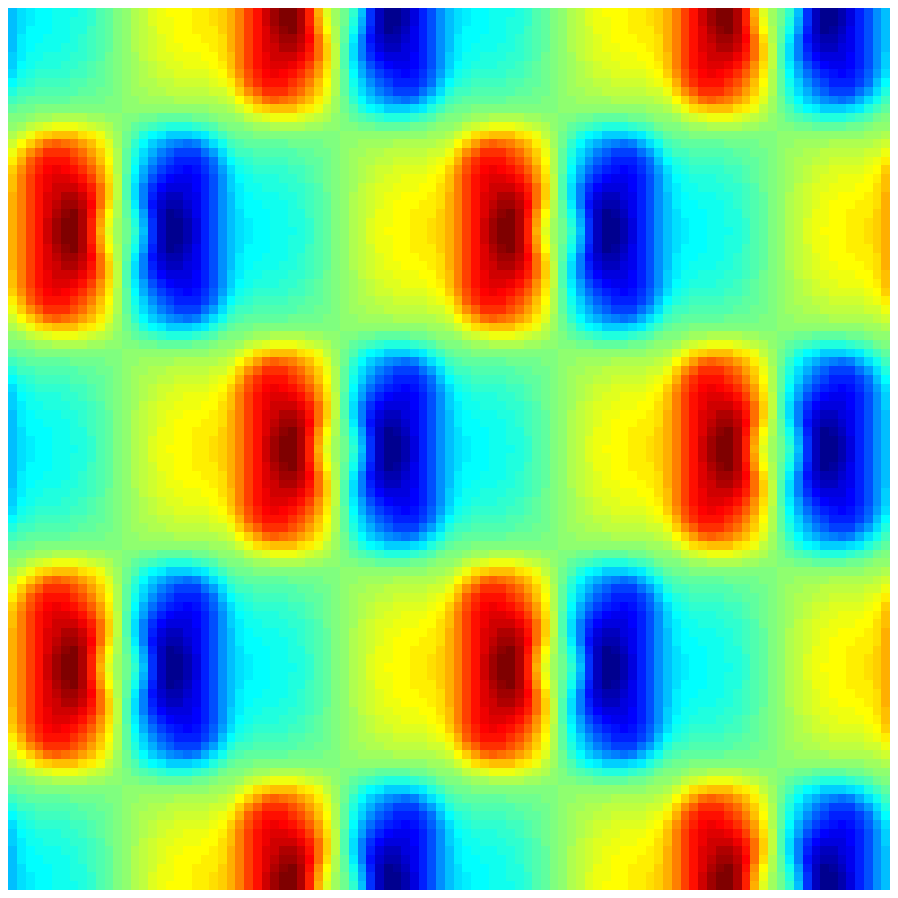}
\includegraphics[width=0.16\columnwidth]{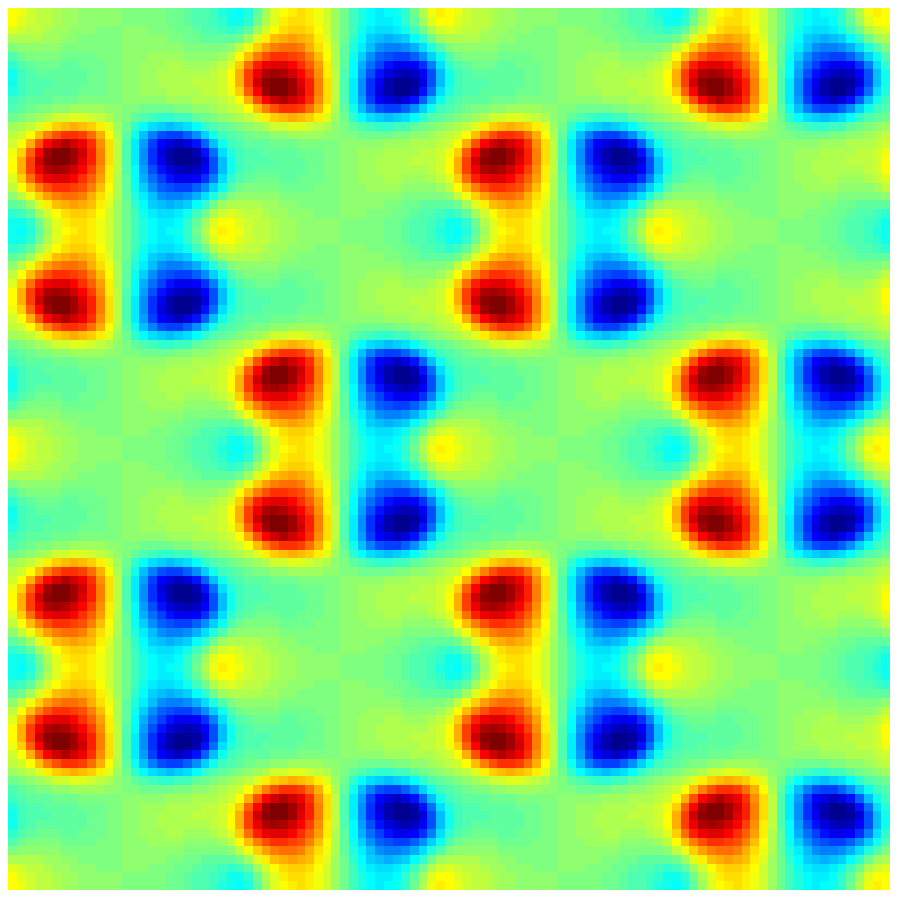}
\includegraphics[width=0.16\columnwidth]{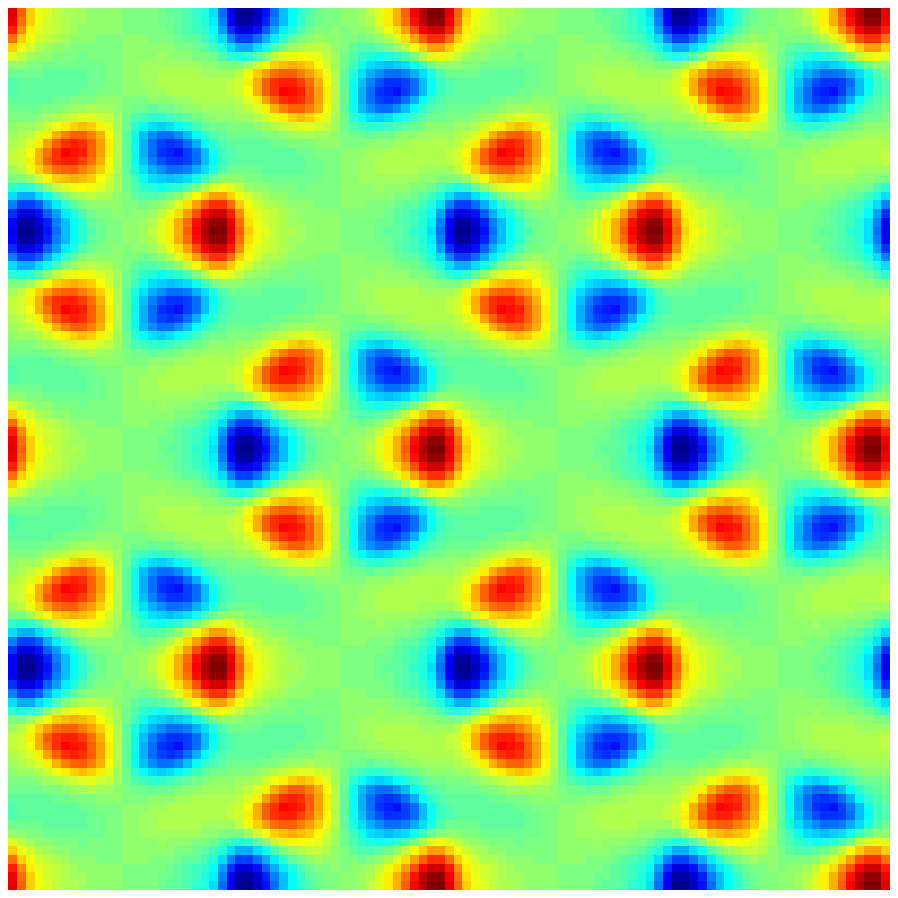}
}
\subfigure[Incompressible] {
\includegraphics[width=0.16\columnwidth]{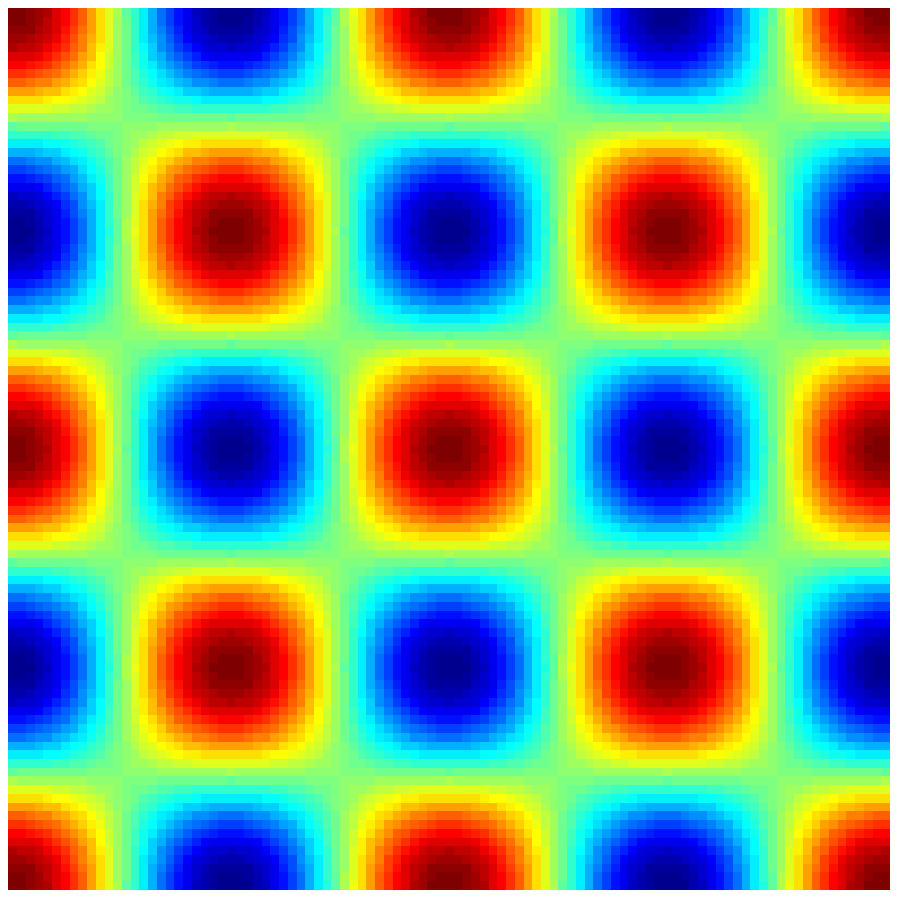}
\includegraphics[width=0.16\columnwidth]{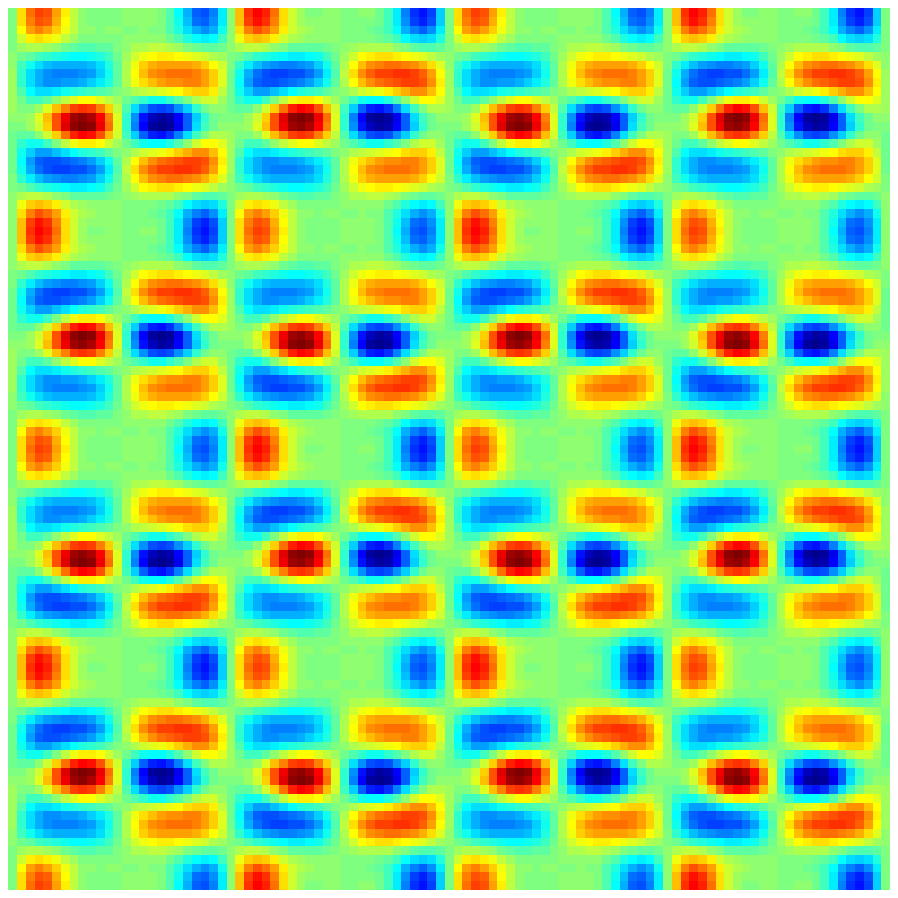}
\includegraphics[width=0.16\columnwidth]{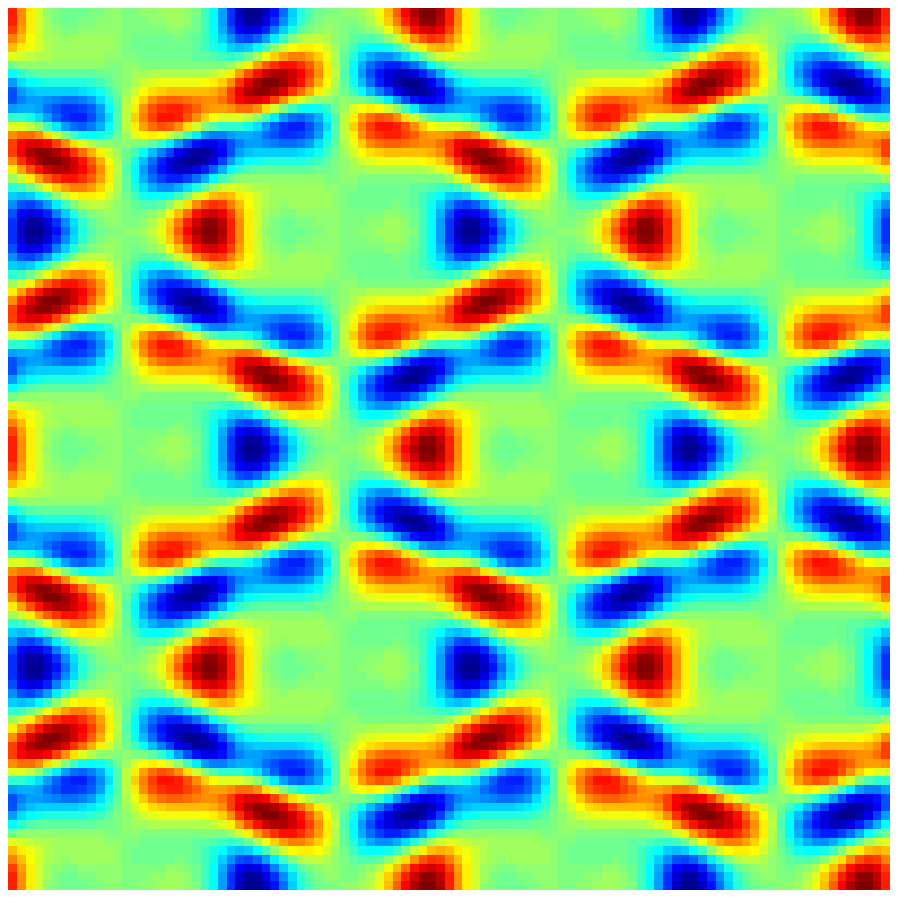}
\includegraphics[width=0.16\columnwidth]{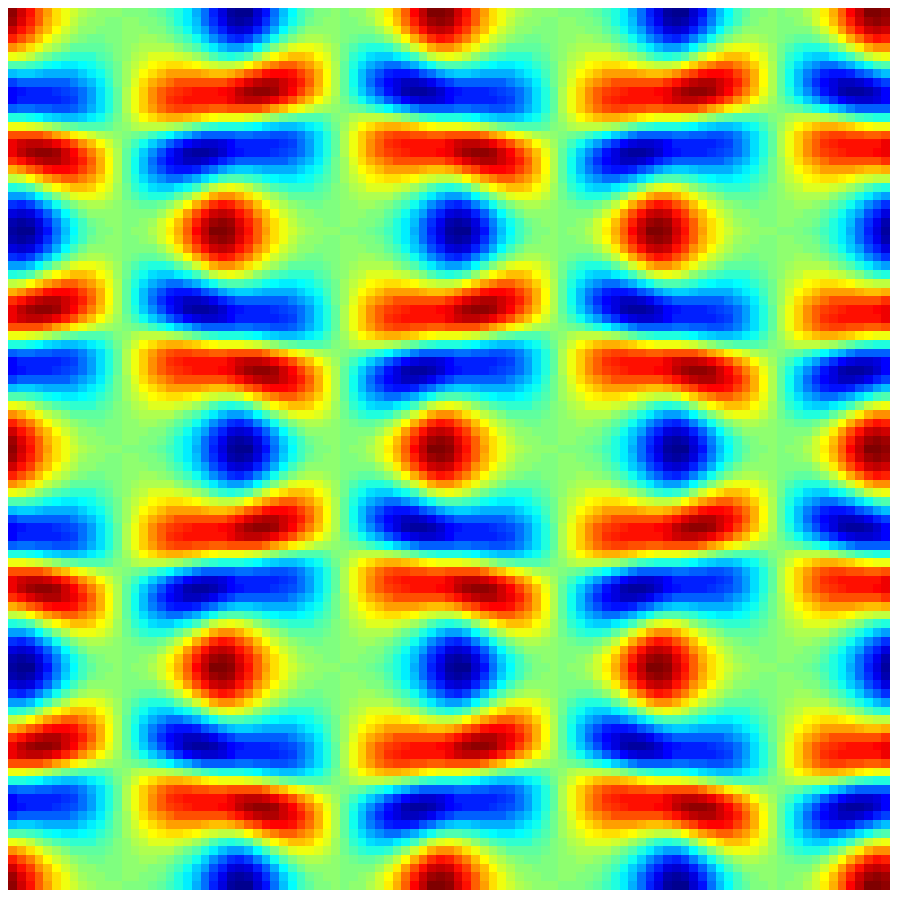}
\includegraphics[width=0.16\columnwidth]{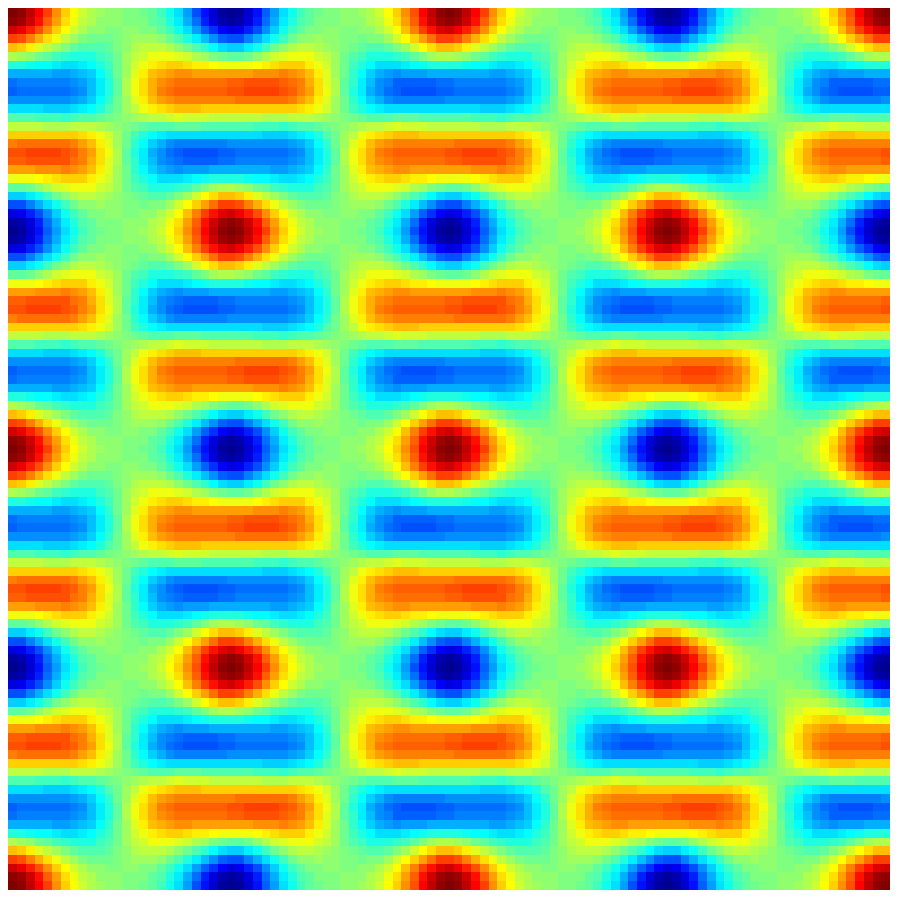}
}
\caption{(Color online) Horizontal velocity with the minimum in blue and the maximum in red at $t/1000=0.1,3,5,7,20$ for the quasi-incompressible model (a) and incompressible model (b) as $\gamma=1$, $Pe=50$, and $Cn=4$.}
\label{fig:figure11}
\end{figure}

\begin{figure}[H] \centering
\subfigure[Quasi-incompressible] {
\includegraphics[width=0.16\columnwidth]{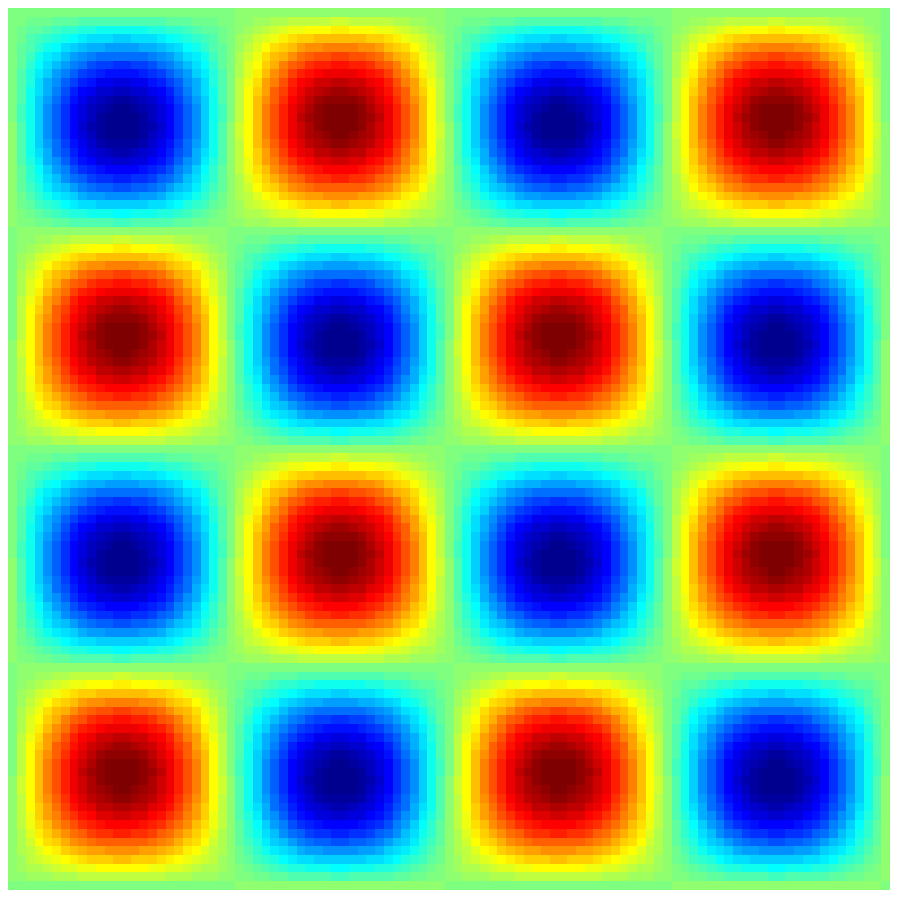}
\includegraphics[width=0.16\columnwidth]{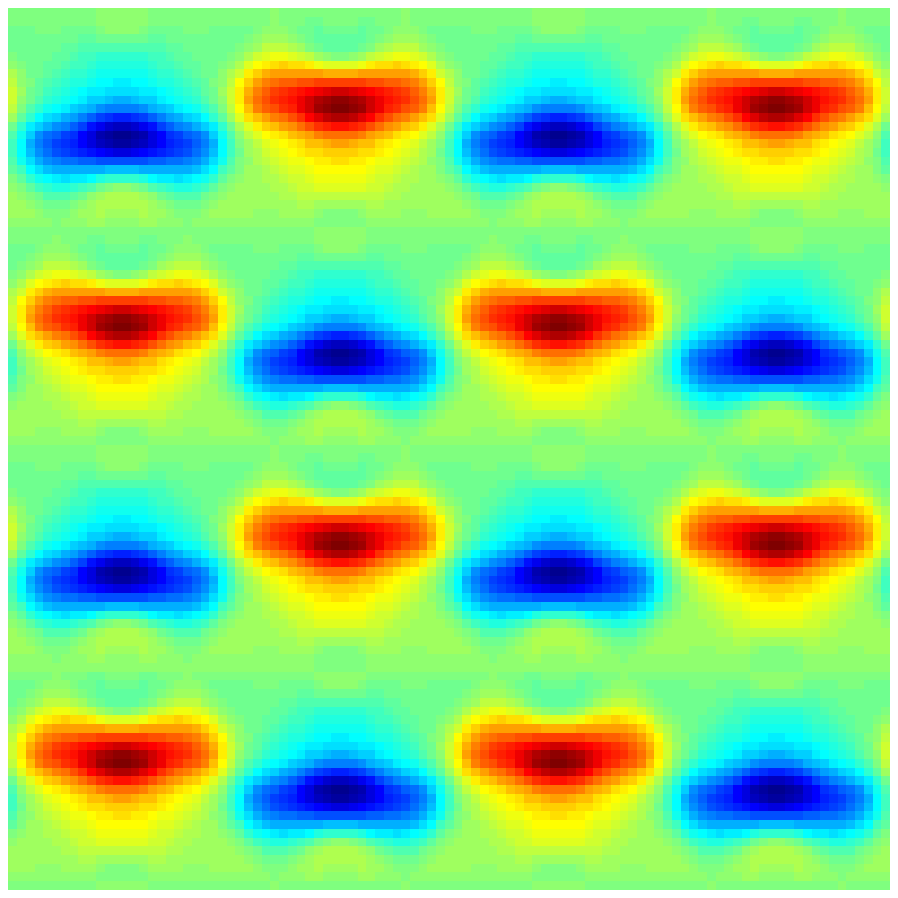}
\includegraphics[width=0.16\columnwidth]{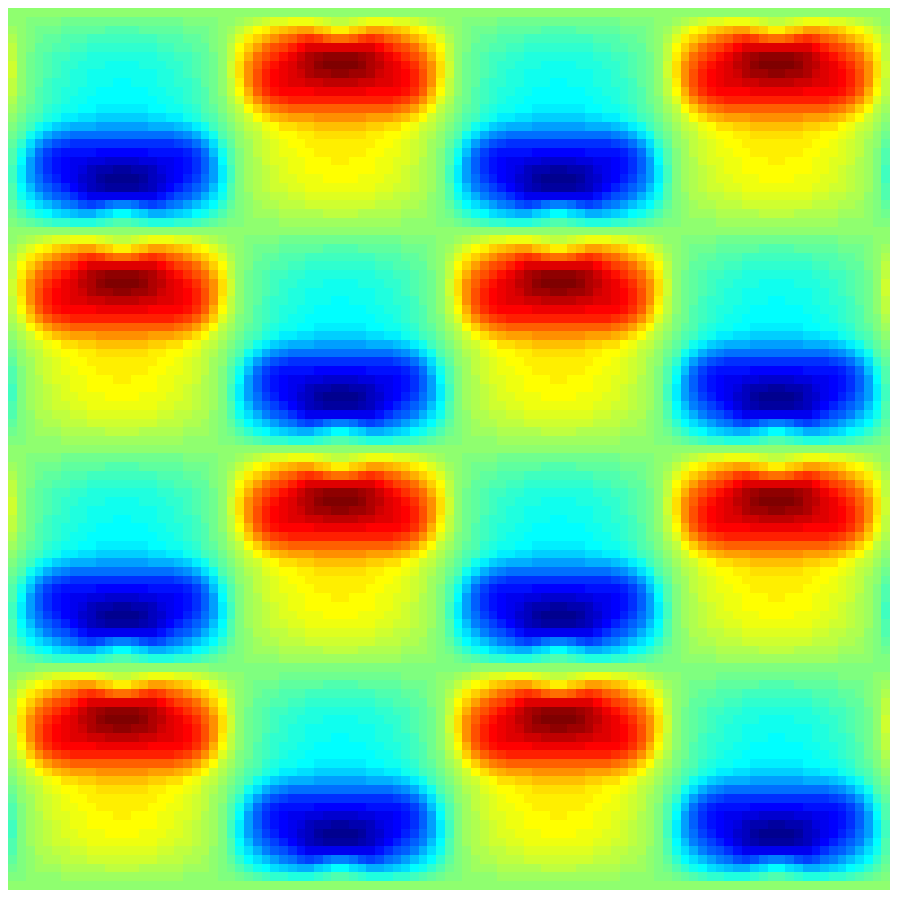}
\includegraphics[width=0.16\columnwidth]{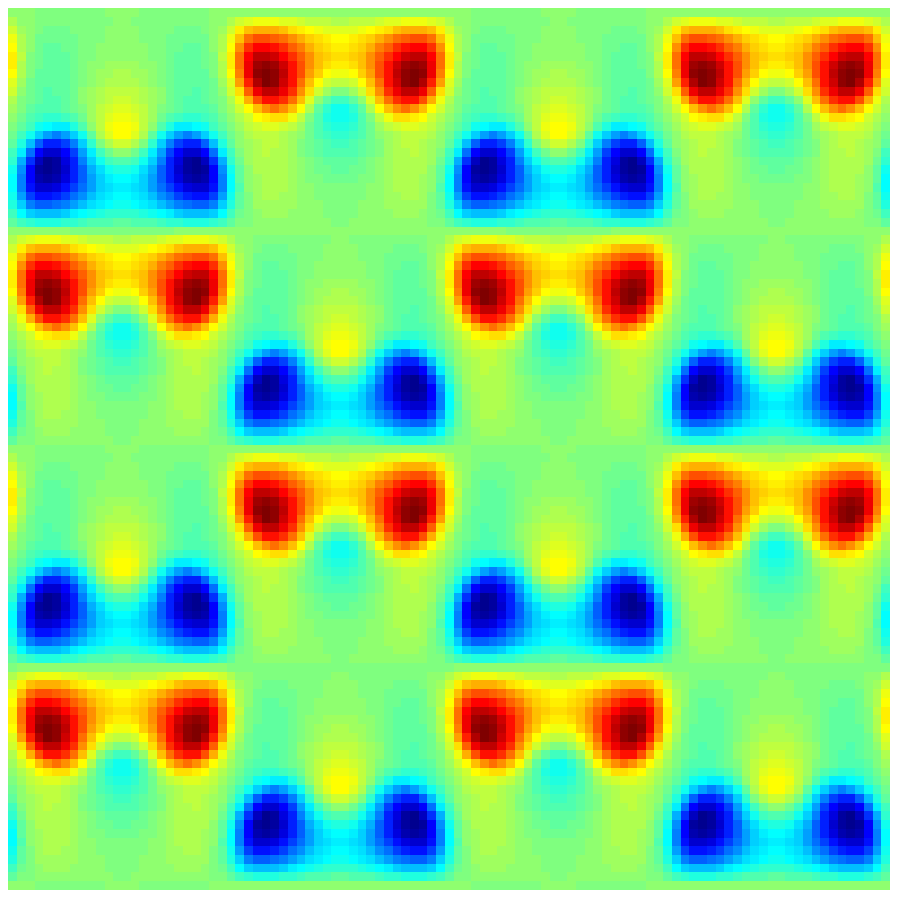}
\includegraphics[width=0.16\columnwidth]{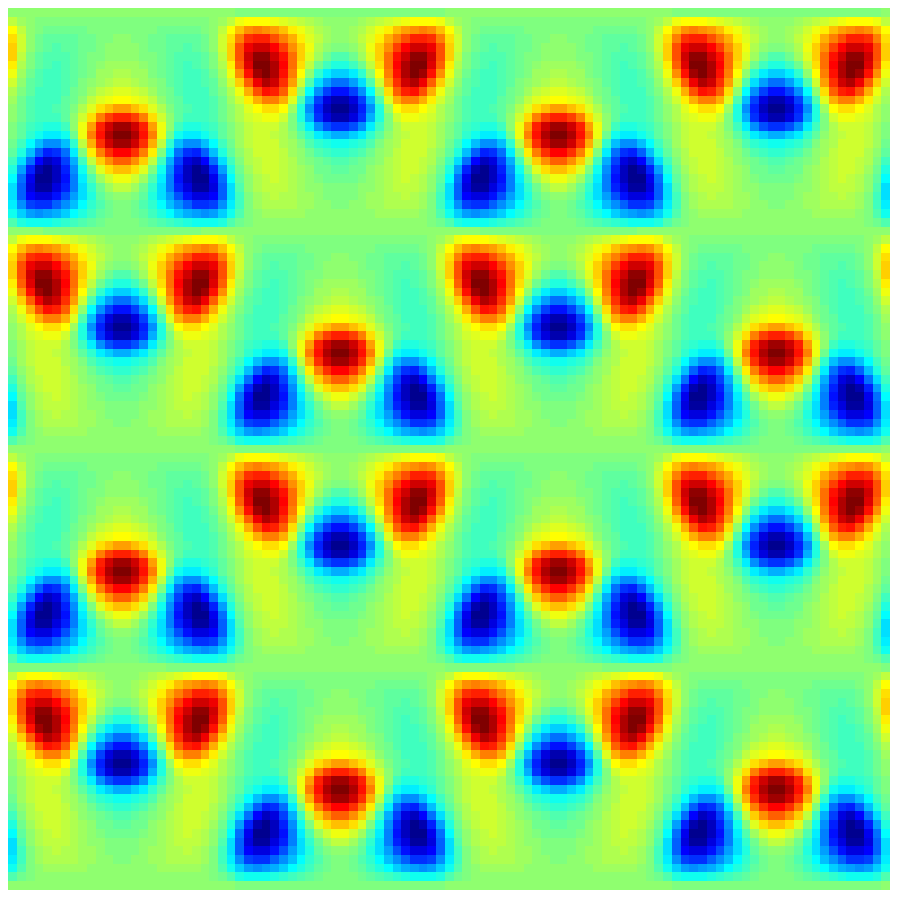}
}
\subfigure[Incompressible] {
\includegraphics[width=0.16\columnwidth]{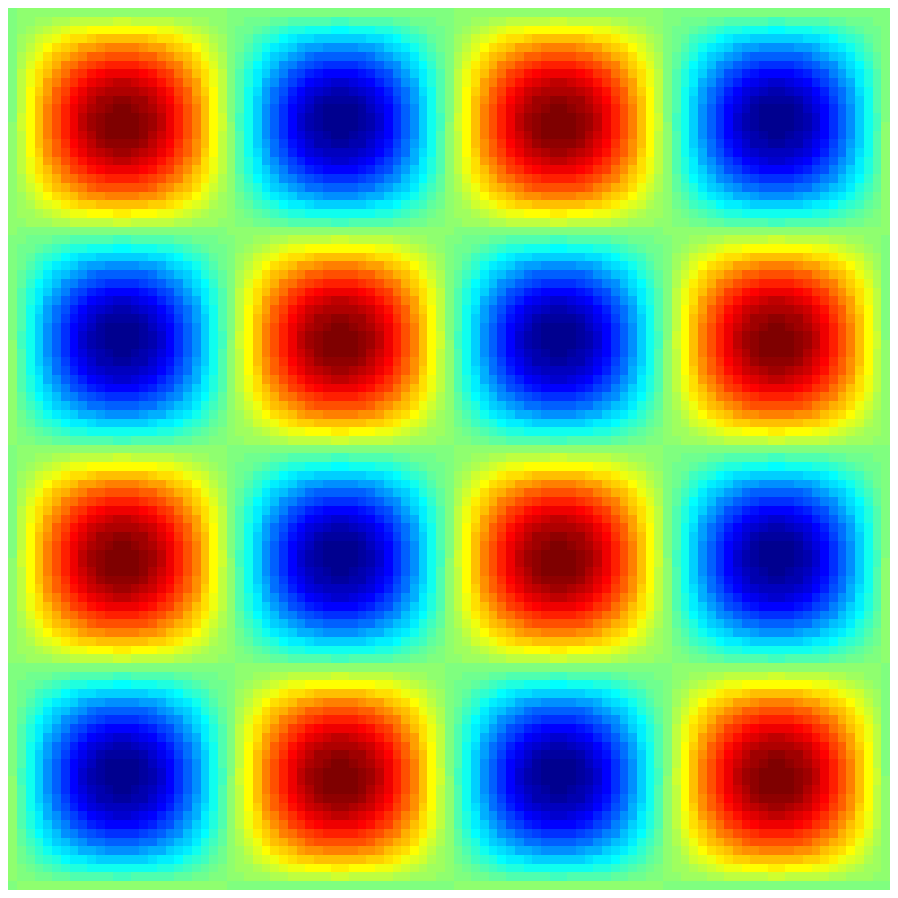}
\includegraphics[width=0.16\columnwidth]{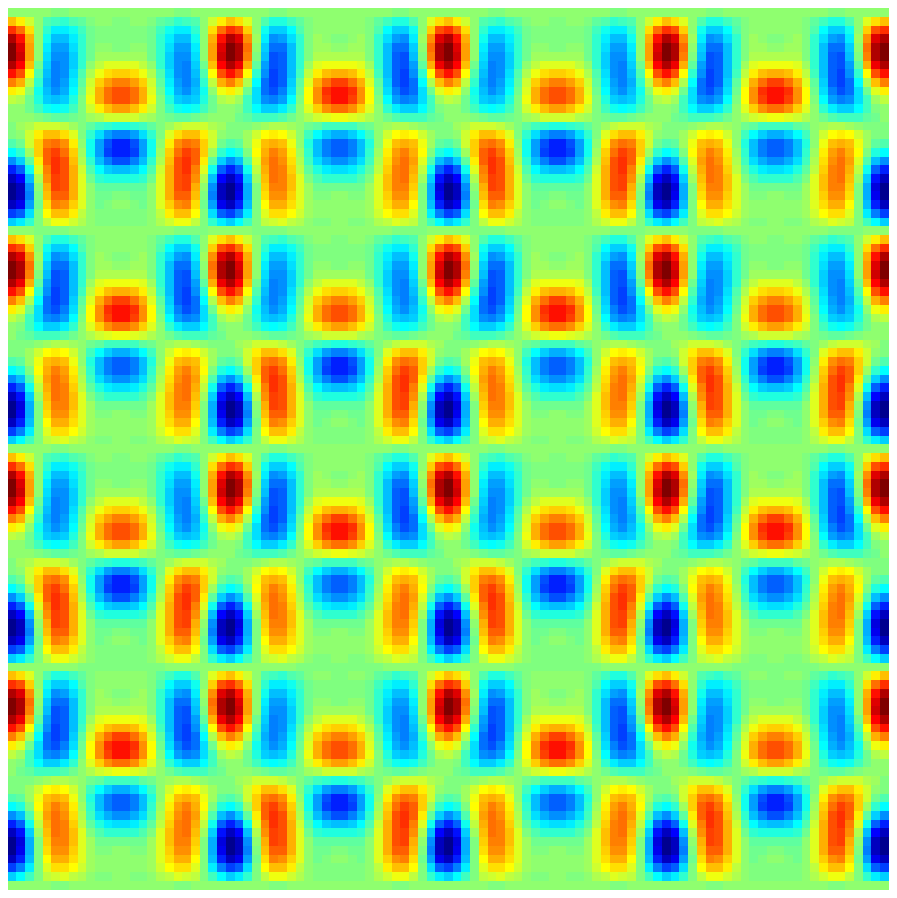}
\includegraphics[width=0.16\columnwidth]{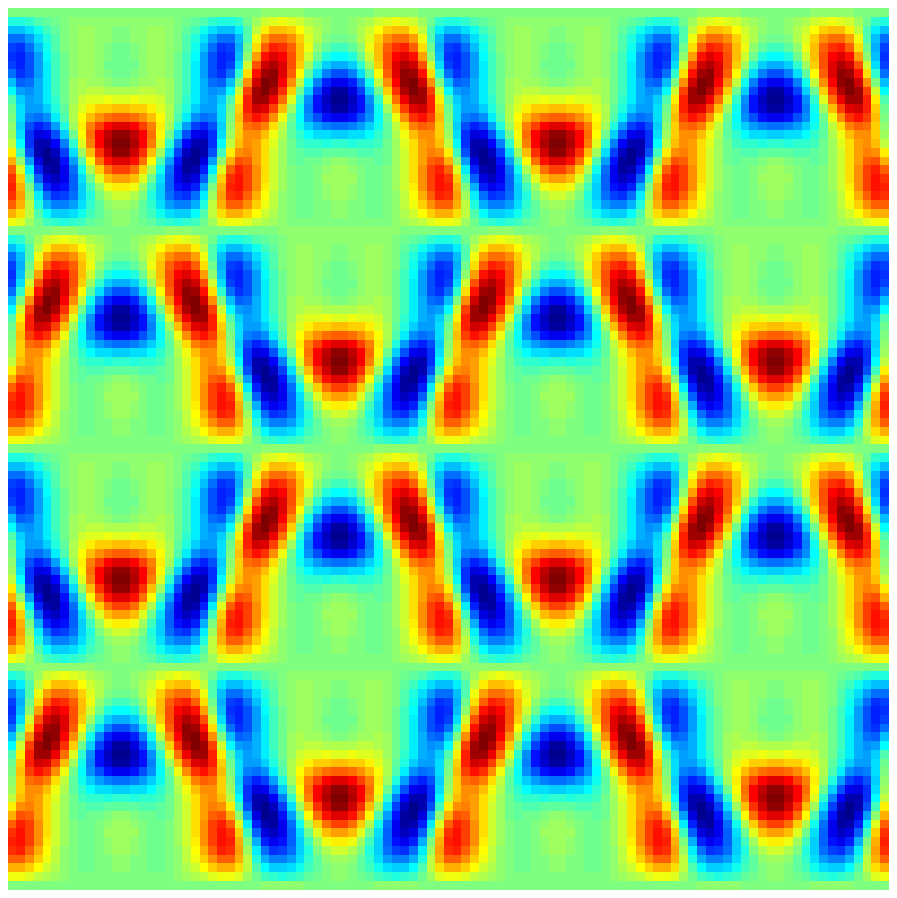}
\includegraphics[width=0.16\columnwidth]{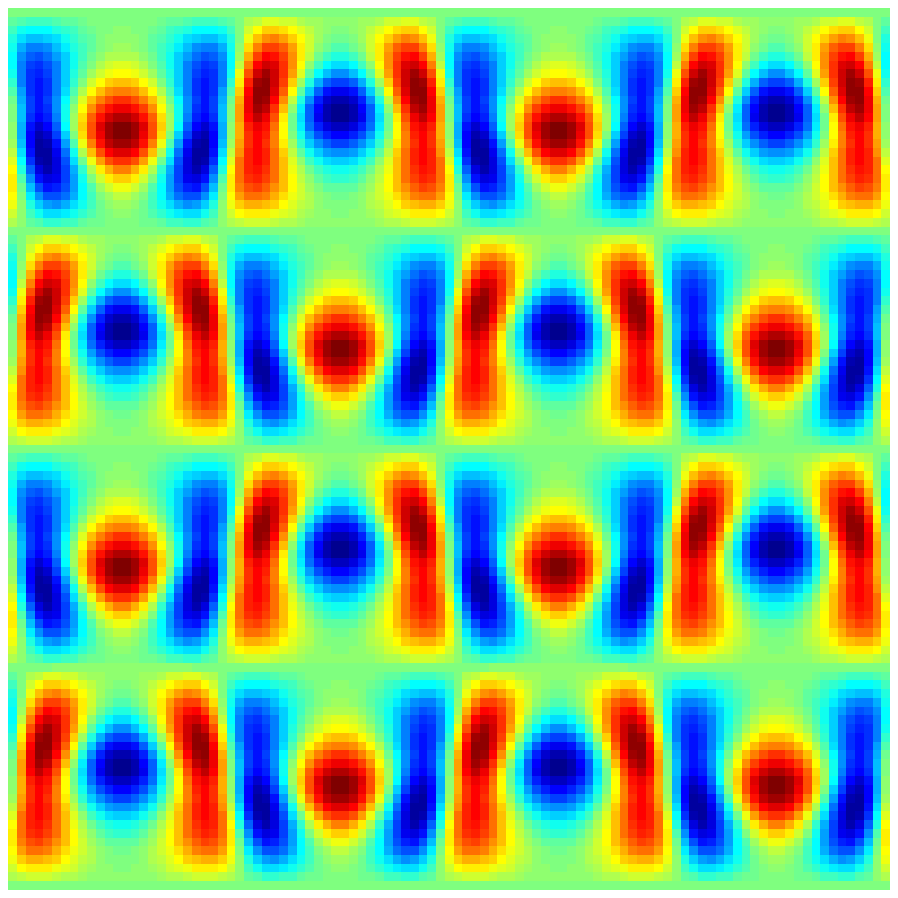}
\includegraphics[width=0.16\columnwidth]{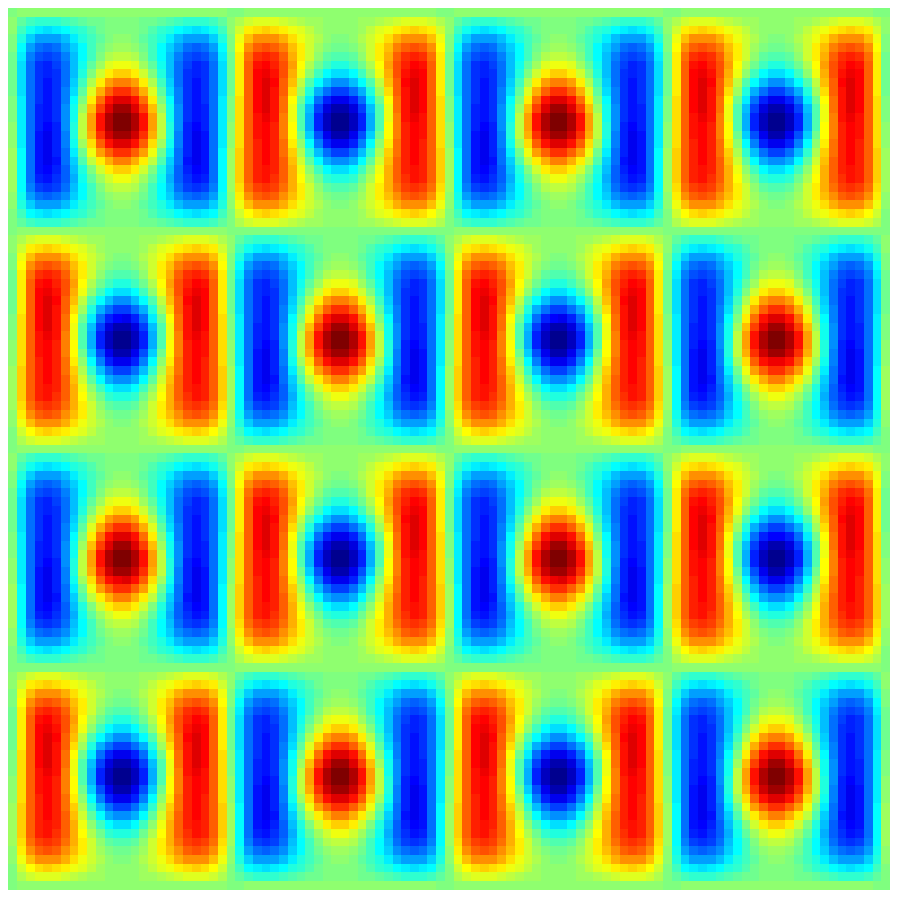}
}
\caption{(Color online) Vertical velocity with the minimum in blue and the maximum in red at $t/1000=0.1,3,5,7,20$ for the quasi-incompressible model (a) and incompressible model (b) as $\gamma=1$, $Pe=50$, and $Cn=4$.}
\label{fig:figure12}
\end{figure}

\begin{figure}[H] \centering
\subfigure[Quasi-incompressible] {
\includegraphics[width=0.16\columnwidth]{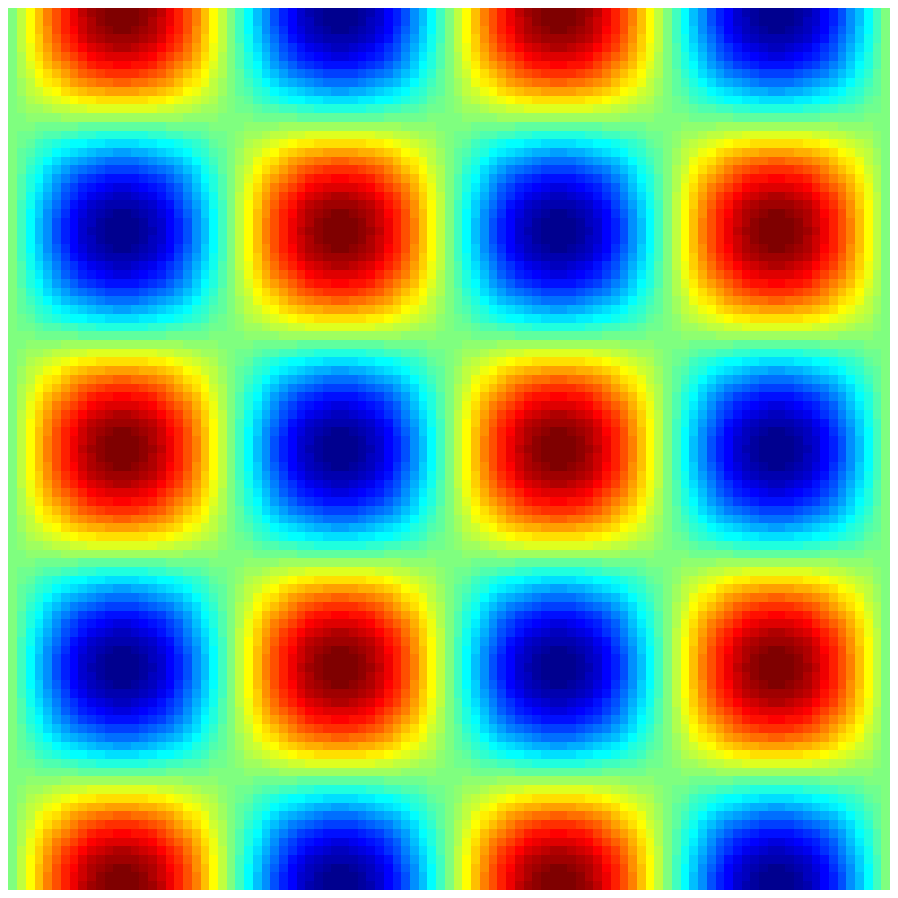}
\includegraphics[width=0.16\columnwidth]{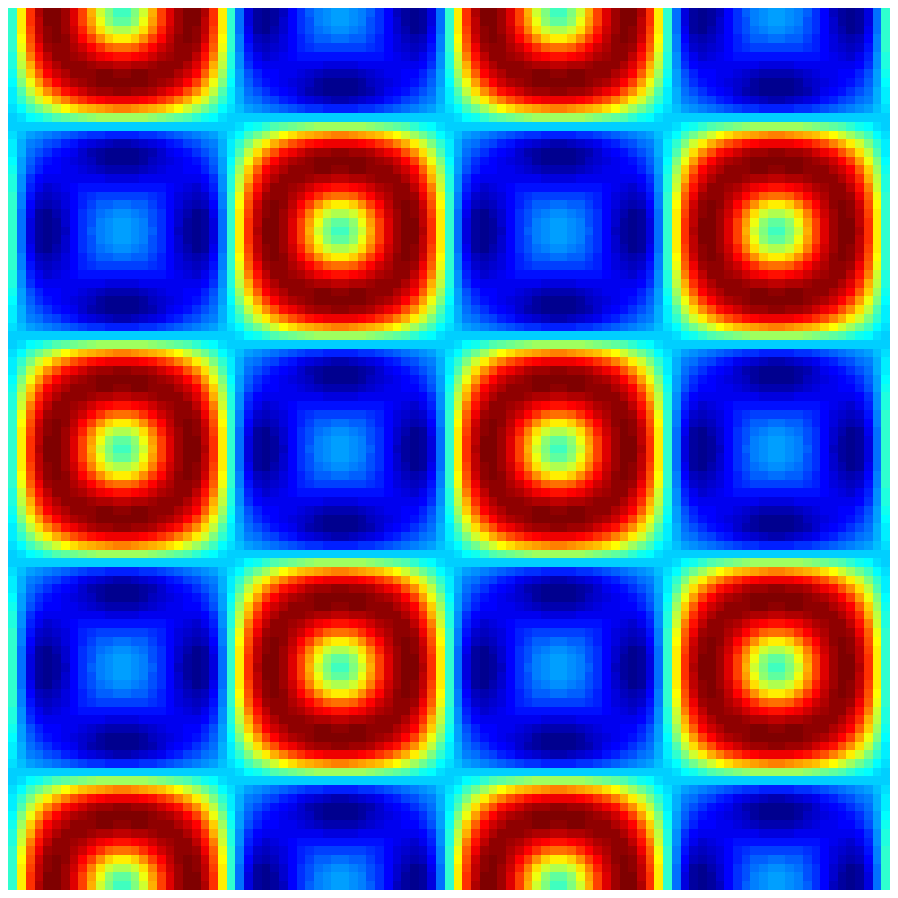}
\includegraphics[width=0.16\columnwidth]{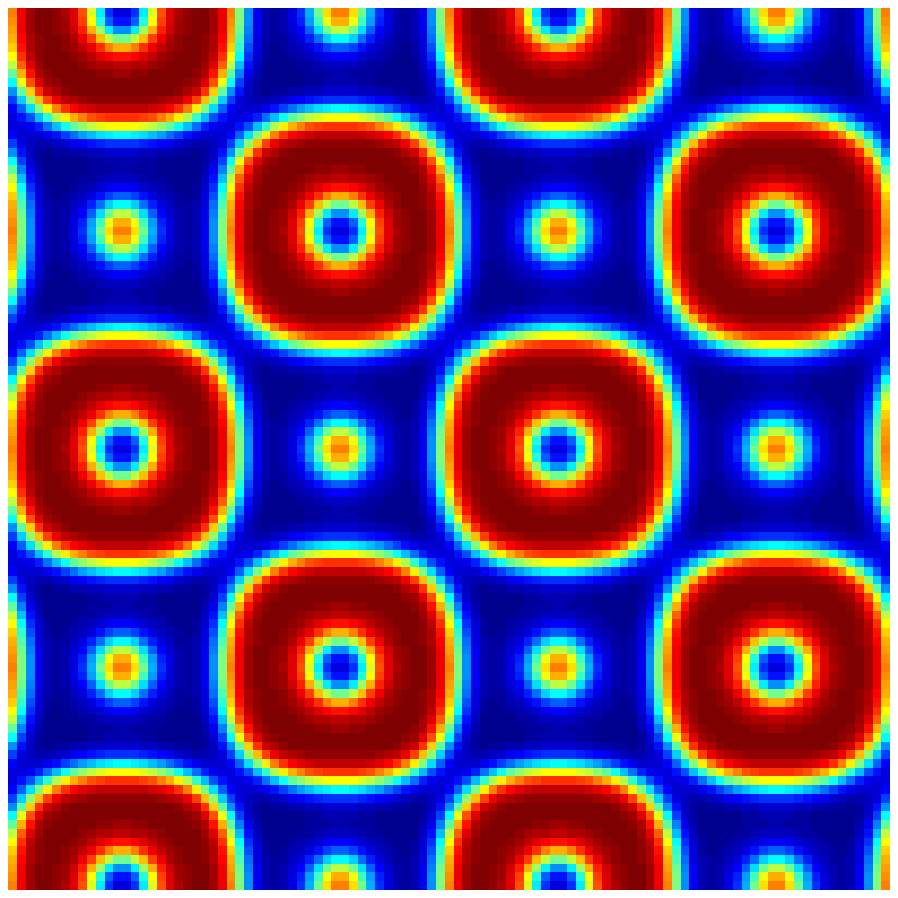}
\includegraphics[width=0.16\columnwidth]{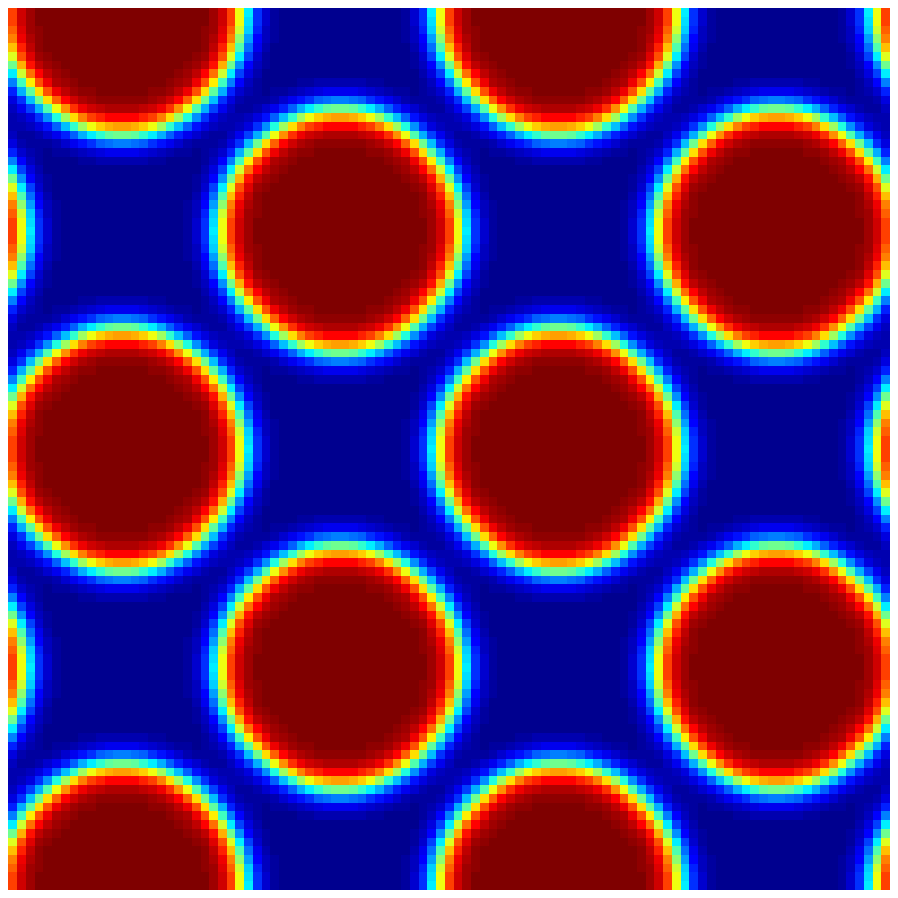}
\includegraphics[width=0.16\columnwidth]{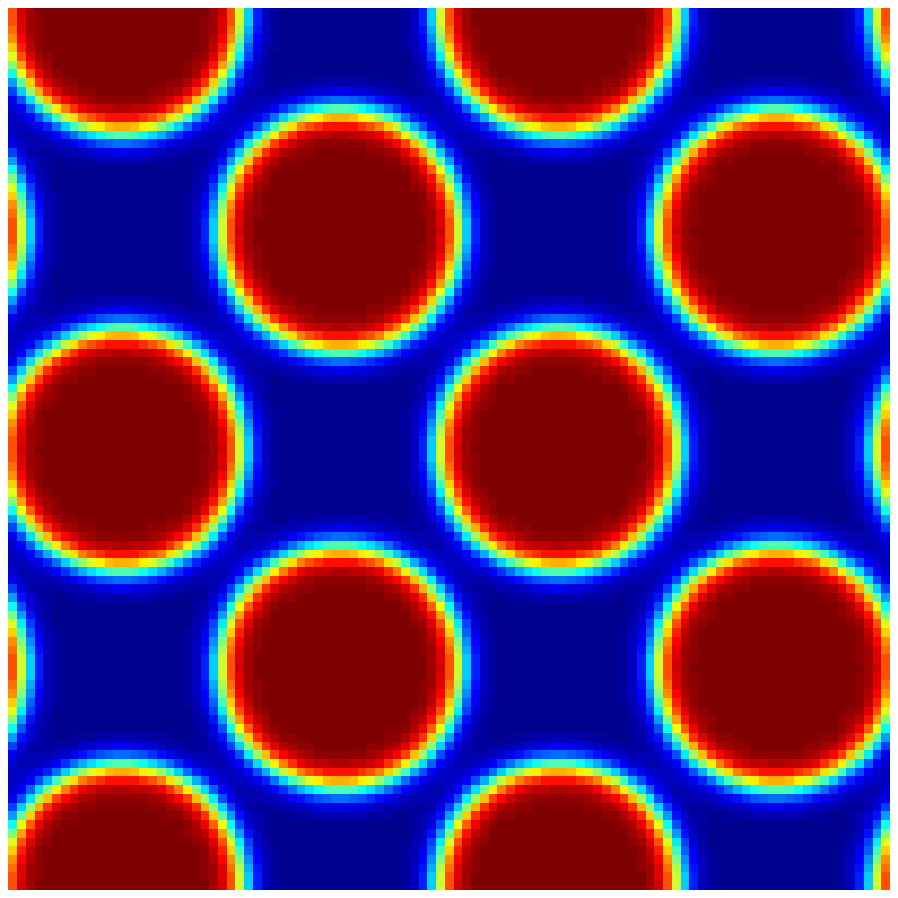}
}
\subfigure[Incompressible] {
\includegraphics[width=0.16\columnwidth]{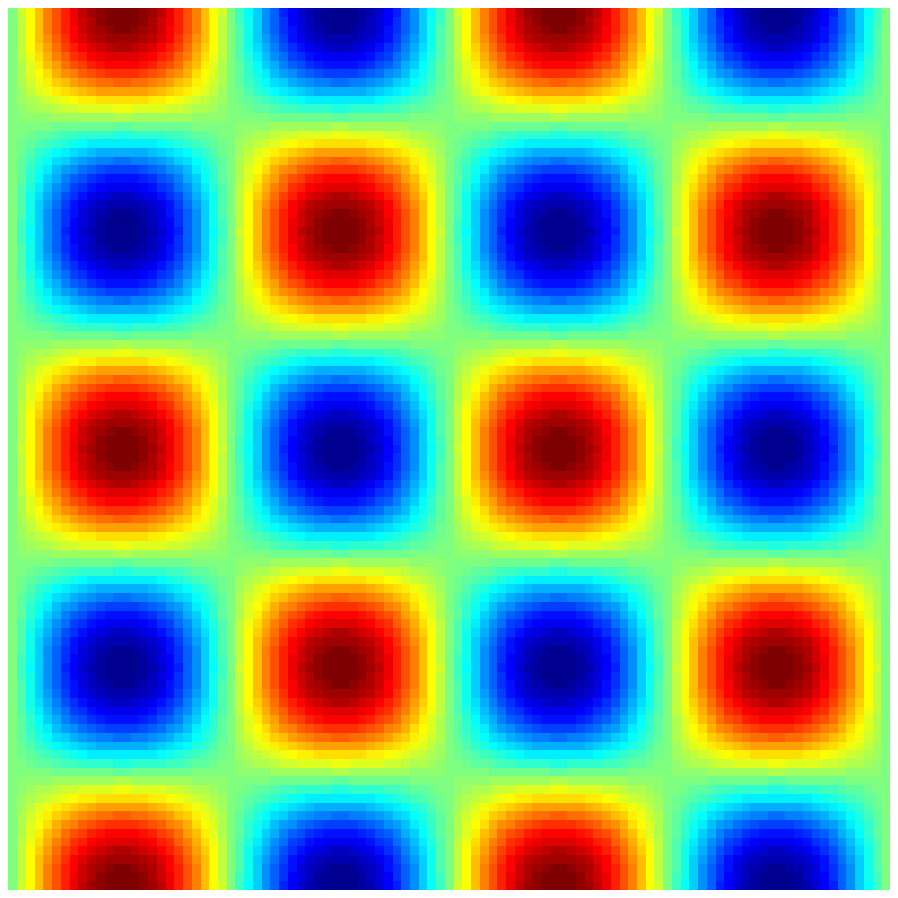}
\includegraphics[width=0.16\columnwidth]{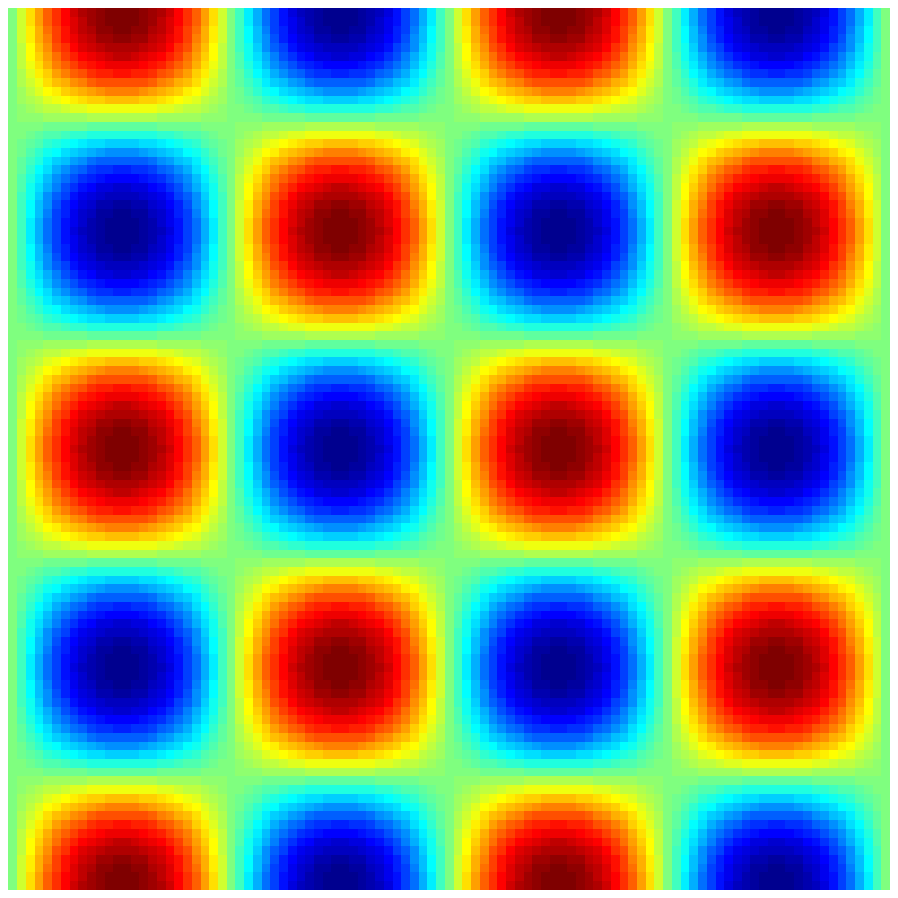}
\includegraphics[width=0.16\columnwidth]{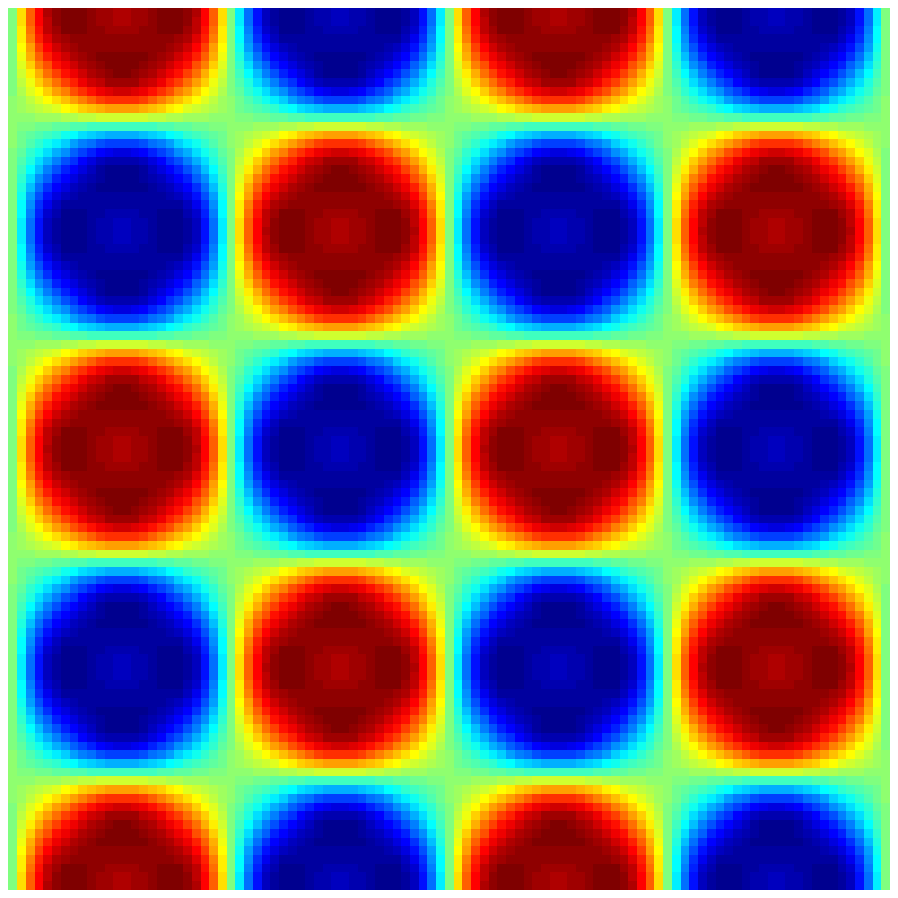}
\includegraphics[width=0.16\columnwidth]{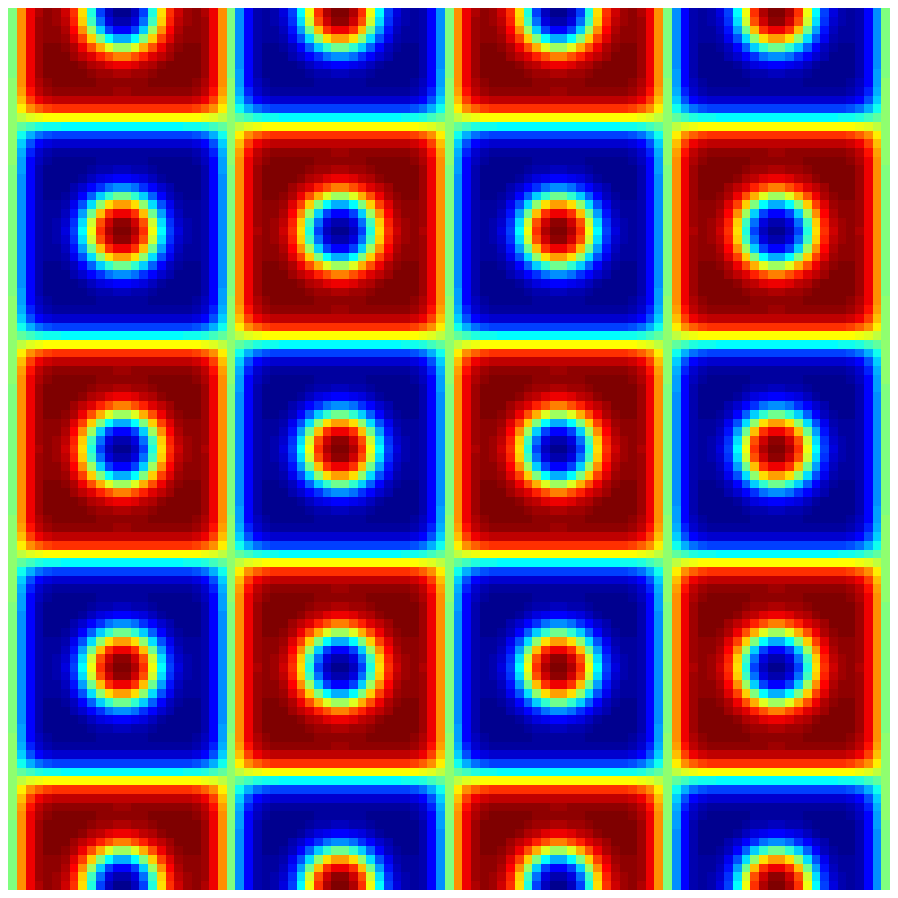}
\includegraphics[width=0.16\columnwidth]{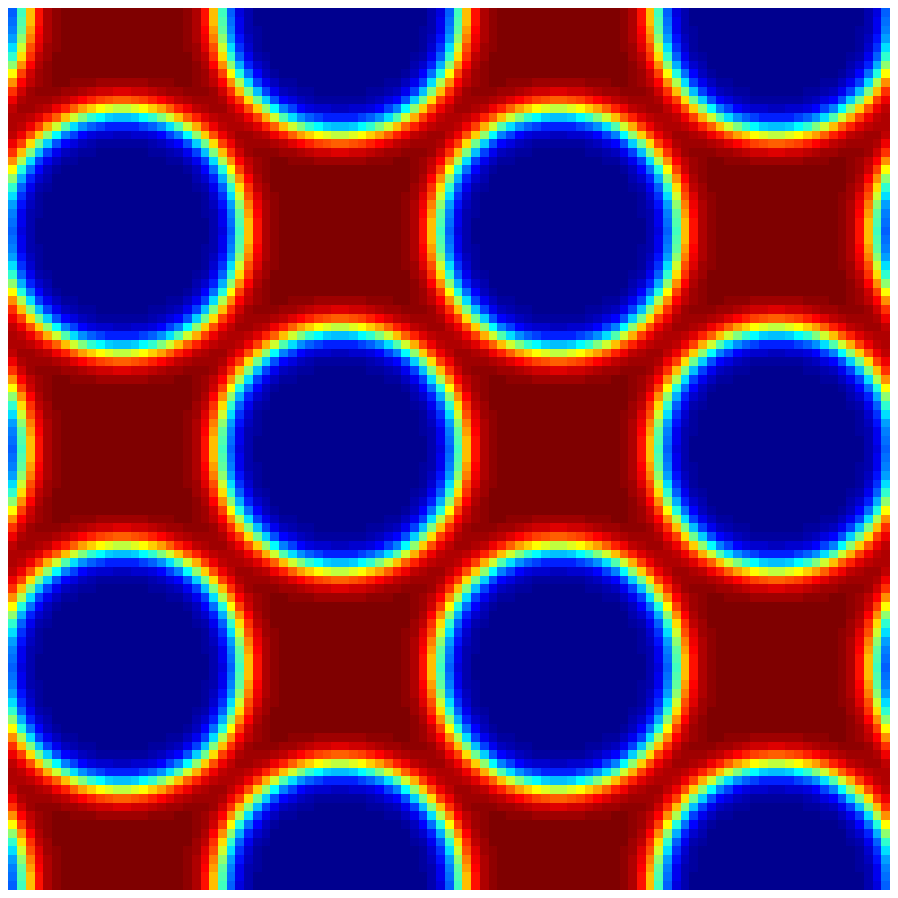}
}
\caption{(Color online) Density configuration with the minimum in blue and the maximum in red at $t/1000=0.1,1,2,4,15$ for the quasi-incompressible model (a) and incompressible model (b) as $\gamma=4$, $Pe=50$, and $Cn=4$.}
\label{fig:figure13}
\end{figure}

\begin{figure}[H] \centering
\subfigure[Quasi-incompressible] {
\includegraphics[width=0.16\columnwidth]{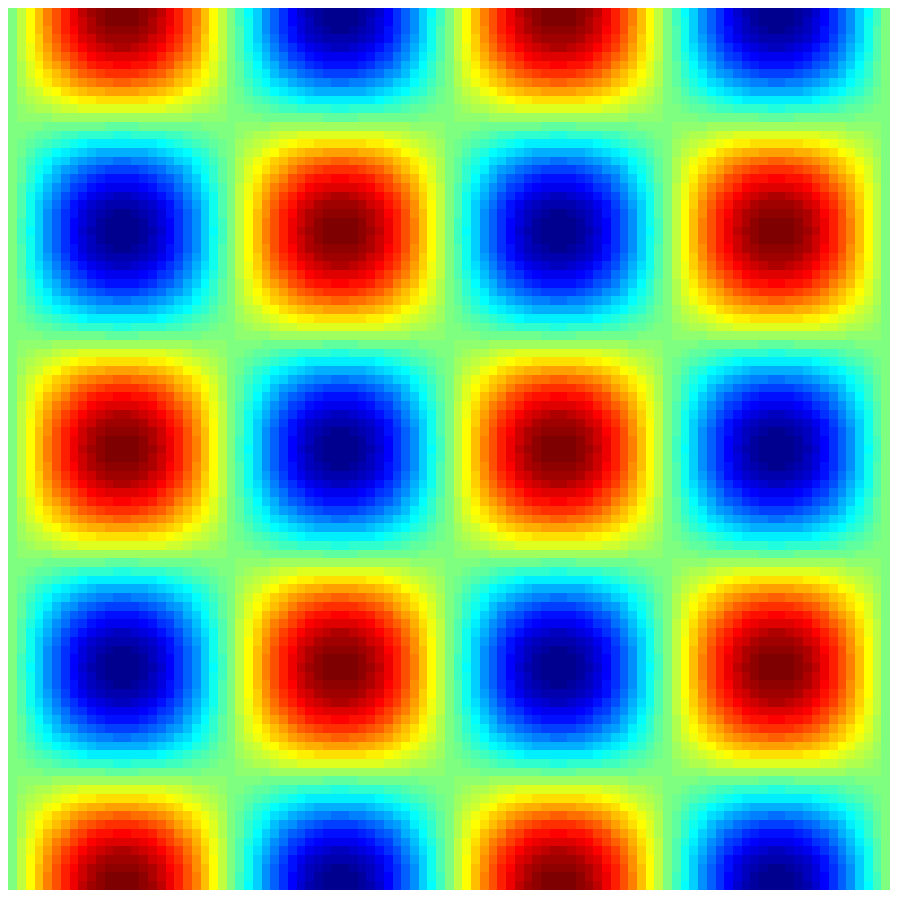}
\includegraphics[width=0.16\columnwidth]{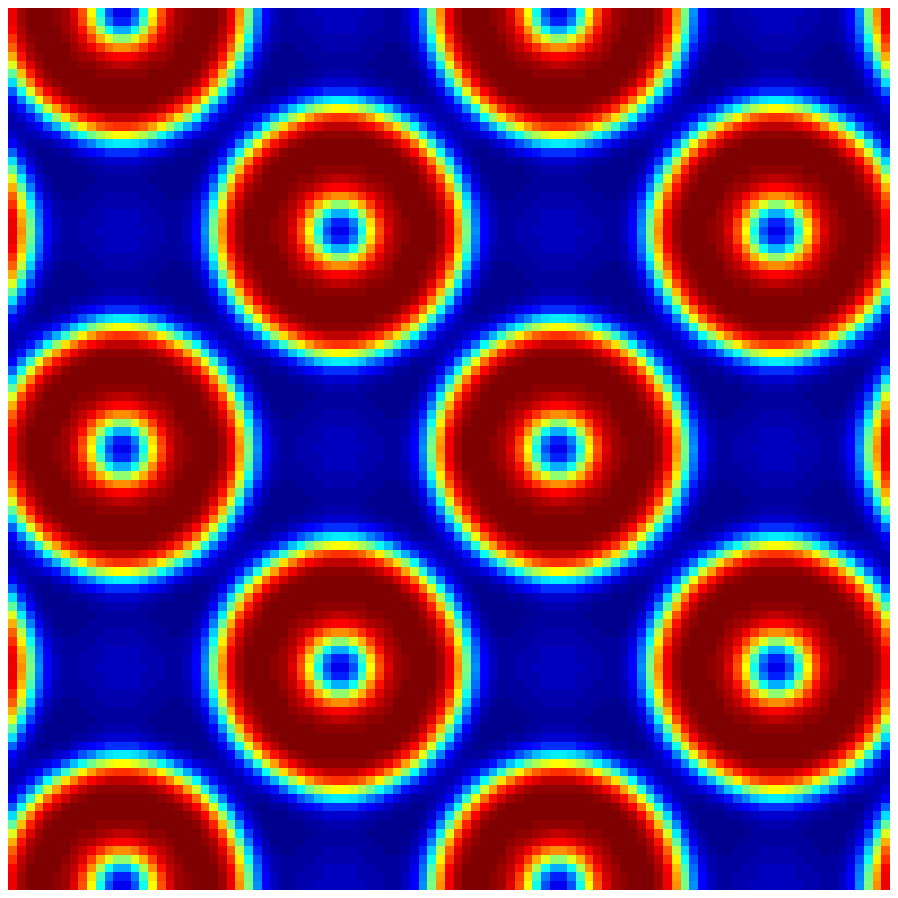}
\includegraphics[width=0.16\columnwidth]{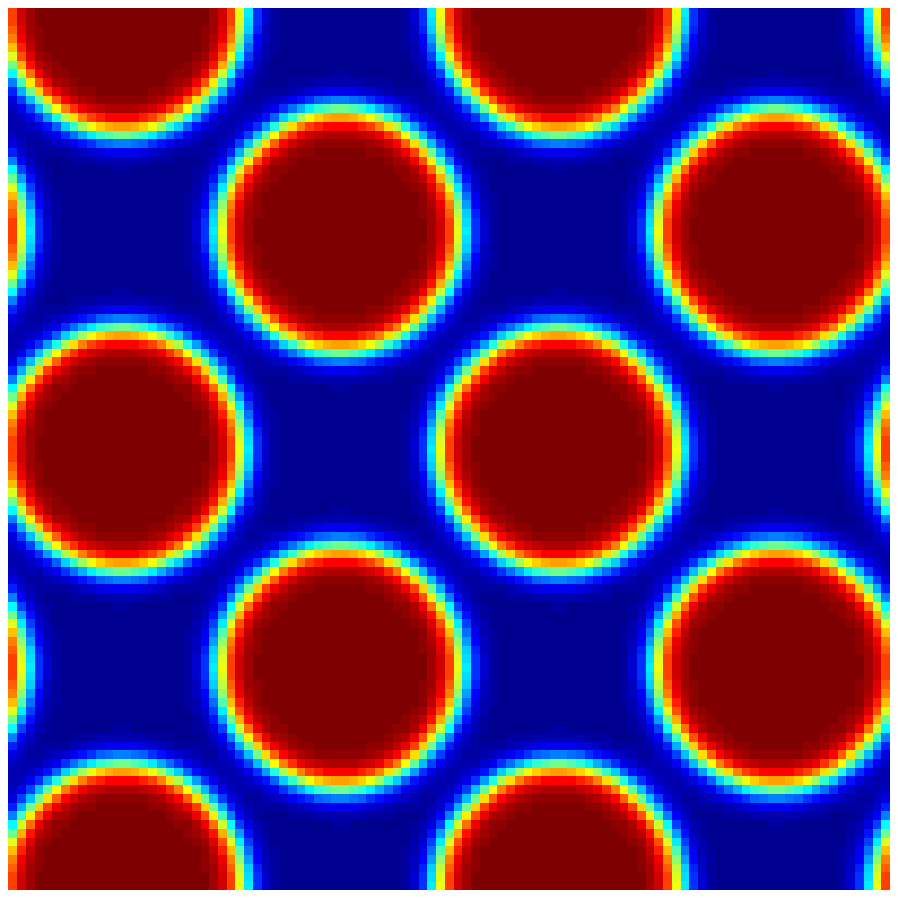}
\includegraphics[width=0.16\columnwidth]{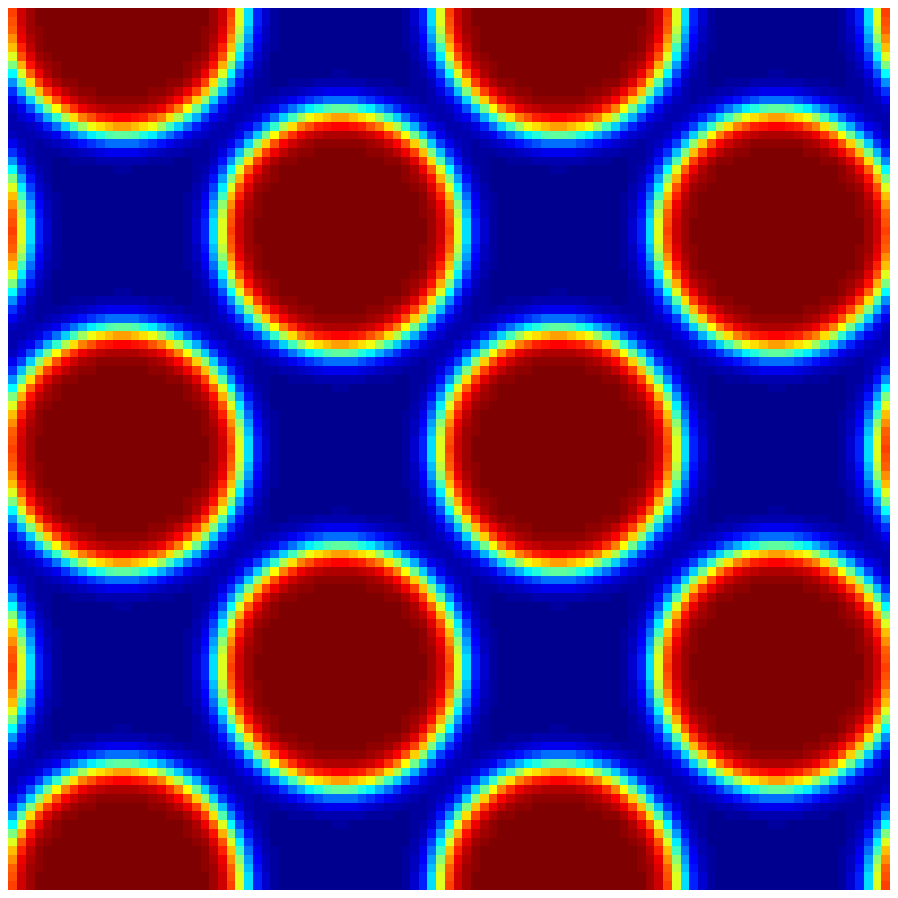}
\includegraphics[width=0.16\columnwidth]{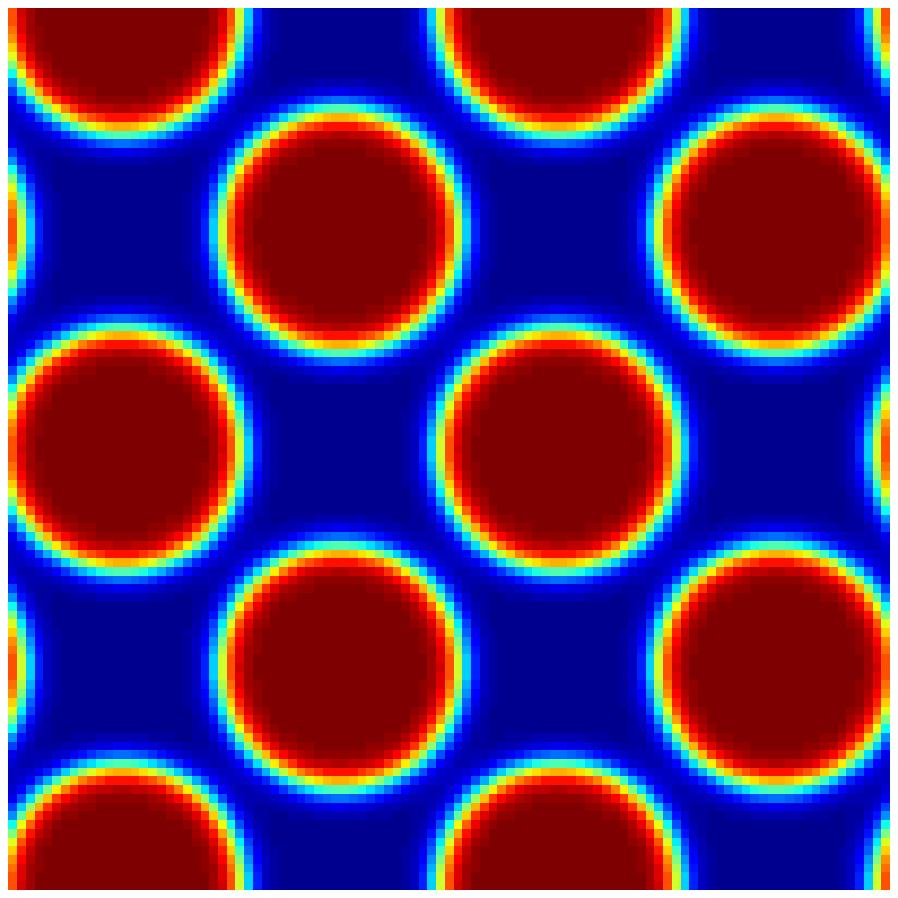}
}
\subfigure[Incompressible] {
\includegraphics[width=0.16\columnwidth]{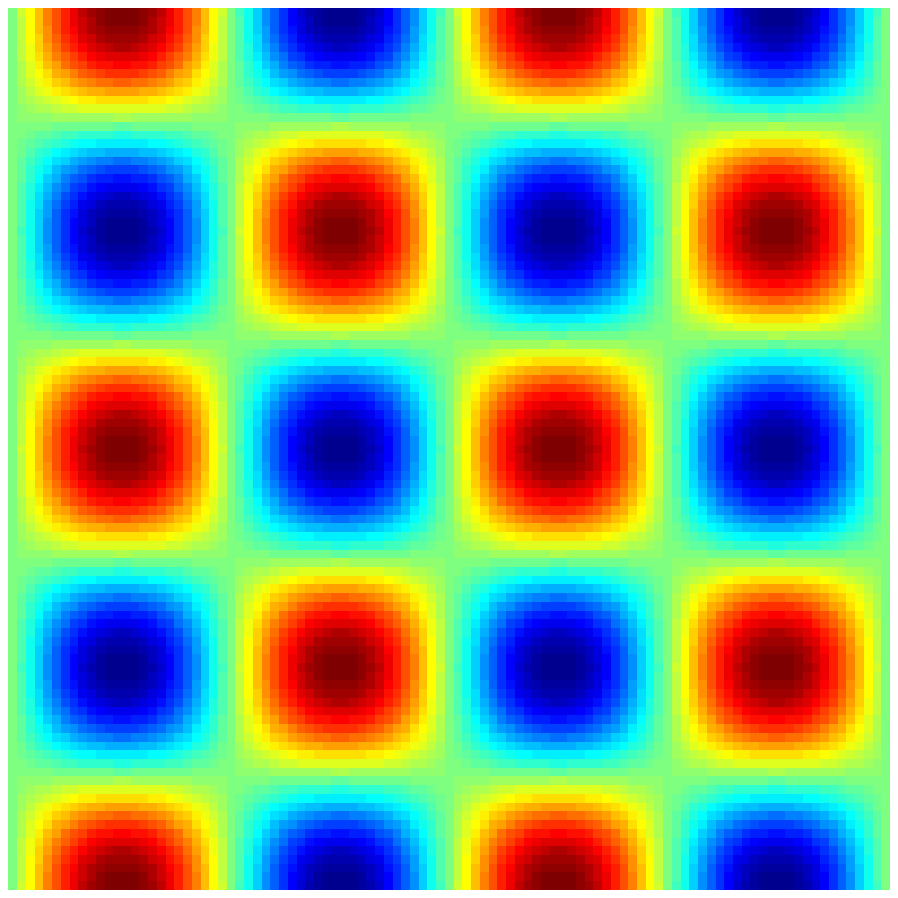}
\includegraphics[width=0.16\columnwidth]{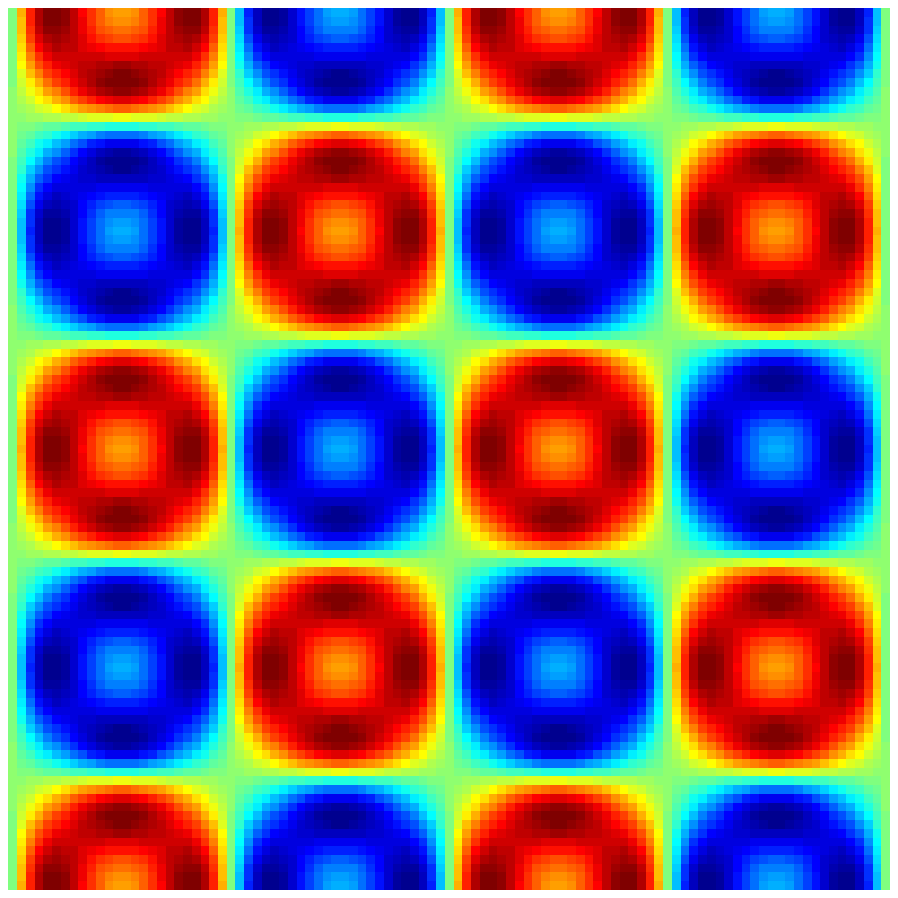}
\includegraphics[width=0.16\columnwidth]{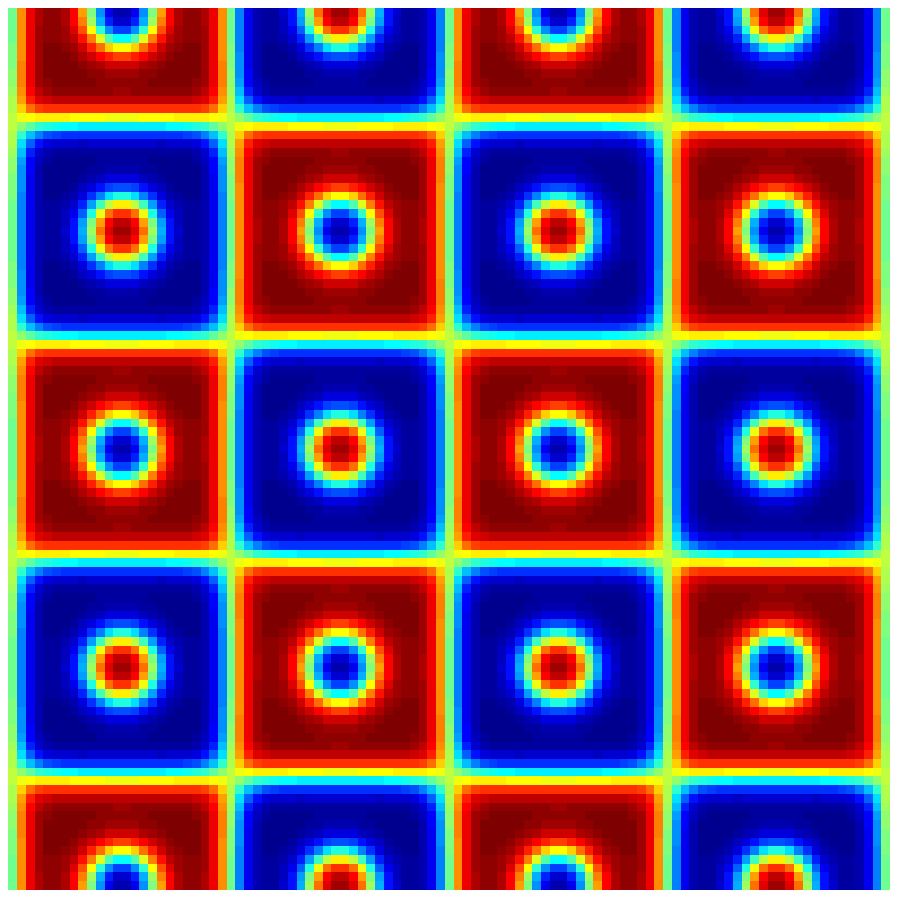}
\includegraphics[width=0.16\columnwidth]{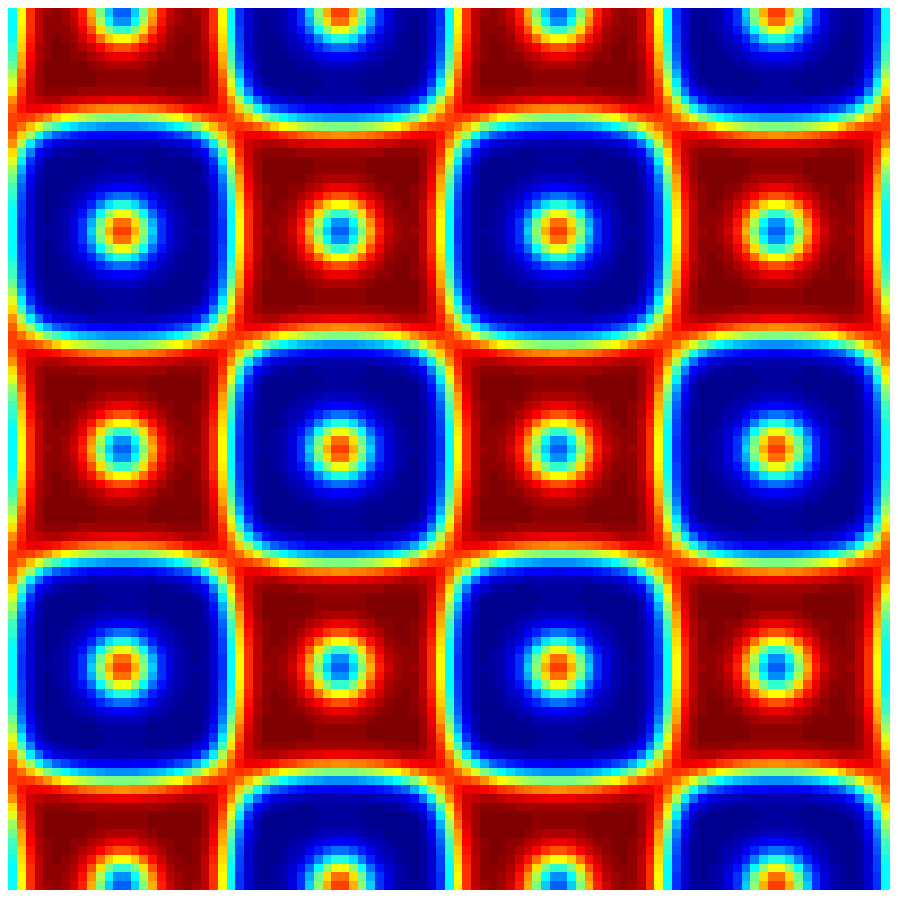}
\includegraphics[width=0.16\columnwidth]{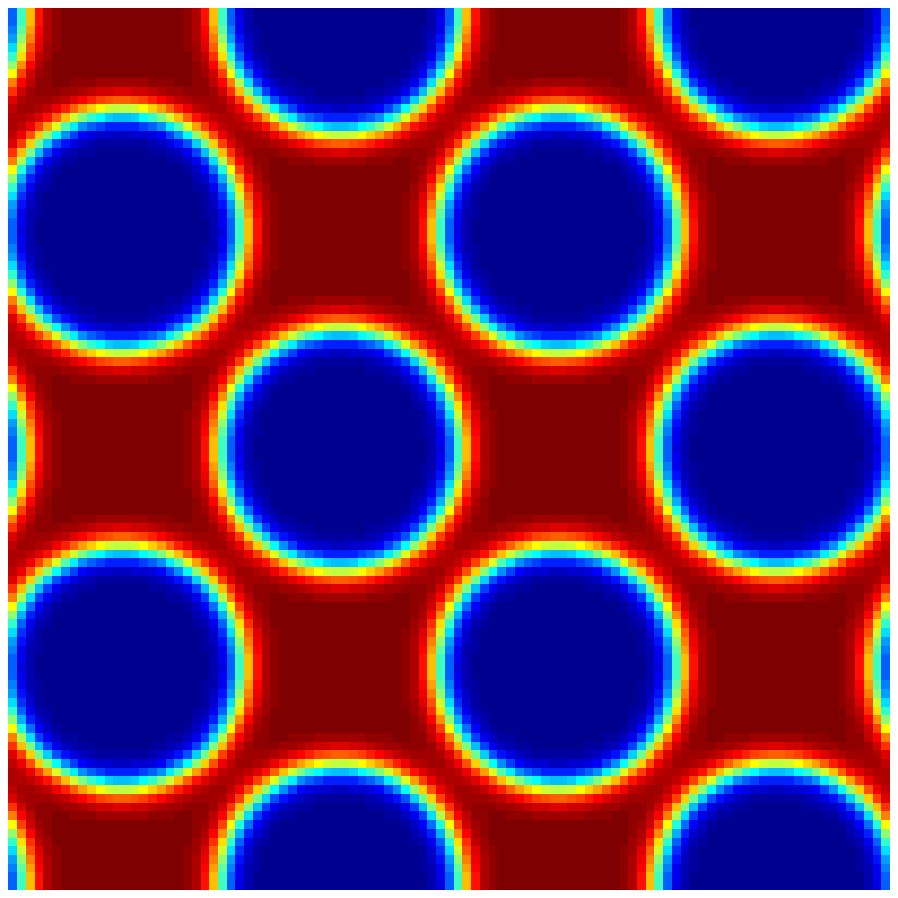}
}
\caption{(Color online) Density configuration with the minimum in blue and the maximum in red at $t/1000=0.1,10,20,25,50$ for the quasi-incompressible model (a) and incompressible model (b) as $\gamma=4$, $Pe=200$, and $Cn=4$.}
\label{fig:figure14}
\end{figure}

\begin{figure}[H] \centering
\subfigure[Quasi-incompressible] {
\includegraphics[width=0.16\columnwidth]{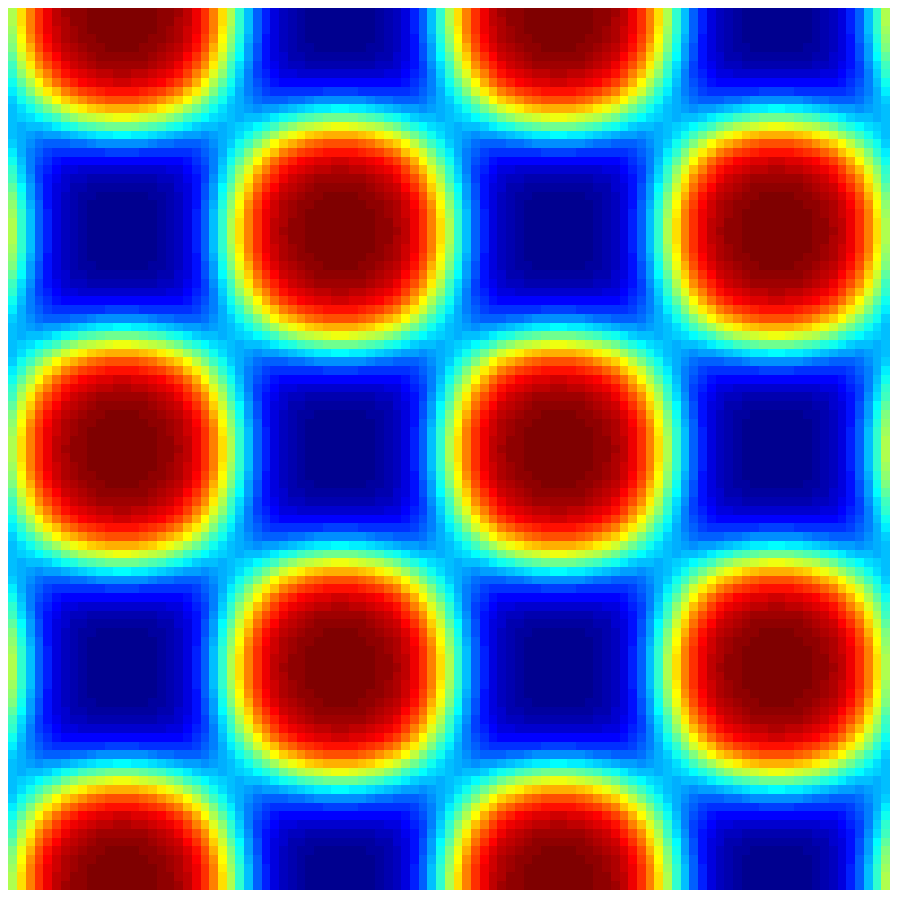}
\includegraphics[width=0.16\columnwidth]{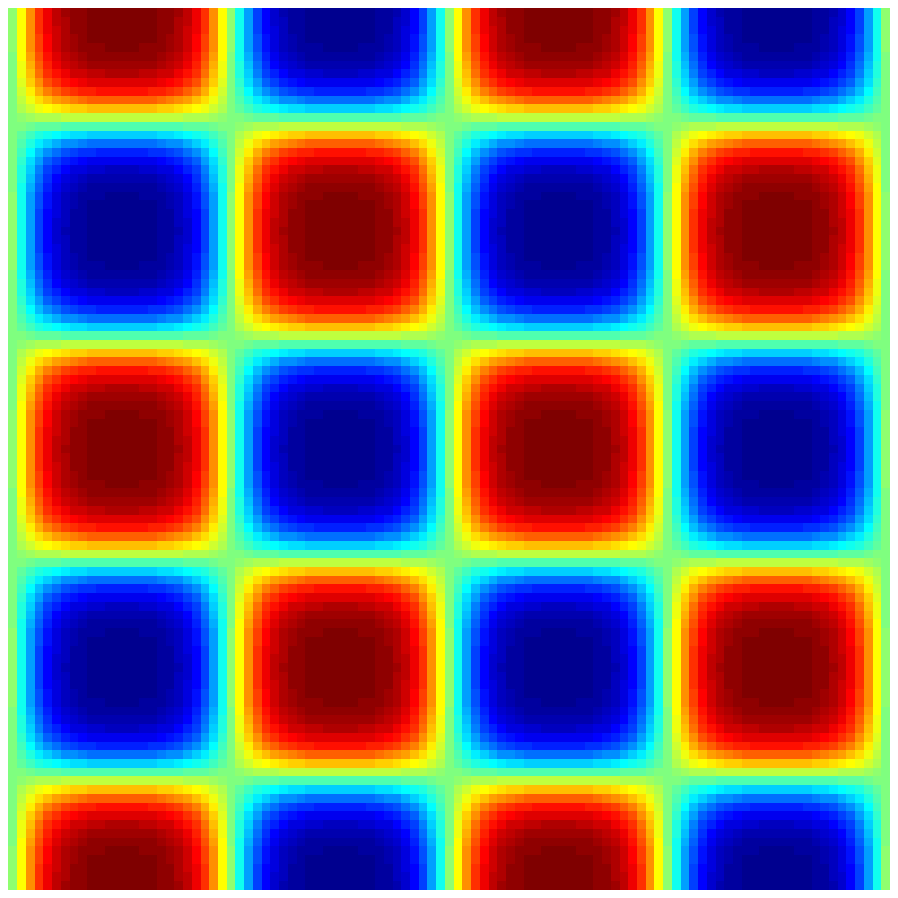}
\includegraphics[width=0.16\columnwidth]{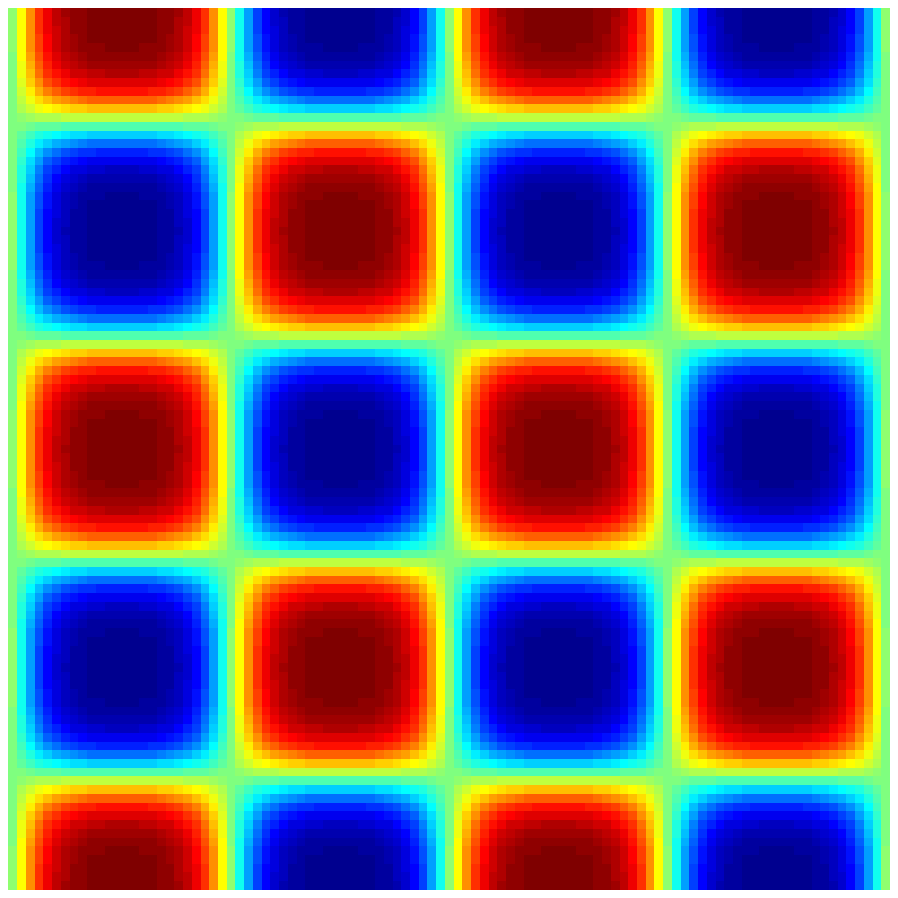}
\includegraphics[width=0.16\columnwidth]{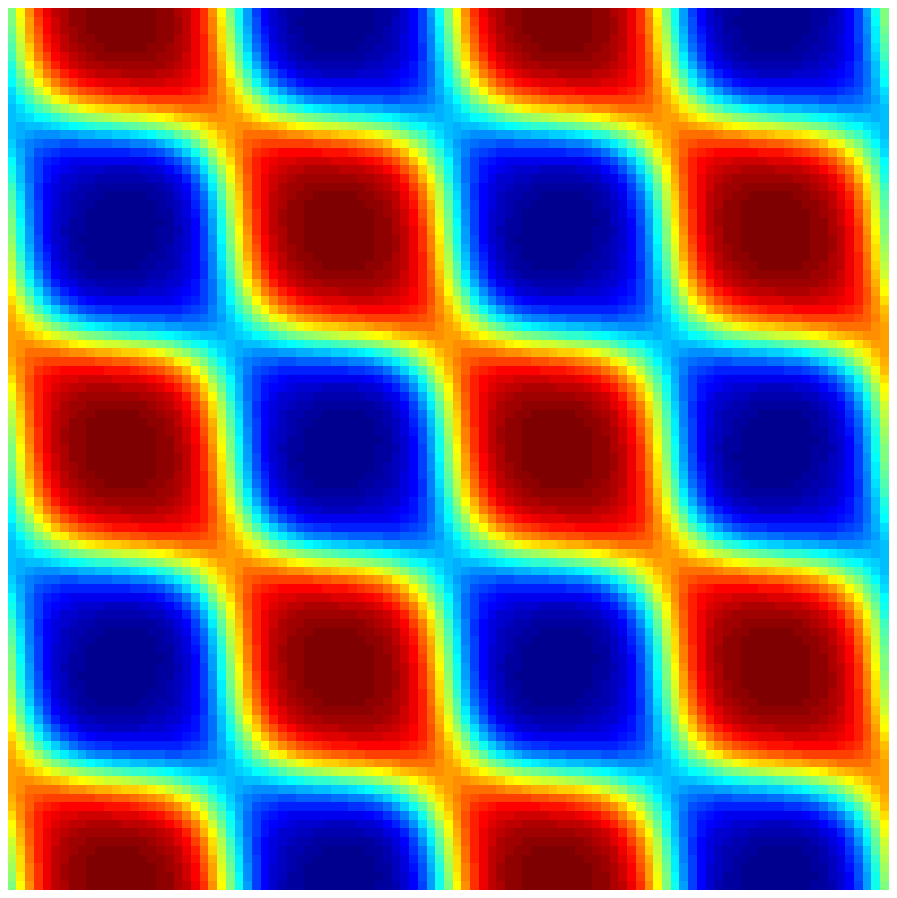}
\includegraphics[width=0.16\columnwidth]{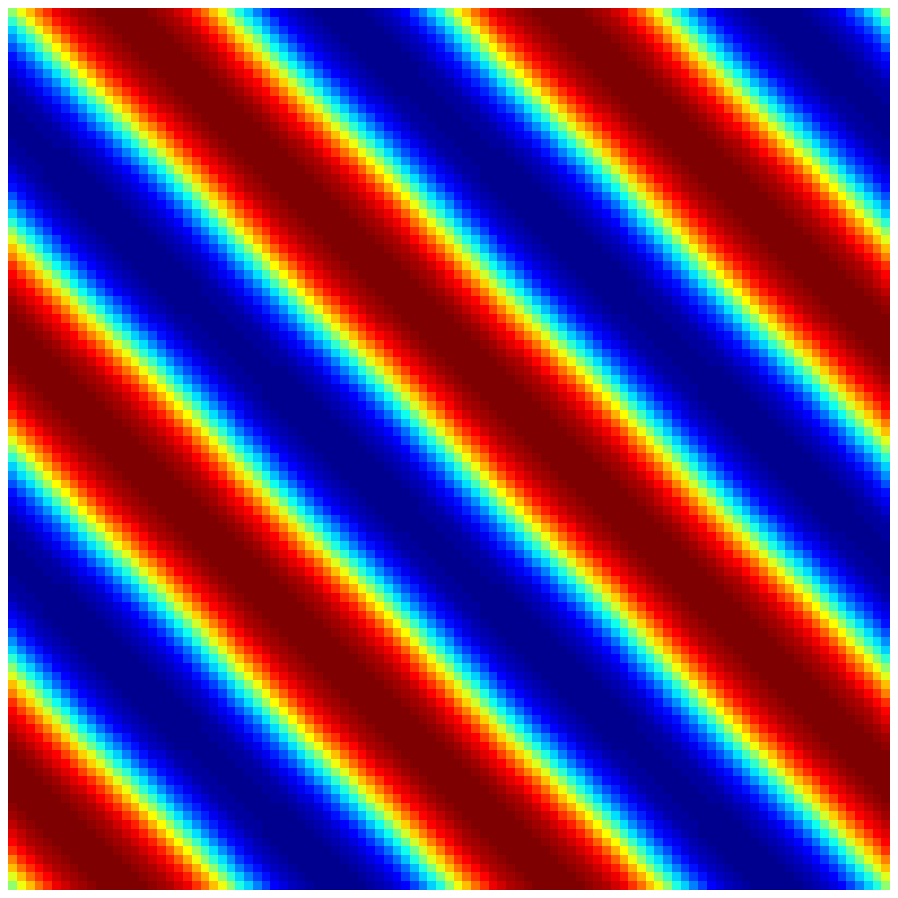}
}
\subfigure[Incompressible] {
\includegraphics[width=0.16\columnwidth]{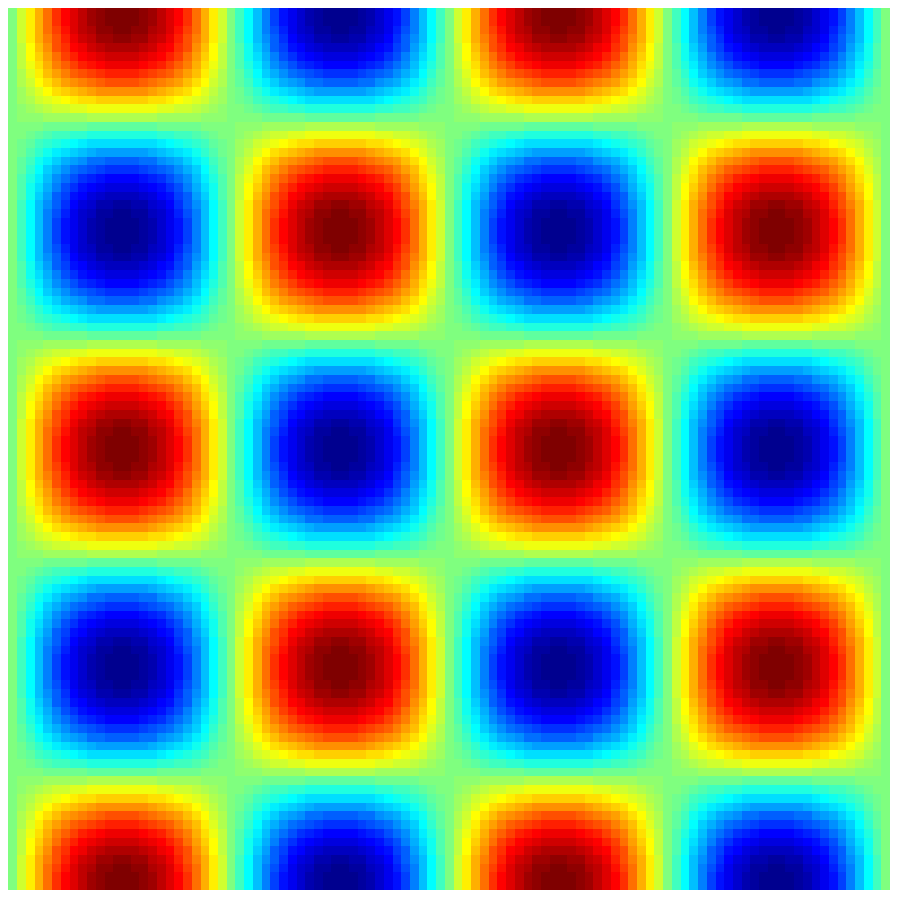}
\includegraphics[width=0.16\columnwidth]{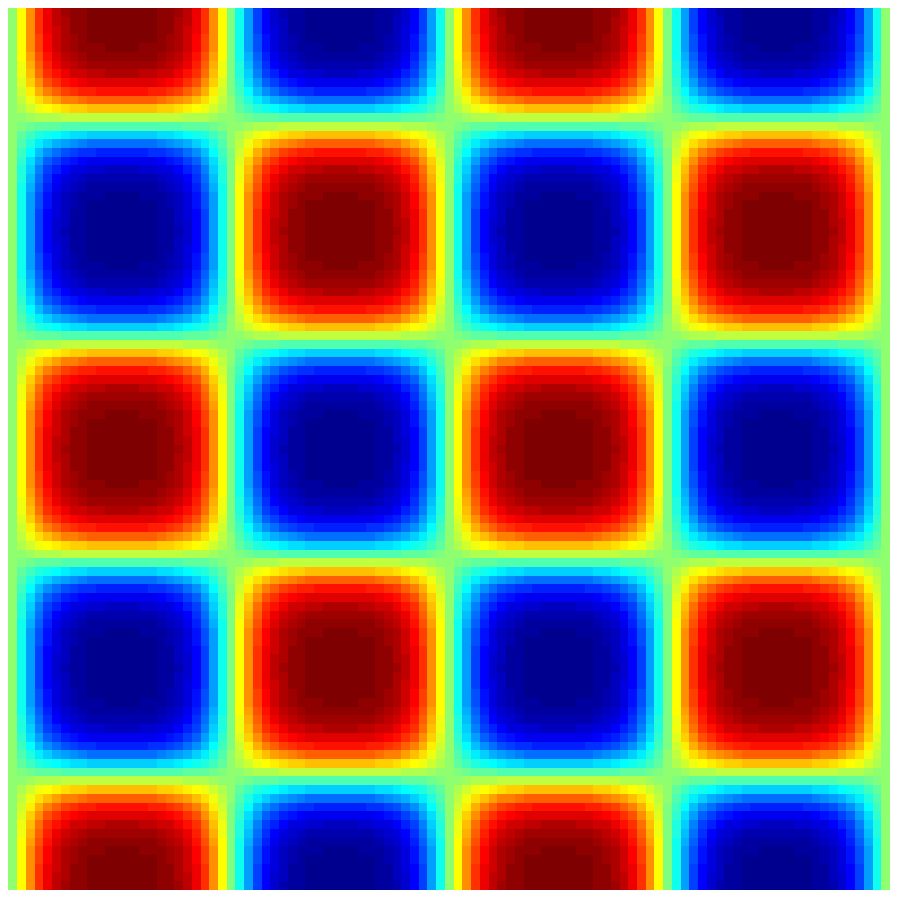}
\includegraphics[width=0.16\columnwidth]{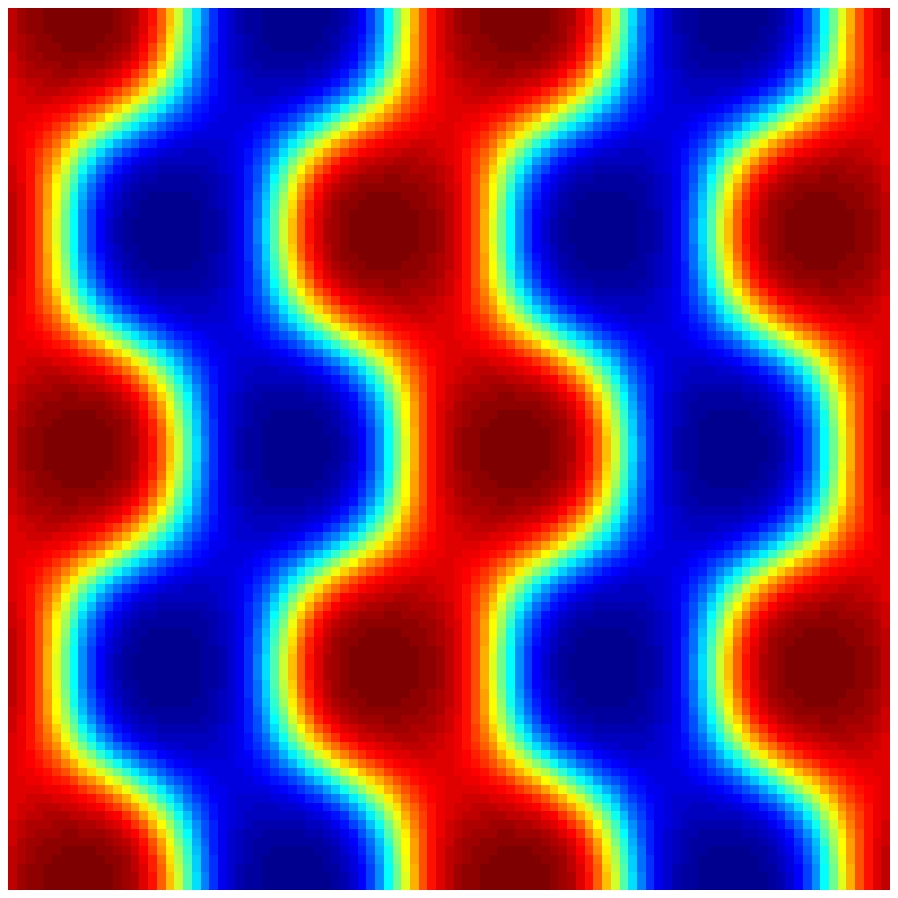}
\includegraphics[width=0.16\columnwidth]{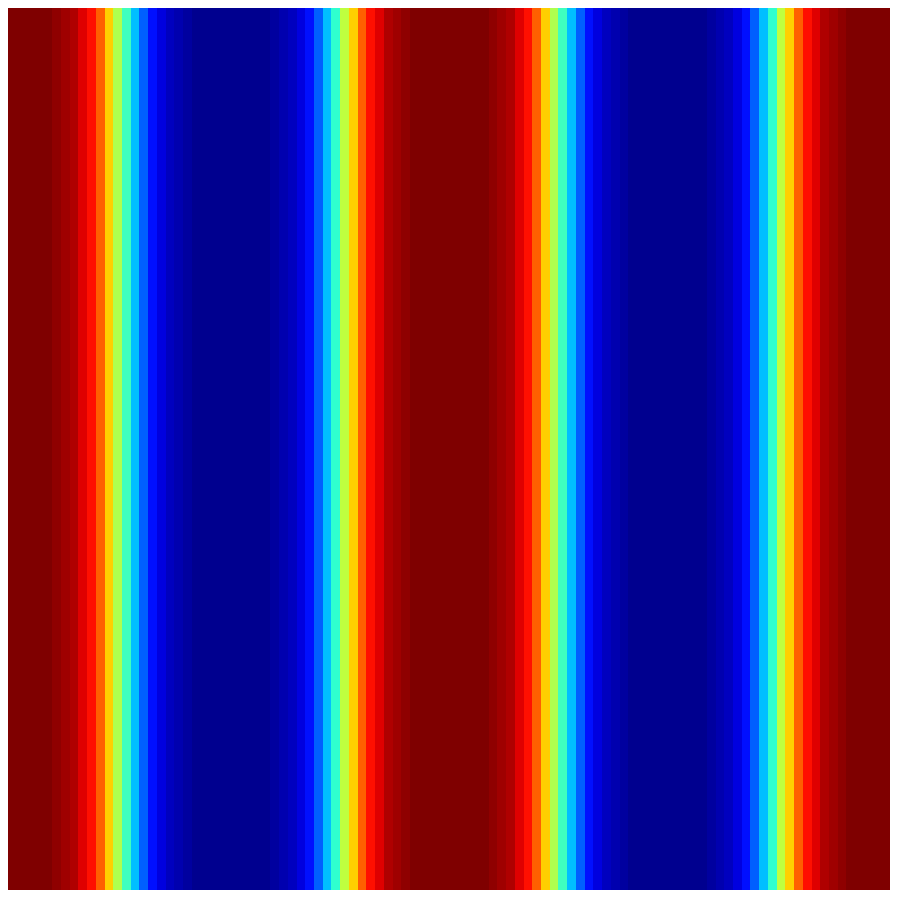}
\includegraphics[width=0.16\columnwidth]{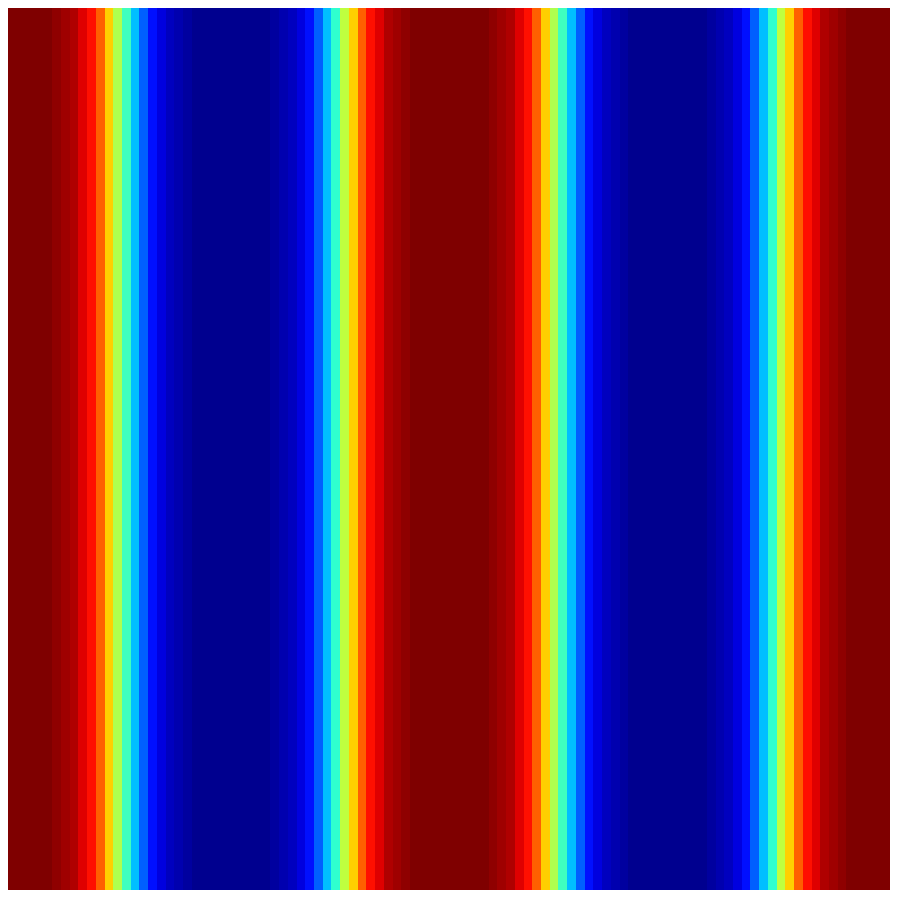}
}
\caption{(Color online) Density configuration with the minimum in blue and the maximum in red at $t/1000=10,60,90,200,240$ for the quasi-incompressible model (a) and incompressible model (b) as $\gamma=4$, $Pe=200$, and $Cn=8$.}
\label{fig:figure15}
\end{figure}

Figures \ref{fig:figure9} to \ref{fig:figure15} depict the density, dynamic pressure and velocity of the mixture with various dimensionless parameters $\gamma$, $Pe$ and $Cn$.
Firstly, we investigate the effect of $\gamma$ as shown in
Figs. \ref{fig:figure9} to \ref{fig:figure13}.
 As $\gamma=1$ (see Figs. \ref{fig:figure9} to \ref{fig:figure12}), in the transient period, the density, pressure and velocity fields predicted by the two models are quite different.
 In the steady state, although the density fields appears to be similar, the dynamic pressure and velocities are remarkably distinct.
 As $\gamma=4$ (see Fig. \ref{fig:figure13}), the density fields predicted by the two LBE models are opposite in the steady state,
 and the phase separation predicted by the present LBE model occurs earlier than that by the incompressible model.
  By comparing the Figs. \ref{fig:figure9} and \ref{fig:figure13}, it can also be found that the density field from the present
   model varies with $\gamma$ while that from  the incompressible model does not.
   Then we further consider the effect of Peclet number. As $Pe$ increases to 200 (see Fig. \ref{fig:figure14}),
   it can be found the phase separation processes predicted by the two LBE models are both slowed down, but
   the distributions of the density fields are in opposite in the steady state.
The effect of $Cn$ on the phase separation process is also investigated. As shown in Fig. \ref{fig:figure15}, when the $Cn$ is increased from 4 to 8, the density fields change greatly. In the transient period ($t=10^4$ to $9 \times 10^4$), the results predicted by the two LBE models present similar configurations; while in the steady state,
 some strips with different angles of inclination appear in the density fields.
 The above phenomena completely exhibit the discrepancy between the two models when the chemical potential is nonuniformly distributed  due to the non-equilibrium order parameter, and the discrepancy is deeply influenced by the dimensionless parameters $\gamma$, $Pe$ and $Cn$.

\section{CONCLUSIONS}
In this study, a LBE model for binary fluids is proposed based on the quasi-incompressible phase-field theory, which overcomes the mass conservation problem in the incompressible phase-field LBM.
 To validate the accuracy of the proposed model and compare its performance with an incompressible model, a series of numerical tests are performed.

Firstly, with the one-dimensional flat interface and stationary droplet tests, it is shown that the proposed LBE model can track the interface accurately and satisfies the Laplace law. Furthermore, the comparison of the stationary droplet shows no obvious difference between the two LBE  model.
 Then, the test of the bubble rising under buoyancy shows that some subtle distinctions exist in the two LBE models but overall they still agree with each other qualitatively. The results of the phase separation problem show the dynamic processes and static structures from the two models can be rather different.
These phenomena indicate that there are some distinctions between the two models involved with the multiphase flows, especially for problems with the nonuniform distribution of chemical potential.
Since the quasi-incompressible LBE model satisfies the fundamental mass conservation law, its results should be more reliable.

\section*{ACKNOWLEDGMENT}
This study is financially supported by the National Natural Science Foundation of China.
(Grant No. 51125024).
\section*{Appendix: Chapman-Enskog analysis of the quasi-incompressible LBE model}
In this section, the proposed LBE model for hydrodynamic equations is firstly analyzed by applying the Chapman-Enskog expansion
\begin{equation}\label{A1-q}
{f_i} = f_i^{(0)} + \varepsilon f_i^{(1)} + {\varepsilon ^2}f_i^{(2)} + ...
\end{equation}
\begin{equation}\label{A2-q}
{\partial _t} = \varepsilon {\partial _{{t_0}}} + {\varepsilon ^2}{\partial _{{t_1}}},\;\;\;\;\;{\bm{\nabla}}  = \varepsilon {{\bm{\nabla}} _0},\;\;\;\;\;\;{F_i} = \varepsilon F_i^{(0)}+\varepsilon^2 F_i^{(1)},
\end{equation}
with
\begin{equation}\label{AA-q}
F_i^{(0)}{\rm{ = }}({{\bf{c}}_i} - {\bf{u}}) \cdot \left[ {  {\omega _i}{\Gamma _i}{\bf{F}} + {\omega _i}{s_i}c_s^2{\bm{\nabla}} \rho } \right],\;\;\;\;F_i^{(1)}{\rm{ = }} - {\omega _i}c_s^2\rho \gamma {\bm{\nabla}} .(\lambda {\bm{\nabla}} \mu ),
\end{equation}
where $\varepsilon$ is a small expansion parameter.
Using the Taylor expansion in Eq.~\eqref{LB2}, one can obtain
\begin{equation}\label{A3-q}
{D_i}{f_i} + \frac{{{\delta t}}}{2}D_i^2{f_i} =  - \frac{1}{{{\tau _c}}}({f_i} - f_i^{eq}) + \left(1-\frac{1}{2\tau_f}\right){F_i},
\end{equation}
where ${\tau _c} = {\tau _f}{\delta t}$, and ${D_i} = {\partial _t} + {{\bf{c}}_i}\cdot{\bm{\nabla}}$.
The substitution of Eqs.~\eqref{A1-q} and \eqref{A2-q} into Eq.~\eqref{A3-q} yields the Chapman-Enskog system as
\begin{equation}\label{A4-q}
O({\varepsilon ^0}):\;\;\;f_i^{(0)} = f_i^{eq},
\end{equation}
\begin{equation}\label{A5-q}
O({\varepsilon ^1}):\;\;\;{D_{0i}}f_i^{(0)} - \left(1-\frac{1}{2\tau_f}\right)F_i^{(0)} =  - \frac{1}{{{\tau _c}}}f_i^{(1)},
\end{equation}
\begin{equation}\label{A6-q}
O({\varepsilon ^2}):\;\;\;{\partial _{t_1}}f_i^{(0)} + {D_{0i}}f_i^{(1)} + \frac{{{\delta t}}}{2}D_{0i}^2f_i^{(0)} =  - \frac{1}{{{\tau _c}}}f_i^{(2)}+\left(1-\frac{1}{2\tau_f}\right)F_i^{(1)}.
\end{equation}
Then, the substitution of Eq.~\eqref{A5-q} into \eqref{A6-q} yields
\begin{equation}\label{A7-q}
{\partial _{t_1}}f_i^{(0)} + \left(1 - \frac{1}{{2\tau_f }}\right){D_{0i}}f_i^{(1)} + \frac{{{\delta t}}}{2}\left(1-\frac{1}{2\tau_f}\right){D_{0i}}F_i^{(0)} =  - \frac{1}{{{\tau _c}}}f_i^{(2)}+\left(1-\frac{1}{2\tau_f}\right)F_i^{(1)}.
\end{equation}
Meanwhile, from the definitions \eqref{feq} and \eqref{AA-q}, it is easy to calculate the following moments:
\begin{equation}\label{A8-q}
\sum\limits_i {f_i^{eq} = p} ,\;\;\;\sum\limits_i {{c_i}f_i^{eq} = c_s^2\rho {\bf{u}},\;\;\;\;}
\end{equation}
\begin{equation}\label{A9-q}
\sum\limits_i {{c_i}{c_i}f_i^{eq} = c_s^2p + c_s^2\rho {\bf{uu}}}, \;
\end{equation}
\begin{equation}
\sum\limits_i {{c_i}{c_i}{c_i}f_i^{eq} = } c_s^4\rho ({u_\alpha }{\delta _{\beta \gamma }} + {u_\beta }{\delta _{\alpha \gamma }} + {u_\gamma }{\delta _{\alpha \beta }}),
\end{equation}
\begin{equation}
\sum\limits_i {F_i^{(0)} = c_s^2{\bf{u}}.{\bm{\nabla}} \rho }, \;\;\sum\limits_i {F_i^{(1)} =  - c_s^2\rho \gamma {\bm{\nabla}} .(\lambda {\bm{\nabla}} \mu )},
\end{equation}
\begin{equation}
\sum\limits_i {{{\bf{c}}_i}F_i^{(0)} = c_s^2{\bf{F}},\;\;} \sum\limits_i {{{\bf{c}}_i}F_i^{(1)} = 0,\;\;}
\end{equation}
\begin{equation}
\sum\limits_i {{c_i}{c_i}F_i^{(0)} = c_s^2\left( {{\bf{F'u}} + {\bf{uF'}}} \right) + c_s^4{\bf{u}}.{\bm{\nabla}} \rho ,} \;\;\sum\limits_i {{c_i}{c_i}F_i^{(1)} = } \; - c_s^4\rho \gamma {\bm{\nabla}} .(\lambda {\bm{\nabla}} \mu ),
\end{equation}
where ${\bf{F'}} = {\bf{F}}  + c_s^2{\bm{\nabla}} \rho$. From  Eqs.~\eqref{velocity-q}, \eqref{pressure-q} and \eqref{A8-q}, we can obtain
\begin{equation}
\sum\limits_i {f_i^{(1)} =  - } \frac{{{\delta t}}}{2}c_s^2{\bf{u}}.{\bm{\nabla}} \rho, \;\sum\limits_i {f_i^{(2)} =   } \frac{{{\delta t}}}{2}c_s^2\gamma \rho {\bm{\nabla}} .(\lambda {\bm{\nabla}} \mu ),\;\sum\limits_i {f_i^{(k)} = 0,\;k > 2} \;
\end{equation}
\begin{equation}
\sum\limits_i {{c_i}f_i^{(1)} =  - \frac{{{\delta t}}}{2}c_s^2{\bf{F}},\;} \;\;\;\;\sum\limits_i {{c_i}f_i^{(k)} = 0.\;\;\;k > 1}
\end{equation}
Taking the zeroth- and first-order moments of Eq.~\eqref{A5-q}, we can obtain
\begin{equation}\label{A10-q}
\frac{1}{{\rho c_s^2}}{\partial _{t_0}}p + {\bm{\nabla}} \cdot{\bf{u}} = 0,
\end{equation}
\begin{equation}\label{A11-q}
{\partial _{t_0}}(\rho {\bf{u}}) + {\bm{\nabla}} \cdot(p + \rho {\bf{uu}}) =  {\bf{F}}.
\end{equation}
Likewise, taking the zeroth- and first-order moments of Eq.~\eqref{A7-q}, we can obtain
\begin{equation}\label{A12-q}
{\partial _{t_1}}p = - c_s^2\rho \gamma {\bm{\nabla}}\cdot(\lambda {\bm{\nabla}} \mu ),\;
\end{equation}
\begin{equation}\label{A13-q}
{\partial _{{t_1}}}(c_s^2\rho {\bf{u}}) + {\nabla _0} \cdot \left[ {\left( {1 - \frac{1}{{2{\tau _f}}}} \right)\sum\limits_i {{{\bf{c}}_i}{{\bf{c}}_i}f_i^{(1)}} } \right] + \frac{{\delta t}}{2}{\nabla _0} \cdot \left[ {\left( {1 - \frac{1}{{2{\tau _f}}}} \right)\sum\limits_i {{{\bf{c}}_i}{{\bf{c}}_i}F_i^{(0)}} } \right] = 0.
\end{equation}
According to the Eqs.~\eqref{A5-q}, and \eqref{A10-q} to \eqref{A12-q},
\begin{equation}
\begin{array}{l}\label{A14-q}
 - \frac{1}{{{\tau _c}}}\sum\limits_i {{{\bf{c}}_i}{{\bf{c}}_i}f_i^{(1)} = } c_s^4{\bf{u}}\cdot{\bm{\nabla}} \rho  + c_s^2({\bf{uF'}} + {\bf{F'u}})+ c_s^4\rho ({\bm{\nabla}} {\bf{u}} + {\bm{\nabla}} {{\bf{u}}^T})\\\;\;\;\;\;\;\;\;\;\;\;\;\;\;\;\;\;\;\; - \sum\limits_i \left(1-\frac{1}{2\tau_f}\right){{{\bf{c}}_i}{{\bf{c}}_i}F_i^{(0)}}  + O(M_a^3).
\end{array}
\end{equation}
Substituting  Eq.~\eqref{A14-q} into Eq.~\eqref{A13-q}, we can obtain
\begin{equation}\label{A15-q}
{\partial _{t_1}}(\rho {\bf{u}}) = {\bm{\nabla}} \cdot\left[ {\rho \nu({\bm{\nabla}} {\bf{u}} + {\bm{\nabla}} {{\bf{u}}^T})} \right],
\end{equation}
where $\nu=c_s^2({\tau _f} - 0.5){\delta t}$.
From Eqs.~\eqref{A10-q} and \eqref{A12-q}, we can obtain the continuity equation
\begin{equation}\label{DV-q}
\frac{1}{{\rho c_s^2}}{\partial _t}p + {\bm{\nabla}} \cdot{\bf{u}} = \; - \gamma {\bm{\nabla}} \cdot(\lambda {\bm{\nabla}} \mu ).
\end{equation}
Similarly, the momentum equation can be derived from Eqs.~\eqref{A11-q} and \eqref{A15-q}
\begin{equation}\label{mass}
{\partial _t}(\rho {\bf{u}}) + {\bm{\nabla}} \cdot(\rho {\bf{uu}}) =  - {\bm{\nabla}} p + {\bm{\nabla}} \cdot \left[ {\rho \nu ({\bm{\nabla}} {\bf{u}} + {\bm{\nabla}} {{\bf{u}}^T})} \right] +{\bf{F}}.
\end{equation}

Next we will derive the CH equation from Eq.~\eqref{LB1} by the Chapman-Enskog expansion. Similarly, the multiscale expansions are given by
\begin{equation}\label{B1-q}
{\textrm g_i} = \textrm g_i^{(0)} + \varepsilon \textrm g_i^{(1)} + {\varepsilon ^2}\textrm g_i^{(2)} + ...,
\end{equation}
\begin{equation}\label{B2-q}
{\partial _t} = \varepsilon {\partial _{{t_0}}} + {\varepsilon ^2}{\partial _{{t_1}}},\;\;\;\;\;{\bm{\nabla}}  = \varepsilon {{\bm{\nabla}} _0},\;\;\;\;\;\;{G_i} = \varepsilon G_i^{(0)}.
\end{equation}
 Using the Taylor expansion in Eq.~\eqref{LB1}, one can obtain
\begin{equation}\label{B3-q}
{D_i}{\textrm g_i} + \frac{{{\delta t}}}{2}D_i^2{\textrm g_i} =  - \frac{1}{{{\tau _c}}}({\textrm g_i} - \textrm g_i^{eq}) + \left(1-\frac{1}{2\tau_{\textrm g}}\right){G_i},
\end{equation}
where ${\tau _c} = {\tau _{\textrm g}}{\delta t}$. Substituting Eqs.~\eqref{B1-q} and \eqref{B2-q} into the Eq.~\eqref{B3-q}, we can obtain the following infinite consecutive series of equations
\begin{equation}\label{B4-q}
O({\varepsilon ^0}):\;\;\;\textrm g_i^{(0)} = \textrm g_i^{eq},
\end{equation}
\begin{equation}\label{B5-q}
O({\varepsilon ^1}):\;\;\;{D_{0i}}\textrm g_i^{(0)} - \left(1-\frac{1}{2\tau_{\textrm g}}\right)G_i^{(0)} =  - \frac{1}{{{\tau _c}}}\textrm g_i^{(1)},
\end{equation}
\begin{equation}\label{B6-q}
O({\varepsilon ^2}):\;\;\;{\partial _{t_1}}\textrm g_i^{(0)} + {D_{0i}}\textrm g_i^{(1)} + \frac{{{\delta t}}}{2}D_{0i}^2\textrm g_i^{(0)} =  - \frac{1}{{{\tau _c}}}\textrm g_i^{(2)}.
\end{equation}
Then substituting Eq.~\eqref{B5-q} into Eq.~\eqref{B6-q}, we can obtain
\begin{equation}\label{B7-q}
{\partial _{t_1}}\textrm g_i^0 + \left( {{\tau _c}-\frac{{{\delta t}}}{2}} \right){D_{0i}}G_i^{(0)} + \left( {\frac{{{\delta t}}}{2} - {\tau _c}} \right)D_{0i}^2\textrm g_i^{(0)} =  - \frac{1}{{{\tau _c}}}\textrm g_i^{(2)}.
\end{equation}
In order to recover the CH equation, we derive the following moments from the Eq.~\eqref{geq} and Eq.~\eqref{force},
\begin{equation}\label{B8-q}
\sum\limits_i {\textrm g_i^{eq} = \phi }, \;\;\;\sum\limits_i {{c_i}\textrm g_i^{eq} = \phi {\bf{u}},\;\;\;\;} \sum\limits_i {{c_i}{c_i}\textrm g_i^{eq} = \phi {\bf{uu}} + {\bf{c}}_s^2\alpha \mu }, \;
\end{equation}
\begin{equation}\label{B9-q}
\sum\limits_i {{G_i} = 0,\;\;\;\;\;} \sum\limits_i {{{\bf{c}}_i}{G_i} =  \frac{\phi }{\rho }{\bf{G}},\;\;\;\;}
\end{equation}
where ${\bf{G}} = -{\bm{\nabla}} p +{\bf{F}} $. Taking the zeroth-order moment of Eqs.~\eqref{B5-q} and \eqref{B7-q}, we can obtain
\begin{equation}\label{B10-q}
{\partial _{t_0}}\phi  + {\bm{\nabla}} \cdot (\phi {\bf{u}}) = 0,
\end{equation}
\begin{equation}\label{B11-q}
{\partial _{t_1}}\phi  + \left( {\frac{{{\delta t}}}{2} - {\tau _c}} \right)\left[ {\partial _{t_0}^2\phi  + 2{\partial _{t_0}}{\bm{\nabla}} \cdot(\phi {\bf{u}}) + {\bm{\nabla}} \cdot{\bm{\nabla}} \cdot(\phi {\bf{uu}} + {\bf{c}}_s^2\alpha \mu )- {\bm{\nabla}} \cdot(\frac{\phi }{\rho }{\bf{G}})}\right] = 0.
\end{equation}
According to Eqs.~\eqref{A11-q} and \eqref{B10-q}, the Eq.~\eqref{B11-q} reduces to the following equation
\begin{equation}\label{B12-q}
{\partial _{t_1}}\phi  = \lambda{\nabla ^2}\mu,
\end{equation}
where $\lambda = {\rm{c}}_s^2\alpha {\delta {\rm{t}}}({\tau _{\textrm g}} - 1/2)$.
Combined Eqs.~\eqref{B10-q} and \eqref{B12-q}, we can obtain the CH equation as follows
\begin{equation}\label{CHA-q}
{\partial _t}\phi  + {\bm{\nabla}} \cdot(\phi {\bf{u}}) = \lambda{\nabla ^2}\mu.
\end{equation}
From Eqs.~\eqref{density}, \eqref{DV-q} and \eqref{CHA-q}, we can derive the following mass conservation equation by neglecting the term ${\partial _t}p$, which is of order $\rm Ma^2$
\begin{equation}
{\partial _t}\rho  + {\bm{\nabla}} \cdot(\rho {\bf{u}}) = 0,
\end{equation}
and then the momentum equation in Eq.~\eqref{mass} could be rewritten as
\begin{equation}
 \rho \left( {\frac{{\partial {\bf{u}}}}{{\partial t}} + {\bf{u}} \cdot {\bm{\nabla}} {\bf{u}}} \right) =  - {\bm{\nabla}} p + {\bm{\nabla}} \cdot  \left[ {\rho \nu \left( {{\bm{\nabla}} {\bf{u}} + {\bm{\nabla}} {{\bf{u}}^T}} \right)} \right] + {\bf{F}}.
\end{equation}

{}

\end{document}